\documentclass[10pt]{article}
\usepackage{amsmath}
\usepackage{graphicx}
\usepackage{amsfonts}
\usepackage{amssymb}
\usepackage{epsf}
\usepackage{latexsym}

\def\be{\begin{equation}}
\def\ee{\end{equation}}
\def\bea{\begin{eqnarray}}
\def\eea{\end{eqnarray}}
\def\pa{\partial}

\def\fn{\footnote}
\def\vc{V^{\frac{2}{3}}}

\def\case#1/#2{\textstyle\frac{#1}{#2}}

\begin{document}
\begin{titlepage}

\begin{center}
\Huge
{\bf Leibniz--Mach Foundations for GR and Fundamental Physics}

\normalsize 

\vspace{.1in}

\Large{\bf Edward Anderson}

\vspace{.1in}

{\em  Astronomy Unit, School of Mathematical Sciences, 

Queen Mary, University of London, E1 4NS, U.K. }

\normalsize

\vspace{0.1in}

\begin{abstract}

\noindent 
Consider the configuration space $Q$ for some physical system, and a continuous group of 
transformations G whose action on the configurations is declared to be physically irrelevant.  
G is to be implemented indirectly by adjoining 1  auxiliary g per independent generator to 
$Q$, {\sl by writing the action for the system in an arbitrary G-frame}, and then passing to 
the quotient $Q$/G thanks to the constraints encoded by variation w.r.t the g's.  I show that 
this arbitrary G-frame principle supercedes (and indeed leads to a derivation of) the 
Barbour--Bertotti best matching principle.  I furthermore consider absolute external time to 
be meaningless for the universe as a whole.  For various choices of $Q$ and G, these 
Leibniz--Mach considerations lead to Barbour--Bertotti's proposed absolute-structure-free 
replacement of Newtonian Mechanics, to Gauge Theory, and to the 3-space approach (TSA) 
formulation of General Relativity (GR).  

For the TSA formulation of GR with matter fields, I clarify how the Special Relativity 
postulates emerge, discuss whether the Principle of Equivalence emerges, and study which 
additional simplicity postulates are required.  I further develop my explanation of how a 
full enough set of fundamental matter fields to describe nature can be accommodated in the 
TSA formulation, and further compare the TSA with the `split spacetime formulation' of 
Kucha\v{r}.  I explain the emergence of broken and unbroken Gauge Theories as a consequence 
of the Principle of Equivalence.  

Whereas for GR one would usually quotient out 3-diffeomorphisms, I also consider as further 
examples of the arbitrary G-frame principle the further quotienting out of conformal 
transformations or volume-preserving conformal transformations.  Depending on which choices 
are made, this leads to York's initial value formulation (IVF) of GR, new alternative 
foundations for the GR IVF, or alternative theories of gravity which are built out of similar 
conformal mathematics to the GR IVF and yet admit no GR-like spacetime interpretation.  

\end{abstract}

\end{center}

\section{Introduction: Relational Principles}

Newtonian Physics is based on absolute space and absolute time existing and playing a physical 
role alongside that of the contents of the universe \cite{Scholium}.  Leibniz disapproved 
\cite{Leibniz}.  He saw both space and time not as additional entities but abstractions from 
the contents of the universe: 

\mbox{ }

``I hold space to be something merely relative, as time is; that I hold it to be an order of 
coexistences, as time is an order of successions."  

\mbox{ }  

\noindent It then only makes sense for objects to be somewhere if `somewhere' is defined by other 
objects, and change only makes sense with respect to other change.  This is the 
{\it (spatiotemporal) relationalist} viewpoint.

The absolutist element of Newton's conceptualization, in contrast, led to distinct status 
being given to universes with physically indistinguishable contents.  This contradicted 
Leibniz's ``identity of indiscernibles": 

\mbox{ }

``But in truth the one would exactly be the same thing as the other, they being absolutely 
indiscernible; and consequently there is no room to enquire after a reason of the preference 
of the one to the other"

\vspace{.3in}
\baselineskip=24pt

\end{titlepage}


\noindent which thus clearly rests on Leibniz' ``principle of sufficient reason".  Similar relational thoughts were 
also expressed by Bishop Berkeley \cite{Berkeley} and particularly by Mach \cite{Mach}, who 
emphasized temporal relationalism alongside its spatial counterpart.  Relationalism was 
moreover significantly restricted due to practical reasons: nobody knew how to formulate 
physical laws that implemented it.  

I will exposit and develop a brand of relationalism within a modern context.  I use the 
notion of configuration space $Q$ of a physical system, i.e is the space of all permissible 
instantaneous values that the canonical coordinates 
$q_{\mbox{\sffamily\scriptsize A\normalfont\normalsize}}$ of a system can take.  Here 
{\sffamily A} is a suitably broad multi-index, covering distinguishable particle labels, 
spatial tensor indices and dependence on spatial position.  Motion is a path in configuration 
space parametrized by a time label $\lambda$.

Leibniz's ``identity of indiscernibles" is to be captured by two relational principles.  

\mbox{ }

\noindent {\bf Temporal relationalism (TR)}: there is no meaningful external time label 
for the universe as a whole \cite{BB82, B94I, EOT, BOF}.  

\mbox{ }

\noindent In other words, we have no access to any divine clock outside the universe, 
so what time such a device might keep should not have any relevance to the physics we perceive our universe to have.  

\mbox{ }

\noindent {\bf (A generalization of) spatial relationalism (GSR)}: given the configuration 
space $Q$ of a physical system, one is entitled to declare that a collection of 
transformations G are to have a physically irrelevant effect on $Q$.     

\mbox{ }

\noindent In this work it is required that the G form a group of continuous transformations 
acting on $Q$.  The physical information in a state 
$q_{\mbox{\sffamily\scriptsize A\normalfont\normalsize}}$ is entirely contained in which 
G-orbit 
$$
\mbox{Orb}_{\mbox{\scriptsize G\normalsize}}(q_{\mbox{\scriptsize\sffamily A\normalfont\normalsize}}) 
= \{gq_{\mbox{\scriptsize\sffamily A\normalfont\normalsize}} | g \in \mbox{G} \}
$$
it pertains to.  
The quotient space $\frac{Q}{\mbox{\scriptsize G\normalsize}}$ is the space of orbits, 
i.e the space of possibly\fn{Redundancy may remain as one may subsequently discover that G is 
but a subgroup of a group K of physically irrelevant motions so that distinct G-orbits may no 
longer be distinct K-orbits.} distinguishable states.  I call 
$\frac{Q}{\mbox{\scriptsize G\normalsize}}$ the {\it relative configuration space} (RCS).  
In order for the notion of G-orbit to make physical sense, G requires group structure: 
the inverse property ensures that G-orbits are disjoint, so that a single configuration 
represents one and not several physical states.  Note that $\frac{Q}{\mbox{\scriptsize G\normalsize}}$ 
itself need not inherit a group structure -- quotient spaces are only occasionally quotient 
groups.  Nor need it be a manifold since different orbits can have different dimension.  
Rather it is a peculiar collection of manifolds of different dimensions, called a 
{\it stratified manifold}.  

\mbox{ } 

\noindent {\bf TR} is implemented by\fn{I denote particular 
implementations of these postulates by bold names followed by what they implement in square 
brackets.} {\bf RI[TR]}: the use of manifestly reparametrization-invariant actions.  

\mbox{ }

\noindent By a {\bf RI} action, I mean  
$\int\textrm{d}\lambda \mbox{\sffamily L\normalfont}
\left(
\lambda, x, \frac{\textrm{d}x}{\textrm{d}\lambda}
\right)$ 
with {\sffamily L} homogeneous linear in $x$ so $\lambda \longrightarrow \lambda^{\prime}$ 
sends it to 
$\textrm{d}\lambda F
\left(
\lambda, x, \frac{\textrm{d}x}{\textrm{d}\lambda^{\prime}}\frac{\textrm{d}\lambda^{\prime}}{\textrm{d}\lambda}
\right) 
= 
d\lambda^{\prime}F
\left(
\lambda, x, \frac{\textrm{d} x}{\textrm{d}\lambda^{\prime}}
\right)$
by the homogeneous linearity.

Now, the adoption of homogeneous linear actions has an immediate consequence.  
They invariably lead to relations between the canonical momenta, by the following argument of Dirac 
\cite{Dirac}.  
The canonical momenta must then be homogeneous of degree $0$.  
Thus they are functions of ratios of velocities alone, but there are only $n - 1$ of these, so that there 
is at least one relation between the $n$ momenta.  Such relations, which arise without use of variation, 
are termed {\it primary constraints}.

\mbox{ }

\noindent{\bf GSR} may be implemented {\sl indirectly} using the {\bf arbitrary G-frame principle AF[GSR]}: 
the action for the theory is to be written in arbitrary frame i.e i.t.o the 
`corrected coordinate' 
\be
{q}^{\prime}_{\alpha\mbox{\sffamily\scriptsize A\normalsize\normalfont}} = 
\sum_{\mbox{\sffamily\scriptsize  X \normalfont$\in $ generators of G \normalsize}} 
\stackrel{\longrightarrow}
         {L_{\mbox{\scriptsize a\normalfont}_{\mbox{\tiny\sffamily X\normalfont\normalsize}}}      }
q_{\alpha\mbox{\sffamily\scriptsize A\normalsize\normalfont}}
\ee 
where greeks are manifold-multi-indices, 
the {\sffamily A}-th object has components $q_{\alpha \mbox{\sffamily\scriptsize A\normalsize\normalfont}}$,   
$L_{\mbox{\scriptsize a\normalsize}_{\mbox{\tiny\sffamily X\normalfont\normalsize}}}$ are the transformations of G 
corresponding to generators associated with the auxiliary variables 
$\mbox{a}_{\mbox{\scriptsize\sffamily X\normalfont\normalsize}}$, and the arrow denotes group action.    

\mbox{ }

Often it is easy to deal with the potential, since a fair collection of G-invariant and 
G-covariant objects are usually available to build it from.    
The arbitrary G-frame treatment of velocities is more interesting.  
The plain velocities $\frac{\pa q_{\mbox{\tiny\sffamily A\normalsize\normalfont}}}{\pa\lambda}$ 
are {\sl not} good G-objects: they are not even G-covariant due to the involvement of the frame itself: 
\be
\frac{               \pa {q}_{\mbox{\sffamily\scriptsize A\normalfont\normalsize}}               }
{            \pa\lambda              } =  
\left(
\frac{               \pa {q}_{\mbox{\sffamily\scriptsize A\normalfont\normalsize}}               }
{            \pa\lambda              }
\right)
^{\prime}   
+ \sum_{                \mbox{\sffamily\scriptsize X\normalsize\normalfont}}
\frac{            \pa\stackrel{\longrightarrow}
                                 {{L}_{\mbox{\scriptsize a\normalsize}_{\mbox{\tiny\sffamily X\normalsize\normalfont}}} }              }
{                 \pa\lambda                  }
{q}_{\mbox{\sffamily\scriptsize A\normalfont\normalsize}} = 
\left(
\frac{               \pa {q}_{\mbox{\sffamily\scriptsize A\normalfont\normalsize}}               }
{            \pa\lambda              }
\right)
^{\prime}   
+  
\sum_{                \mbox{\sffamily\scriptsize X\normalsize\normalfont}}
                     \stackrel
                     {\longrightarrow}
                     {{L}_{\dot{\mbox{\scriptsize a\normalsize}}_{\mbox{\tiny\sffamily X\normalfont\normalsize}}} }    
{q}_{\mbox{\sffamily\scriptsize A\normalfont\normalsize}} 
\label{stackedon}
\mbox{ } .
\ee   
Thus use of an arbitrary frame in building the action amounts to requiring to pass from the 
plain velocites $\frac{\pa q_{\mbox{\tiny\sffamily A\normalsize\normalfont}}}{\pa\lambda}$, 
which are now perceived as `bare' or `stacked', to arbitrary G-frame corrected velocities 
$$
\left(
\frac{\pa {q}_{\mbox{\sffamily\scriptsize A\normalfont\normalsize}} }{\pa \lambda}
\right)
^{\prime}
\equiv \&_{\dot{\mbox{\scriptsize a\normalsize}}}
q_{\mbox{\sffamily\scriptsize A\normalsize\normalfont}} 
\mbox{ } ,
$$ 
which are also Barbour and Bertotti's \cite{BB82} so-called {\it best-matched} or 
`unstacked' velocities 
$$
\&_{\dot{\mbox{\scriptsize a\normalsize}}_{\mbox{\tiny\sffamily X\normalsize\normalsize}}}
q_{\mbox{\sffamily\scriptsize A\normalfont\normalsize}} 
\equiv \mbox{\ss}_{{\mbox{\scriptsize b\normalsize}}_{\mbox{\tiny\sffamily X\normalsize\normalsize}}}
q_{\mbox{\sffamily\scriptsize A\normalfont\normalsize}} 
\mbox{ } , 
$$
if one identifies 
$\mbox{b}_{\mbox{\scriptsize\sffamily X\normalfont\normalsize}} = 
\dot{\mbox{a}}_{\mbox{\scriptsize\sffamily X\normalsize\normalsize}}$.  
These velocities permit adjustments w.r.t unphysical auxiliaries; their form may be simply 
deduced by rearranging (\ref{stackedon}).  Thus {\bf AF[GSR]} leads to 

\mbox{ }

\noindent {\bf BM[GSR]}: the best matching implementation whereby the velocities in the 
action are to be corrected by auxiliaries 
$\mbox{b}_{\mbox{\sffamily\scriptsize X\normalsize\normalfont}} 
= \dot{\mbox{a}}_{\mbox{\sffamily\scriptsize X\normalsize\normalfont}}$  corresponding to the 
(infinitesimal) group action of the generators of G.

\mbox{ }

The arbitrary G-frame derivation both {\it proves} that {\bf BM} auxiliaries 
$\mbox{b}_{\mbox{\scriptsize X\normalsize}}$ are velocities 
$\dot{\mbox{a}}_{\mbox{\scriptsize X\normalsize}}$, and identifies these with the G-frame 
velocities.  It is important that they are established to be velocities in order for the 
passage to a arbitrary G-frame action not to spoil the adopted {\bf RI} property of the 
original action.  It is not physically relevant that the frame nature of the auxiliary 
velocities is unveiled, but this is technically convenient for the computation of the form of 
the best matching in each specific example.  

{\bf AF[GSR]} brings in generators which {\sl infinitesimally drag} in all unphysical directions.  
This usually corresponds to keeping one configuration fixed and shuffling a second one around by means of the 
G-transformations so as to cast it into as similar a form as possible to the first one's.  This visualization as dragging 
requires continuity: both Barbour--Bertotti and I consider only G-motions that can be built out of 
infinitesimal motions, rather than discrete motions\fn{Examples of discrete motions include reflections 
and `large' diffeomorphisms; the exclusion of such transformations means that G is being taken 
to be connected.}.
  
Prima facie, the adoption of {\bf AF} leads to actions additionally containing variables in 
one-to-one correspondence with independent generators 
$\mbox{a}_{\mbox{\sffamily\scriptsize X\normalsize\normalfont}}$ of G.  Thus these actions 
are on an {\sl enlarged} configuration space $Q$ $\times$ G.  Thus the G-redundancy of the 
physics has not been removed, but rather doubled!  This reflects that {\bf AF} is an 
{\sl indirect} implementation of {\bf GSR}.  That it is an implementation at all rests on the 
next procedure: variation w.r.t each of these introduced 
$\mbox{a}_{\mbox{\sffamily\scriptsize X\normalfont\normalsize}}$ gives one {\it secondary 
constraint}.  These are linear in the momenta throughout the examples below.  If one can 
sucessfully take these into account, then one passes from the doubly degenerate $Q$ $\times$ 
G to the quotient space $\frac{Q}{\mbox{\scriptsize G\normalsize}}$ which has no G-redundancy.   

\mbox{ }

The standard variational viewpoint in physics would be to regard the 
$\mbox{b}_{\mbox{\sffamily\scriptsize X\normalsize\normalfont}}$, which are usually the 
only {\sl manifest} auxiliaries, as Lagrange multiplier canonical coordinates 
$l_{\mbox{\scriptsize\sffamily X\normalfont\normalsize}}$.  This is undesirable in this 
chapter because {\bf RI} would be violated.  I rather exploit a distinct 
{\it free endpoint} (FEP) variational viewpoint \cite{PD, ABFO} that 
permits the $\mbox{b}_{\mbox{\sffamily\scriptsize X\normalsize\normalfont}}$ to be interpreted as G-frame velocities 
$\dot{\mbox{a}}_{\mbox{\sffamily\scriptsize X\normalsize\normalfont}}$.   
It suffices until Sec 14
to consider the $\mbox{a}_{\mbox{\sffamily\scriptsize X\normalsize\normalfont}}$ to be {\it cyclic} coordinates 
$c_{\mbox{\scriptsize\sffamily X\normalfont\normalsize}}$, i.e coordinates such that 
$\dot{\mbox{c}}_{\mbox{\sffamily\scriptsize X\normalsize\normalfont}}$ occurs in the action but 
$\mbox{c}_{\mbox{\sffamily\scriptsize X\normalsize\normalfont}}$  does not.  The new variational viewpoint 
is that the variations w.r.t 
$c_{\mbox{\scriptsize\sffamily X\normalfont\normalsize}}$ are to be 
permitted to have freely-flapping endpoints.  This is not a cause for concern because the relevant 
canonical coordinates are, after all, {\sl auxiliary}, so the usual physical demands for fixed endpoints become irrelevant.  
Starting from the standard variational expression 
$$
0 = \delta \mbox{\sffamily I\normalfont} = \int \textrm{d}\lambda
\left[
\frac{\pa \mbox{\sffamily L\normalfont}}{\pa c_{\mbox{\scriptsize\sffamily X\normalfont\normalsize}}} 
- \frac{\pa}{\pa \lambda}
\left(
\frac{\pa \mbox{\sffamily L\normalfont}}{\pa \dot{c}_{\mbox{\scriptsize\sffamily X\normalfont\normalsize}}}
\right)
\right]
\delta c_{\mbox{\scriptsize\sffamily X\normalfont\normalsize}} + 
\left[
\frac{    \pa\mbox{\sffamily L\normalfont}    }{    \pa \dot{c}_{\mbox{\scriptsize\sffamily X\normalfont\normalsize}}    }
\delta c_{\mbox{\scriptsize\sffamily X\normalfont\normalsize}}
\right]_{e_1}^{e_2} \mbox{ } \Rightarrow
$$
$$
\left.
\begin{array}{l}
\mbox{ } 
\frac{        \mbox{\normalsize $\pa$\sffamily L\normalfont}        }
     {        \mbox{\normalsize $\pa \dot{c}$\normalsize}_{\mbox{\scriptsize\sffamily X\normalfont\normalsize}}        } 
= p^{\mbox{\scriptsize\sffamily X\normalfont\normalsize}}_{\mbox{\scriptsize c\normalsize}} \mbox{ } ,  \\
\left.
\frac{        \mbox{\normalsize$\pa$\sffamily L\normalfont}         }
     {        \mbox{\normalsize $\pa \dot{c}$\normalsize}_{\mbox{\scriptsize\sffamily X\normalfont\normalsize}}         }
\right|
_{e_1, \mbox{ } e_2} = 0
\end{array}
\right\}  
\Rightarrow
\frac{\pa \mbox{\sffamily L\normalfont} }{\pa c_{\mbox{\scriptsize\sffamily X\normalfont\normalsize}}} = 0 \mbox{ } . 
$$
The first step uses the definition of cyclic coordinate; for point particles 
$p^{\mbox{\scriptsize\sffamily X\normalfont\normalsize}}_{\mbox{\scriptsize c\normalsize}}$ 
is constant, while more generally it is permitted to be a function of position but not of 
label time $\lambda$ in field theory.  The second step uses the freely-flapping endpoints.   
The third step uses the endpoint condition to fix the value of the 
$p^{\mbox{\scriptsize\sffamily X\normalfont\normalsize}}_{\mbox{\scriptsize c\normalsize}}(x)$ 
for all $\lambda$.  

I also note that best matching a {\bf RI} Lagrangian corresponds precisely to the 
Hamiltonian Dirac-appending \cite{Dirac} of constraints with Lagrange multipliers according to
$$
\mbox{\sffamily H\normalfont} = \mbox{\sffamily T\normalfont}\mbox{+}
\mbox{\sffamily V\normalfont} + 
l^{\mbox{\scriptsize\sffamily X\normalfont\normalsize}}
C_{\mbox{\scriptsize\sffamily XA\normalfont\normalsize}}
p^{\mbox{\scriptsize\sffamily A\normalfont\normalsize}} 
\mbox{$\longrightarrow$}
\mbox{\sffamily L\normalfont}
= p^{\mbox{\scriptsize\sffamily A\normalfont\normalsize}}
\dot{q}_{\mbox{\scriptsize\sffamily A\normalfont\normalsize}} 
- [\mbox{\sffamily T\normalfont} +
\mbox{\sffamily V\normalfont}\mbox + 
l^{\mbox{\scriptsize\sffamily X\normalfont\normalsize}}
C_{\mbox{\scriptsize\sffamily XA\normalfont\normalsize}}
p^{\mbox{\scriptsize\sffamily A\normalfont\normalsize}}]
$$
$$
 = 
(\dot{q}_{\mbox{\scriptsize\sffamily A\normalfont\normalsize}} - 
l^{\mbox{\scriptsize\sffamily X\normalfont\normalsize}} 
C_{\mbox{\scriptsize\sffamily AX\normalfont\normalsize}})
p^{\mbox{\scriptsize\sffamily A\normalfont\normalsize}} 
- \mbox{\sffamily T\normalfont}  - \mbox{\sffamily V\normalfont}      
= 
\mbox{\sffamily T\normalfont}
(\&_{\dot{\mbox{\scriptsize c\normalsize}}_{\mbox{\tiny\sffamily X\normalfont\normalsize}} }
q_{\mbox{\scriptsize\sffamily A\normalfont\normalsize}})
- \mbox{\sffamily V\normalfont}(q_{\mbox{\scriptsize\sffamily A\normalfont\normalsize}})
\longrightarrow 
\mbox{\sffamily L\normalfont} 
= 2\sqrt{    \mbox{\sffamily T\normalfont}(\mbox{\ss}_{\dot{\mbox{\scriptsize c\normalsize}}_{\mbox{\tiny\sffamily X\normalfont\normalsize}} }
\dot{q}_{\mbox{\scriptsize\sffamily A\normalfont\normalsize}})
\mbox{\sffamily V\normalfont}(q_{\mbox{\scriptsize\sffamily A\normalfont\normalsize}})    }
$$
for the common case below of an action homogeneous-quadratic in its velocities.  
The first three steps in this working form a Legendre transform while the fourth step 
is a Jacobi-type passage \cite{Lanczos} to a {\bf RI} form.  

Note that in addition to the primary constraint(s) from {\bf RI[TR]} and the secondary 
constraints from {\bf AF[GSR]}, which I denote collectively by 
${\cal C}_{\mbox{\scriptsize\sffamily S\normalfont\normalsize}_1}$, there may also be 
additional secondary constraints resulting from applying the Dirac procedure \cite{Dirac} to 
the set of all these constraints.  That is, the evolution equations may produce additional 
(functionally-independent) constraints 
${\cal C}_{\mbox{\scriptsize\sffamily S\normalfont\normalsize}_2}$ required in order for the 
${\cal C}_{\mbox{\scriptsize\sffamily S\normalfont\normalsize}_1}$ to continue to hold along 
the configuration space curve away from initial value of $\lambda$.  One can take a `discover 
and encode' attitude to this: standardly this would involve appending the discovered 
constraints using new constraint-encoding multipliers and thus build toward a `total 
Hamiltonian' \cite{Dirac}.  In the arbitrary G-frame viewpoint, the encoding would likewise 
involve introducing some form of new auxiliaries; furthermore as these are new generators, 
this means that consistency is {\it enforcing} G to be more extensive than hitherto assumed.  
The choice of G is not necessarily free!  The Dirac procedure is furthermore recursive: the 
evolution of the ${\cal C}_{\mbox{\scriptsize\sffamily S\normalfont\normalsize}_2}$ might 
similarly give new ${\cal C}_{\mbox{\scriptsize\sffamily S\normalfont\normalsize}_3}$  and so 
on.  But if the system has just a few d.o.f's and the Dirac procedure does not quickly 
terminate, the d.o.f's will be used up and the action will be demonstrated to be inconsistent, 
or at least to have a greatly undersized solution space.  The Dirac procedure therefore lends 
itself to proof by exhaustion (see e.g \cite{AB, Thanderson}).  

In Secs 2-4, I consider point particle mechanics, Gauge Theory and the 
Baierlein--Sharp--Wheeler geometrodynamical formulation of GR \cite{BSW, MTW} as examples of relational theories.  In 
Sec 5 I explain how this last formulation arises exhaustively as one of a few possibilities 
following from the `3-space approach' (TSA) relational principles \cite{BOF, Vanderson} 
{\sl making use of no prior knowledge of GR}.  In Sec 6 I begin to explain how this may be 
extended to include fundamental matter \cite{BOF, BOF2, AB, Thanderson}.  This inclusion of 
matter makes it possible to explain how the conventional Special Relativity principles arise 
in the TSA (Sec 7).  I then relate \cite{Vanderson} the TSA to the `split spacetime framework' 
(SSF) \cite{KucharI, KucharII, KucharIII, KucharIV} of geometrodynamics with matter in Sec 8.  
The SSF usually involves assuming more structure (GR spacetime structure) than is assumed in 
the TSA, but I use the SSF to show that the TSA by itself has enough structure to include all 
the usual fundamental bosonic fields (Sec 9) and fermionic fields (Sec 10).  Thus the TSA 
permits accommodation of the full standard set of fundamental matter fields that describe 
nature as we know it.  In Sec 11 I moreover present further evidence that, contrary to 
previous suggestions, the TSA does not particularly pick out this full standard set.  In Sec 
12 I correlate what the TSA does and does not pick out with the possible emergence of the 
Principle of Equivalence.  Sec 13 briefly suggests an alternative viewpoint on the origen of 
Gauge Theory.  In Sec 14 I give final examples of relational theories: geometrodynamics in 
which conformal transformations are additionally held to be irrelevant \cite{conformal, ABFO, ABFKO}.  
These are closely tied to the GR initial value formulation (IVF) \cite{CBY}, and include both new approaches 
to the GR IVF and new alternative theories of gravity based on conformal mathematics similar to that of the IVF.  
I conclude in Sec 15.

Appendix A contains what I mean by all the principles and simplicities used in the standard approach to GR.  
Appendix B gives the standard geometrodynamical split of GR.  Appendix C interprets this, and instigates the 
search for underlying first principles for it.  Appendix D contains the standard IVF.

\section{Example 1: Absolute or Relative Motion in Particle Mechanics} 

The $(n)$-particle configuration space is the 3$n$-dimensional 
\be
Q = 
\left\{
\mbox{\b{q}}_{(i)} | (i) = (1) \mbox{ to } (n) 
\right\} 
\mbox{ } .
\ee
The choice of 
\be
G_{\mbox{\scriptsize N\normalsize}} = \mbox{ id }
\ee
gives absolute Newtonian Mechanics with configuration space Q, while the choice 
\be
G_{\mbox{\scriptsize L\normalsize}} = \mbox{Eucl} \equiv \{\mbox{Translations and Rotations}\} 
\ee
gives some `Leibnizian mechanics'.  The configuration space for this is the (3$n$ - 6)-d RCS 
\cite{BB82, B94I, EOT, Gergely}
$\frac{Q}{\mbox{\scriptsize Eucl\normalsize}}$.  For example, for the unit mass 3-body problem, 
it is \it Triangle Land \normalfont, the space of all triangular shapes with one 
particle at each vertex.   Triangle Land is a simple example of stratified manifold, 
constituting of tetrahaedron along with 3 faces; 3 edges, and the vertex at which they meet.  
The lower-d strata correspond to highly symmetric situations: the faces are colinear configurations, the edges double collisions, 
while the vertex is the infamous triple collision.
\begin{figure}[h]
\centerline{\def\epsfsize#1#2{0.5#1}\epsffile{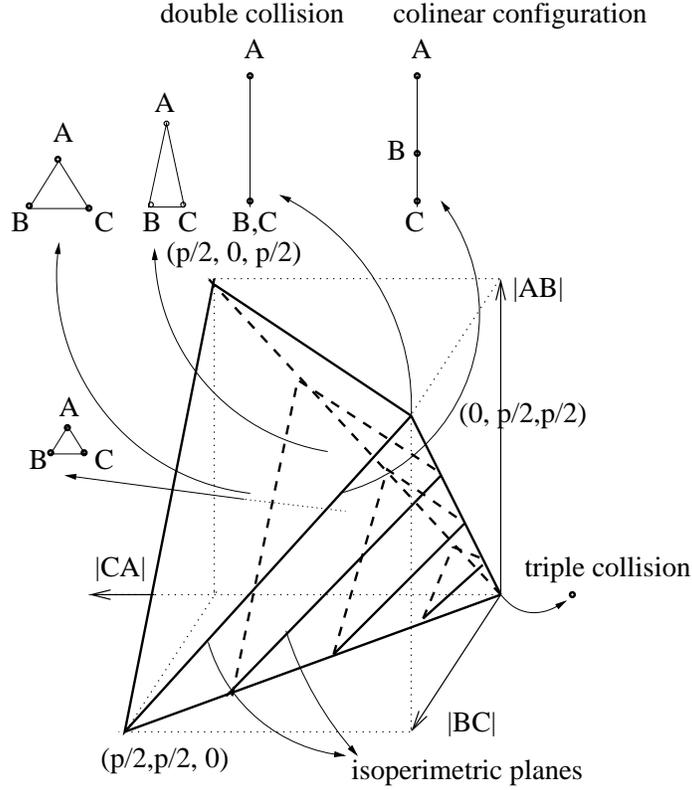}}
\caption[]{\label{TO7.ps}
\scriptsize The stratified manifold {\it Triangle land}: 
a representation of the relative configuration space for the Newtonian 3-body 
problem. 
\normalsize}
\end{figure}

Thus the handling of the choice of redundant group represents the classical absolute or 
relative motion debate.  In Ehlers' study \cite{Ehlers} of miscellaneous idealized notions of 
space, time and spacetime, he pointed out that Leibnizian spacetime was disregarded because 
the relationalists had not been able to provide physical laws which realized their ideals.  
Moreover, unlike e.g in Cartan's spacetime \cite{Cartansp} or GR, Leibnizian spacetime did not 
seem to have enough structure. 

Choosing $\mbox{G}_{\mbox{\scriptsize N\normalsize}}$ gives absolute Newtonian Mechanics as 
follows. For homogeneous quadratic mechanics with time-independent potential, 
the usual action is equivalent to the {\bf RI} Jacobi action 
\be
\mbox{\sffamily I\normalfont}_{\mbox{\scriptsize J\normalsize}} 
= \int \textrm{d}\lambda 2\sqrt{(\mbox{\sffamily E\normalfont} - \mbox{\sffamily V\normalfont}) 
\mbox{\sffamily T\normalfont}     }
\mbox{ } , 
\label{Jacky}
\ee
\be
 \mbox{\sffamily T\normalfont} = M^{\mbox{\scriptsize\sffamily AB\normalfont\normalsize}}
\dot{x}_{\mbox{\scriptsize\sffamily A\normalfont\normalsize}}
\dot{x}_{\mbox{\scriptsize\sffamily A\normalfont\normalsize}}
\ee
as follows.  Adjoin $t$ to the canonical coordinates via parametrization i.t.o $\lambda$.  Then 
note that $\dot{t} = \frac{\pa t}{\pa\lambda}$ alone occurs in the action.  Thus $t$ is a 
cyclic coordinate
\be
\frac{\pa \mbox{\sffamily L\normalfont}}{\pa\dot{t}} = p^t = -\mbox{\sffamily E\normalfont} 
\mbox{ } , 
\mbox{ const} 
\Rightarrow 
\dot{t} = \sqrt{\frac{\mbox{\sffamily T\normalfont}}{\mbox{\sffamily E\normalfont} - \mbox{\sffamily V\normalfont}}} 
\mbox{ } ,
\label{eottf}
\ee
and $\dot{t}$ is eliminable from the action by {\it Routhian reduction} 
[i.e using (\ref{eottf}) in $\mbox{\sffamily L\normalfont}(
q_{\mbox{\sffamily\scriptsize A\normalsize\normalfont}}, 
\dot{q}_{\mbox{\sffamily\scriptsize A\normalsize\normalfont}}\dot{t)} - p^{t}\dot{t}$] \cite{Lanczos} 
to form $I_{\mbox{\scriptsize J\normalsize}}$. 

Choosing $G_{\mbox{\scriptsize L\normalsize}}$ = Eucl and applying the arbitrary Eucl-frame 
principle\fn{One could also attempt to implement `Leibnizian mechanics' directly.  See \cite{BB77} for this and its 
lack of success.  The indirect style here is that of the Barbour--Bertotti 1982 paper (BB82) \cite{BB82}.  These works 
have been criticized in \cite{Earman}, while supportive arguments have appeared in \cite{PooleyBrown}.} 
to the Jacobi action (\ref{Jacky}) yields the BB82 relational point particle mechanics.
Writing the action in the arbitrary Eucl-frame entails use of 
\be
\mbox{\b{q}}_{(i)}^{\prime} 
= \stackrel{\longrightarrow}{{E}_{\mbox{\scriptsize l\normalsize}, \Theta}}\mbox{\b{q}}_{(i)}  
\mbox{ } , 
\ee 
for $q_{(i)k}$ the components of the $i$th particle and  
$E_{\mbox{\scriptsize l\normalsize}, \Theta}$ the coordinate transformations of Eucl.  The potential should then be built 
to be Eucl-invariant from the start, i.e translation- and rotation-invariant.    On the other hand,
\be
\frac{\pa \mbox{\b{q}}_{(i)}}{\pa\lambda} =  
\left(
\frac{\pa\mbox{\b{q}}_{(i)}}{\pa\lambda}
\right)
^{\prime}   +  \frac{\pa \stackrel{\longrightarrow}
                                  {E_{      \mbox{\scriptsize l\normalsize}, \Theta} }     }
                   {      \pa\lambda      }
\mbox{\b{q}}_{(i)}     = 
\left(
\frac{  \pa\mbox{\b{q}}_{(i)}  }{  \pa\lambda  }
\right)
^{\prime}   +  
\stackrel{  \longrightarrow  }{  E_{\dot{\mbox{\scriptsize l\normalsize}}, \dot{\Theta}}  }
\mbox{\b{q}}_{(i)} 
\mbox{ } .
\ee 
So one should use not $\frac{\pa q}{\pa\lambda}$ but
$\left(
\frac{    \pa\mbox{\scriptsize\b{q}\normalsize}_{\mbox{\tiny(i)\normalsize}}    }
     {    \pa\lambda}
\right)
^{\prime}$ which is the Eucl-{\bf AF} velocity
$\&_{\dot{\mbox{\scriptsize l\normalsize}}, \dot{\Theta}}\mbox{\b{q}}_{(i)} 
\equiv
\dot{\mbox{\b{q}}}_{(j)} - \dot{\mbox{\b{l}}} - 
\mbox{\b{$\dot{\Theta}$} \scriptsize $\times$ \normalsize \b{q}}_{(j)}$ (or equivalently the 
Eucl-{\bf BM} velocity
$\mbox{\ss}_{\mbox{\scriptsize v\normalsize}, \Omega}\mbox{\b{q}}_{(i)} 
\equiv \dot{\mbox{\b{q}}}_{(j)} - \mbox{\b{v}} - 
\mbox{\b{$\Omega$} \scriptsize $\times$ \normalsize \b{q}}_{(j)} 
\equiv $ for $\mbox{\b{v}} = \dot{\mbox{\b{l}}}$ and 
$\mbox{\b{$\Omega$}} = \dot{\mbox{\b{$\Theta$}}}$).  
This amounts to bringing in Eucl-generators which infinitesimally drag in all unphysical 
directions (translations and rotations of the whole universe).  It is a derivation of BB82's 
{\bf BM} for point particles.  This corresponds to comparing each pair of particle 
configurations $C_1$, $C_2$ by w.l.o.g keeping $C_1$ fixed and shuffling $C_2$ around by means 
of translations and rotations as a means of casting it into as similar a form as possible 
to the first one.  The corrections are Lie derivatives corresponding to draggings in the 
unphysical directions.  Thus one can construct a Leibnizian theory by use of Lie derivatives, 
which indeed require no additional structure on the flat manifold of space.  
\begin{figure}[h]
\centerline{\def\epsfsize#1#2{0.4#1}\epsffile{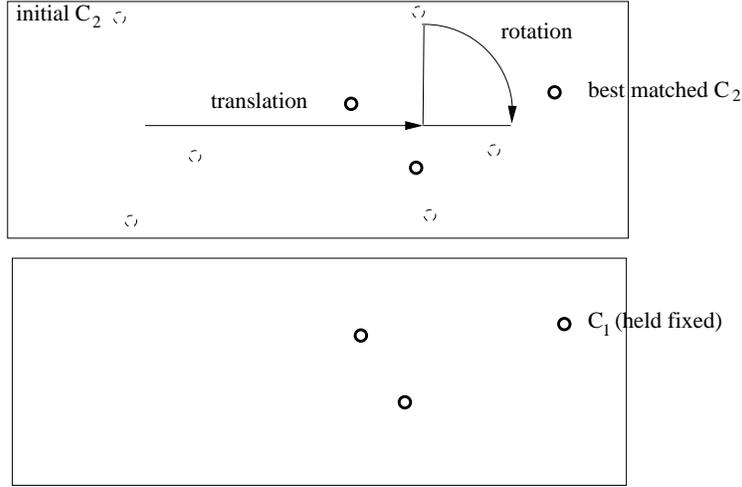}}
\caption[]{\label{TO7.ps}
\scriptsize Best matching of point particle configurations.
\normalsize}
\end{figure}

Thus the BB82 action may be written as 
\be
\mbox{\sffamily I\normalfont}
= 2\int \textrm{d} \lambda \sqrt{     [  \mbox{\sffamily E\normalfont}     -     
                                         \mbox{\sffamily V\normalfont}(\mbox{\b{q}}_{(k)})  ]
\mbox{\sffamily T\normalfont}(
{\&}_{\dot{\mbox{\scriptsize l\normalfont}}, \dot{\Theta}} q_{(k)})    }   
\mbox{ } , \mbox{ } \mbox{ } 
\mbox{\sffamily T\normalfont} = \sum_{(j) = (1)}^{(n)} \frac{  m_{(j)}  }{  2  }
{\&}_{\dot{\mbox{\scriptsize l\normalsize}}, \dot{\Theta}}\mbox{\b{q}}_{(i)}  
\cdot
{\&}_{\dot{\mbox{\scriptsize l\normalsize}}, \dot{\Theta}}{\mbox{\b{q}}}_{(j)}        
\mbox{ } . 
\ee
The particle momenta are 
\be
\mbox{\b{p}}^{(i)} \equiv \frac{\pa \mbox{\sffamily L\normalfont}}{\pa \dot{\mbox{\b{q}}}_{(i)}} =  
\sqrt{      \frac{ \mbox{\sffamily E\normalfont}  -     \mbox{\sffamily V\normalfont}   } 
{   \mbox{\sffamily T\normalfont}   }      }m_{(i)}
{\&}_{\dot{\mbox{\scriptsize l\normalsize}}, \dot{\Theta}}{\mbox{\b{q}}}^{(i)} 
\mbox{ } . 
\label{BB82mom}
\ee
Now, this square root Lagrangian is clearly homogeneous of degree $1$ in the velocities, so 
there will be at least one primary constraint.  In this case, the square root action 
gives rise to precisely one, 
\be 
{\cal P} \equiv \sum_{(i) = (1)}^{(n)}  \frac {\mbox{\b{p} \normalfont  }^{(i)} \cdot  \mbox{\b{p} \normalfont  }^{(i)}} {2m_{(i)}} -(\mbox{\sffamily E\normalfont} 
- \mbox{\sffamily V\normalfont}) = 0 \mbox{ } ,    
\label{BB82prim}
\ee 
by the following form of Pythagoras' theorem: 
$$ 
\sum_{(i) = (1)}^{(n)}  \frac {\mbox{\b{p} \normalfont  }^{(i)} \cdot \mbox{\b{p} \normalfont  }^{(i)}} 
{m_{(i)}} = \sum_{(i) = (1)}^{(n)}  \frac{1}{m_{(i)}}
\left(
\sqrt{\frac{\mbox{\sffamily E\normalfont} - \mbox{\sffamily V\normalfont}}{\mbox{\sffamily T\normalfont}}} 
m_{(i)}{\&}_{\dot{\mbox{\scriptsize l\normalsize}}, \dot{\Theta}}{\mbox{\b{q}}}_{(i)} 
\right)
\cdot
\left(
\sqrt{\frac{\mbox{\sffamily E\normalfont} - \mbox{\sffamily V\normalfont}}{\mbox{\sffamily T\normalfont}}}
m_{(i)}{\&}_{\dot{\mbox{\scriptsize l\normalsize}}, \dot{\Theta}}{\mbox{\b{q}}}_{(i)} 
\right) 
$$
$$
= \frac{\mbox{\sffamily E\normalfont} - \mbox{\sffamily V\normalfont}}{\mbox{\sffamily T\normalfont}} 
\sum_{(i) = (1)}^{(n)} m_{(i)} 
{\&}_{\dot{\mbox{\scriptsize l\normalsize}}, \dot{\Theta}}{\mbox{\b{q}}}_{(i)} 
\cdot
{\&}_{\dot{\mbox{\scriptsize l\normalsize}}, \dot{\Theta}}{\mbox{\b{q}}}_{(i)}  
=\frac{\mbox{\sffamily E\normalfont} - \mbox{\sffamily V\normalfont}}{\mbox{\sffamily T\normalfont}}2\mbox{\sffamily T\normalfont} = 2(\mbox{\sffamily E\normalfont} - 
\mbox{\sffamily V\normalfont}) 
\mbox{ } . 
$$ 

FEP variation w.r.t the auxiliaries \b{l} and \b{$\Theta$}, 
or equivalently, standard variation w.r.t \b{v} and \b{$\Omega$}, one obtains respectively that the total 
momentum and angular momentum of the whole $n$-particle universe must be zero:  
\be
\mbox{\b{${\cal M}$}} \equiv \sum_{(i) = (1)}^{(n)}\mbox{\b{p}}^{(i)} = 0 
\mbox{ } , \mbox{ } \mbox{ } 
\mbox{\b{${\cal L}$}} \equiv \sum_{(i) = (1)}^{(n)}\mbox{\b{q}}_{(i)} \mbox{\scriptsize $\times$\normalsize} \mbox{ } 
\mbox{\b{p}}^{(i)} = 0 \mbox{ } .
\label{BBcons}
\ee
So both the {\bf AF[SR]} and {\bf RI[TR]} implementations lead to constraints.  

The particle ELE's are 
\be
{\&}_{\dot{\mbox{\scriptsize l\normalsize}}, \dot{\Theta}}\mbox{\b{p}}^{(i)}
= \sqrt{      \frac{    \mbox{\sffamily T\normalfont}     }{    \mbox{\sffamily V\normalfont}    }      }\frac{\pa \mbox{\sffamily V\normalfont}}{    \pa \mbox{\b{q}}_{(i)}    }
\mbox{ } . 
\label{BB82EL}
\ee
Coupling (\ref{BB82mom}) and (\ref{BB82EL}), if one picks the unique distinguished 
choice of label time such that 
$\sqrt{\mbox{\sffamily T\normalfont}} = \sqrt{\mbox{\sffamily V\normalfont}}$, one recovers 
Newton's second Law.  This choice corresponds to the total energy of the universe 
also being zero.  By Dirac's procedure, the ELE's ensure that there are no more constraints.

\section{Example 2: Gauge Theory}

\subsection{Electromagnetism}

The configuration space is the space of 1-forms 
\be
Q = \{A_i(x)\} 
\mbox{ } , 
\ee
which has 3 d.o.f's per space point.  The redundant motions are
\be
A_i \longrightarrow A_i - \pa_i\Lambda \mbox{  } ,
\ee
which form the {\sl gauge group}, which is here  
\be
\mbox{G} = \mbox{U}(1) \mbox{ } .
\ee
This corresponds to an {\sl internal} symmetry of the configurations.  
The RCS is then 
\be
\mbox{\sffamily RCS\normalfont}({\mbox{em}}) = \frac{\{A^i(x)\}}{\mbox{U(1)}} \mbox{ } ,
\ee
the gauge-invariant 1-forms.  Note that $A_{i} = A_{i}^{\prime}$ implies that 
$\Lambda$ is constant, regardless of the form of $A_i$, so there is a single stratum.  

The {\bf AF} implementation of {\bf GSR} is to consider gauge invariant potentials such as 
the `Maxwellian curl combination', and furthermore to deduce the U(1)-{\bf BM} corrections 
to the 1-form velocities.  These arise because 
\be
\frac{         \pa(A^{i\prime})        }
     {         \pa\lambda              } = \frac{    \pa(A^i - \pa^i\Lambda)      }{\pa\lambda} = 
\left(
\frac{\pa A^i}{\pa\lambda}
\right)
^{\prime} - \pa^i\dot{\Lambda} \mbox{ } , 
\ee
so one should use not $
\left(
\frac{\pa A^i}{\pa\lambda}
\right) 
$
but 
$
\left(
\frac{\pa A^i}{\pa\lambda}
\right)
^{\prime}$ which is the U(1)-{\bf AF} velocity $\&_{\dot{\Lambda}}A^i$ (or equivalently the 
U(1)-{\bf BM} velocity  $\mbox{\ss}_{\Phi}A^i$ for $\Phi = \dot{\Lambda}$). 
The latter interpretation is a derivation of standard gauge theory.  As a best matching, 
its interpretation is that for any 2 1-form fields on a flat space, one w.l.o.g keeps one 
in a fixed gauge and changes the gauge of the other as generated by $\Phi$ until the 2 
1-form fields are as close as possible.

N.B so far this is an alternative approach to standard electromagnetism based on 
an {\sl internal} part of {\bf generalized spatial relationalism}. One could attempt futhermore to 
have {\bf temporal} and/or bona fide {\bf spatial relationalism}.  
Using a {\bf RI} action additionally imposes a fragment condition: a 
constant energy condition \cite{BB82}.  I first use the global square root 
reparametrization-invariant implementation {\bf RI[TR] } 
\be
\mbox{\sffamily I\normalfont} = \int \textrm{d} \lambda 
\sqrt{                
\left(
{\mbox{\sffamily E\normalfont}}     -    \int\textrm{d}^3x         
F_{pq}
F^{pq}
\right)
\int \textrm{d}^3x
{\&}_{\dot{\Lambda}}
A^i
{\&}_{\dot{\Lambda}}
A_i
                      } \mbox{ } , 
\ee
for $F_{pq} \equiv \pa_qA_p - \pa_pA_q$.  
Defining 
$2N \equiv \sqrt{    \frac{  \mbox{\sffamily\scriptsize E\normalsize\normalfont}    
                             - \int\mbox{\scriptsize d\normalsize}^3xF_{pq}F^{pq} }
                          {  \int\mbox{\scriptsize d\normalsize}^3x
                             \mbox{\scriptsize\&\normalsize}_{\mbox{\tiny$\dot{\Lambda}$\normalsize}}A^i
                             \mbox{\scriptsize\&\normalsize}_{\mbox{\tiny$\dot{\Lambda}$\normalsize}}A_i  }    }$, 
the field momenta are $\pi^i = 2N{\&}_{\dot{\Lambda}}A^i.$
The global square root gives a single primary constraint:
\be
{\cal P} \equiv \int \textrm{d}^3x
\left(
\pi^i\pi_i +   
F_{pq}
F^{pq}
\right)
= \int{\textrm{d}^3x}\mbox{\sffamily E\normalsize} \mbox{ } . 
\ee

Variation w.r.t the auxiliaries $\Lambda(x)$ 
yields the Gauss constraint 
\be 
\pa_i\pi^i = 0 
\mbox{ } , 
\label{ymaux1}
\ee

The ELE's are 
$
\dot{\pi}_{i} 
= \pa_j
(2N F^{ij}) = 0 
$.
These guarantee the propagation of the constraints.  Now, it is required for the time-label 
$\lambda$ to be such that
$$
{\mbox{\sffamily E\normalfont}}  =  \int \textrm{d}^3x
\left(  
{      F_{pq}F^{pq}      }      +
     {      
{\&}_{\dot{\Lambda}}A_i
{\&}_{\dot{\Lambda}}A^i  
     }
\right) \mbox{ } ,
$$
in order to recover the Maxwell equations.  

\mbox{ }

I next consider the local square root ordering.
\be
\mbox{\sffamily I\normalfont} = \int \textrm{d} \lambda \int \textrm{d}^3x   ({\mbox{\sffamily E\normalfont}}      
- F_{pq}F^{pq})
{\&}_{\dot{\Lambda}}A_i
{\&}_{\dot{\Lambda}}A^i
\mbox{ } .
\ee
Defining 
$2N = \sqrt{     \frac{ \mbox{\scriptsize\sffamily E\normalfont\normalsize} - F_{pq}F^{pq} }
                      { \mbox{\scriptsize\&\normalsize}_{\mbox{\tiny$\dot{\Lambda}$\normalsize}}A^i
                        \mbox{\scriptsize\&\normalsize}_{\mbox{\tiny$\dot{\Lambda}$\normalsize}}A_i }   }$, 
the field momenta are now $\pi^i = 2N{\&}_{\dot{\Lambda}}A^i$.
There is now one primary constraint per space point, 
$
{\cal P}(x_k) \equiv \pi_{i}\pi^{i} 
+ F_{pq}F^{pq}
= \mbox{\sffamily E\normalfont}
$
due to the local square root. 
As above, variation w.r.t the auxiliary variables yields the Gauss constraint.   

The ELE's are now
$
\dot{\pi}^b = \pa_a\left(
2N                 
\pa^{[a}A^{b]}
\right) 
$.
These propagate by standard energy--momentum conservation: $\dot{{\cal P}}$ gives  
$$
0 = (\mbox{Poynting vector})_p \equiv (\mbox{\b{E}} \mbox{ } \mbox{\scriptsize $\times$ \normalsize} \mbox{\b{B}})_p 
= \pi^{q}F_{qp} 
$$
as a secondary, whose propagation involves no new constraints.  It is true that now this is locally 
rather than globally restrictive and thus only gives a small fragment of conventional electromagnetism.  
One can remove absolute space at the cost of more fragmentation (zero momentum and angular momentum).

\subsection{Comments on Yang--Mills theory}

The configuration space is now the space of $N$ 1-forms 
\be
Q = \{A^i_{\mbox{\bf\scriptsize I\normalsize\normalfont}}(x)\} \mbox{ } ,
\ee
which has $3N$ d.o.f's per space point.  The redundant motions are more complicated: 
\be
A^{\mbox{\bf\scriptsize I\normalsize\normalfont}}_{i} \longrightarrow 
A^{\mbox{\bf\scriptsize I\normalsize\normalfont}}_{i}  
- \pa_i\dot{\Lambda}^{\mbox{\bf\scriptsize I\normalsize\normalfont}}+
i\mbox{\sffamily g\normalfont}
{f^{\mbox{\bf\scriptsize I\normalsize\normalfont}}}_{\mbox{\bf\scriptsize JK\normalsize\normalfont}}
\dot{\Lambda}^{\mbox{\bf\scriptsize J\normalsize\normalfont}}
A^{\mbox{\bf\scriptsize K\normalsize\normalfont}}_{A} \mbox{ } ,
\ee
and form some internal gauge group G which is a particular kind of Lie group with structure constants 
${f^{\mbox{\bf\scriptsize I\normalsize\normalfont}}}_{\mbox{\bf\scriptsize JK\normalsize\normalfont}}$.   
Examples of this are the SU(2) of the weak force, 
SU(2) $\times$ U(1) of electroweak unification, SU(3) of the strong force, 
\noindent SU(3) $\times$ SU(2) $\times$ U(1) of the 
Standard model, or the SU(5), S0(10) or exceptional groups underlying speculations of grand unification.  
When one has a direct product group, one is permitted a distinct coupling constant {\sffamily g} per 
constituent group.  The RCS is then 
\be
\mbox{\sffamily RCS\normalfont}(\mbox{YM}) 
= \frac{A^i_{\mbox{\bf\scriptsize I\normalsize\normalfont}}(x)}{\mbox{G}} 
\mbox{ } , 
\ee
which does now have $A_i^{\mbox{\bf\scriptsize I\normalsize\normalfont}}$-dependent 
stratification since 
$A_i^{\mbox{\bf\scriptsize I\normalsize\normalfont}} = 
 A_i^{\mbox{\bf\scriptsize I\normalsize\normalfont}\prime}$ 
implies that $\dot{\Lambda}^{\mbox{\bf\scriptsize I\normalsize\normalfont}}$ satisfies 
$\pa_i\dot{\Lambda}^{\mbox{\bf\scriptsize I\normalsize\normalfont}} = i\mbox{\sffamily g\normalfont}
{f^{\mbox{\bf\scriptsize I\normalsize\normalfont}}}_{\mbox{\bf\scriptsize JK\normalsize\normalfont}}
\Lambda^{\mbox{\bf\scriptsize J\normalsize\normalfont}}
A^{\mbox{\bf\scriptsize K\normalsize\normalfont}}_{A}$.

The {\bf AF} implementation of {\bf GSR} and the emergent Yang--Mills G-{\bf BM} are 
directly analogous to the above working and interpretation for electromagnetism.  The 
fragment theories obtained from the analogous local and global {\bf RI} actions also follow 
suit.

\subsection{The Standard Approach to Gauge Theory}

It is worth mentioning that the traditional route of arriving at Gauge Theory is to start in 
flat spacetime with a complex scalar $\varsigma$ or fermion $\psi$ with a global G-symmetry.   
Then two choices and one argument are involved that are relevant to this chapter.  First, 
there is a choice whether this G-symmetry should be promoted to being a local G-symmetry, 
which exemplifies the theoretician's choice of what is to be the group of irrelevant 
transformations in nature.  The argument is that having chosen to have a local G-symmetry, 
1-forms then appear in order for the derivatives of $\varsigma$ or $\psi$ to be good 
G-objects, and that these 1-forms should be themselves dynamical.  Then simultaneous 
imposition of G-symmetry, the flat spacetime's Lorentz symmetry and parity-symmetry enforces 
electromagnetism for a single 1-form, or, alongside the first order in derivatives and 
interaction-limiting na\"{\i}ve renormalizability simplicities\fn{In formulating physics, 
one will enconter many {\it simplicities}, albeit there is a tendancy for these to be tacit 
or underplayed.  These are mathematical statements, and may be subjective.  In this chapter 
I point out a number of simplicities in the TSA formulation of GR with matter `added on'.  
This state of affairs should be compared with this subsection and with the standard 
formulation of GR in App A.} (no more than 4 1-form fields interacting at a point), 
Yang--Mills theory \cite{Weinberg}.   The second choice is whether then to break the 
G-symmetry, e.g so that massive 1-forms such as weak bosons may be described.  I will 
present a rather different perspective on Gauge Theory and broken Gauge Theory in Sec 13.

\section{Example 3: Vacuum General Relativity}

The configuration space is (see also App B, C)
\be
Q = \mbox{Riem} = \{\mbox{Riemannian 3-metrics $h_{ij}$ on M}\} 
\mbox{ } , 
\ee
where M is some manifold of fixed topology, taken here to be compact without boundary (CWB).  
It has 6 d.o.f's per space point.  

The redundant motions are the 3-coordinate transformations given by
\be
h_{ij} \longrightarrow h_{ij} - \pounds_{s}h_{ij} = h_{ij} - 2D_{(i}s_{j)} \mbox{ } . 
\label{iluv}
\ee
$\pounds_s$ is the {\it Lie derivative} w.r.t $s_i$, which is based on solid 
{\sl dragging first principles}.  See \cite{Stewart} for what 
these are for scalars and vectors, and how these lead to the computational formulae
\be
\pounds_{s}\psi = s^i\pa_i\psi 
\mbox{ } , \mbox{ } 
\pounds_{s}V^a = s^i\pa_iV^a - V^i\pa_is^a
\mbox{ }.  
\ee
Assuming that the manifold is affine, one can clearly furthermore write these i.t.o covariant 
derivatives $D_i$, 
as
\be
\pounds_{s}\psi = s^iD_i\psi 
\mbox{ } , \mbox{ } 
\pounds_{s}V^a = s^iD_iV^a - V^iD_is^a
\mbox{ }.  
\ee
For other tensors, the Leibniz rule can be used to deduce the form of the Lie derivative 
starting either form of the above two Lie derivatives.  Note that the exhibited form of the 
metric Lie derivative is somewhat special, due to $D_kh_{ij} = 0$. 

The 3-coordinate transformations $\ref{iluv}$ form the group   
\be
\mbox{G} = \mbox{Diff} \equiv \{\mbox{3-diffeomorphisms}\} 
\mbox{ }. 
\ee
$h_{ij}$ contains information both about the 3-geometry ${\cal G}$ (shape, including scale) 
and about the coordinate grid one might choose to paint on that shape.  Quotienting out the diffeomorphisms removes the 
grid, leaving Wheeler's \cite{Wheeler} RCS
\be
\mbox{RCS}({\cal G}) \equiv \mbox{Superspace} = \frac{\mbox{Riem}}{\mbox{Diff}} 
\mbox{ } .
\ee

Fischer \cite{Fischer70} studied the stucture of Superspace; it is a considerably more complicated example of stratified 
manifold than Triangle Land!  Here is a simple demonstration that different strata indeed exist and why.   
In quotienting out 3-diffeomorphisms, $h_{ij} = h_{ij}^{\prime}$ is clearly relevant.  But this implies 
$D_{(i}s_{j)} = 0$ i.e the Killing equation whose solutions are the Killing vectors associated with the symmetries of the 
metric.  Thus different 3-metrics have Diff-orbits of different dimension depending on what symmetries (isometries) they 
possess.  For example, dim(Isom($\delta_{ij}$)) = dim(Eucl) = 6, while dim(Isom(generic $h_{ij}$)) = 0.  The 
stratification is furthermore related to Superspace not being geodesically-complete 
\cite{DeWitt67, DeWitt70}.  DeWitt proposed 
resolving this by continuation by reflection off the strata \cite{DeWitt70} while Fischer proposed a nonsingular extended 
space built using the theory of fibre bundles \cite{Fischer86}.  Spaces such as Triangle Land may serve as useful toys in 
exploring these proposals, and more generally to gain intuition about the nature of the gravitational configuration space 
\cite{DeWitt70, EOT}.  Another difficulty with Superspace follows from how quotienting out the 
3-diffeomorphisms only goes part of the way toward isolating a representation of the true dynamical d.o.f's of GR 
(see below).  This means that one is concerned not with single trajectories on Superspace but rather with families of them 
(sheaves) \cite{DeWitt70}.   

To implement {\bf TR}, one can choose the Baierlein--Sharp--Wheeler (BSW) \cite{BSW} 
{\bf RI} action 
\be
\mbox{\sffamily I\normalfont} = \int \textrm{d}t \int \textrm{d}^3x \sqrt{h} 
\sqrt{R\mbox{\sffamily T\normalfont}^{\mbox{\scriptsize g\normalsize}}} 
\mbox{ } ,
\label{VBashwe}
\ee
where  
\be
\mbox{\sffamily T\normalfont}^{\mbox{\scriptsize g\normalsize}} = 
\frac{1}{\sqrt{h}}G^{abcd}(\dot{h}_{ij} - \pounds_{\beta}{h}_{ab})(\dot{h}_{ij} - \pounds_{\beta}{h}_{cd}) 
\label{tpot}
\ee 
is the gravitational kinetic term
and
$
G^{abcd} = \sqrt{h}(h^{ac}h^{bd} - h^{ab}h^{cd})
$ 
is the inverse of the DeWitt \cite{DeWitt67} supermetric
$
G_{abcd} = \frac{1}{\sqrt{h}}(h_{ac}h_{bd} - h_{ab}h_{cd})
$.
$h$ is the determinant of the 3-metric and $R$ is the 3-d Ricci scalar.

Note that this BSW form is equivalent to the standard split GR Lagrangian of App B as follows.  
Use the $\alpha$-multiplier equation 
$
2\alpha = \pm \sqrt{ \frac{\mbox{\sffamily\scriptsize T\normalsize\normalfont}^{\mbox{\scriptsize g\normalsize}}}{R}} 
$
as an expression to to \sl algebraically \normalfont eliminate $\alpha$ from the BSW action 
(notice the analogy \cite{BOF} with setting up the homogeneous Jacobi principle).  
Thus (assuming $R \neq 0$ everywhere in the region of interest) one arrives at the BSW action 
\ref{VBashwe}.  Also note that the BSW action has the local square root as opposed to the 
global square root ordering \cite{BB82, B94I, BOF}
$$
\mbox{\sffamily I\normalfont} 
= \int\textrm{d}\lambda\sqrt{\int\textrm{d}^3x
\sqrt{h}\mbox{\sffamily T\normalfont}^{\mbox{\scriptsize g\normalsize}}      }
\sqrt{\int\textrm{d}^3x\sqrt{h}R} 
\mbox{ } .
$$

Note furthermore that (as done here and as opposed to the presentation in App B), this may be regarded as an action 
which is already constructed to meet {\bf GSR} by use of the {\bf AF} implementation.  
The potential $R$ is good 3-diff object. $\dot{h}_{ij}$ is not: 
\be
\frac{\pa}{\pa\lambda}h_{ij}^{\prime} = \frac{\pa s^a}{\pa s^{\prime i}}\frac{\pa s^b}{\pa s^{\prime j}}
\left(
\frac{\pa h_{ab}}{\pa\lambda}  - 2D_{(a}\frac{    \pa s_{b)}    }{    \pa\lambda    }
\right)
\mbox{ } . 
\ee
Thus the action already uses a gravitational Diff-{\bf AF}
\be
\dot{h}_{ij} \longrightarrow \&_{\dot{\mbox{\scriptsize s\normalsize}}}h_{ij} \equiv 
\dot{h}_{ij} - \pounds_{\dot{\mbox{\scriptsize s\normalsize}}}h_{ij} = 
\dot{h}_{ij} - 2D_{(i}\dot{s}_{j)}\mbox{ } , 
\label{precombo}
\ee
or equivalently constitutes a derivation of gravitational Diff-{\bf BM} 
\be
\dot{h}_{ij} \longrightarrow \mbox{\ss}_{{\xi}}h_{ij} \equiv 
\dot{h}_{ij} - \pounds_{\xi}h_{ij} = 
\dot{h}_{ij} - 2D_{(i}\xi_{j)} 
\label{combo}
\mbox{ } ,
\ee
for $\xi^i = \dot{s}^i$.  
Note that the above involves rewriting $\beta^i$ (the Arnowitt--Deser--Misner (ADM) 
\cite{ADM} split shift of App B) as the emergent auxiliary $\xi^i$ or $\dot{s}^i$. 
Likewise below I use $N$ the emergent lapse rather than the ADM split lapse $\alpha$.  
The corresponding interpretation of Diff-{\bf BM} is that for any two configurations 
$\Sigma_1$, $\Sigma_2$ (3-metrics on topologically-equivalent 3-geometries), one w.l.o.g 
keeps the coordinates of $\Sigma_1$ fixed whilst shuffling around those of $\Sigma_2$ until 
they are as `close' as possible to those of $\Sigma_1$ (Fig 3).
\begin{figure}[h]
\centerline{\def\epsfsize#1#2{0.4#1}\epsffile{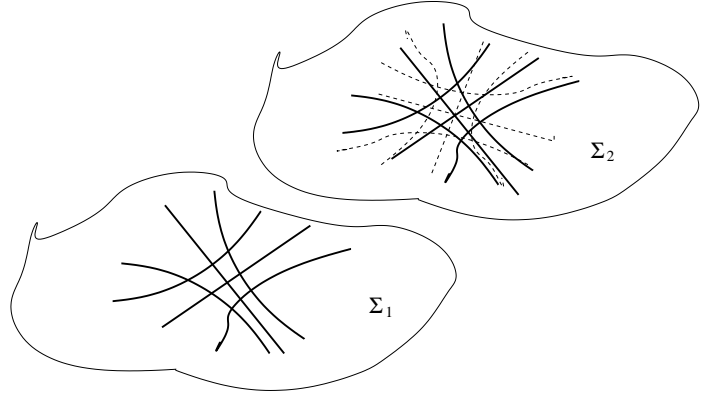}}
\caption[]{\label{TO7.ps}
\scriptsize Best matching of coordinate grids.
\normalsize}
\end{figure}

\mbox{ }

The canonical momenta (defined at each space point) are 
\be
p^{ij} \equiv \frac{  \partial\mbox{\sffamily{L}\normalfont}  }{ \partial \dot{h}_{ij}  }
 = \sqrt{    \frac{  h  }{  2N  }    }G^{ijcd} {\&}_{\dot{\mbox{\scriptsize s\normalsize}}} {h}_{cd} 
\label{GRmom}
\ee
for $2N \equiv \sqrt{\frac{\mbox{\sffamily\scriptsize T\normalsize\normalfont}^{\mbox{\tiny g\normalfont}}}{R}}$.
The Lagrangian is homogeneous of degree $1$ in the velocities.  Thus primary constraints exist 
by Dirac's argument; specifically here the {\sl local} square root gives 1 constraint 
{\sl per space point}: 
$$
G_{ijkl}p^{ij}p^{kl} = G_{ijkl}
G^{ijcd}\frac{1}{2N}{\&}_{\dot{\mbox{\scriptsize s\normalsize}}}h_{cd}
G^{klab}\frac{1}{2N}{\&}_{\dot{\mbox{\scriptsize s\normalsize}}}h_{ab}  
= \sqrt{h}\frac{  \mbox{\sffamily T\normalfont}^{\mbox{\scriptsize g\normalsize}}  }{  (2N)^2  } 
= \sqrt{h}R 
\mbox{ } ,
$$
so
\be
{\cal H} \equiv G_{ijkl}p^{ij}p^{kl} - \sqrt{h}R = 0 
\mbox{ } . 
\mbox{\hspace{0.3in}}
\label{GRHam} 
\ee 
This may be identified as the {\it Hamiltonian constraint of 
GR} (c.f App B), and is the reason why Superspace still contains redundancy.  

In addition, the FEP variation w.r.t the cyclic variable $s^i$ (or equivalently the 
standard variation w.r.t the `multiplier' $\xi^i$) gives as a secondary constraint
\be
{\cal H}_i \equiv -2D_j{p^{j}}_i = 0 
\mbox{ } .
\label{mommy}
\ee 
This may be identified as the {\it momentum constraint of GR} (c.f App B).     
Whereas the BSW action is associated with curves on the space 
Riem $\times$ Diff (where the $\xi^i$ generate Diff), if the momentum constraint can be 
solved as a p.d.e for $\xi^i$ (the thin sandwich conjecture \cite{tslit}), the action will 
depend only on the curve in Superspace. This follows from the constraints being free of 
$\xi^i$, and the three components of the momentum constraint reducing the number of d.o.f's 
from the 6 of Riem to the $3$ per space point in a $3$-geometry.  

The ELE's are (\ref{GRmom}) and  
\be
\dot{p}^{ij}  = \frac{   \delta\mbox{\sffamily L\normalfont}   }{   \delta h_{ij}   } 
= \sqrt{h}N(h^{ij}R - R^{ij}) - \frac{2N}{\sqrt{h}}
\left(
p^{im}{p_m}^j -\frac{1}{2}p^{ij}p
\right) 
+ \sqrt{h}(D^iD^jN -h^{ij}{D}^2 N) + \pounds_{\dot{\mbox{\scriptsize s\normalsize}}}p^{ij} 
\mbox{ } .
\label{GReleq}
\ee
This is indeed GR, for which it is well-known by the contracted Bianchi identity that both constraints 
propagate without recourse to further secondary constraints.  Furthermore, in this chapter, 
the momentum constraints are automatically propagated as a further consequence of the action 
being deliberately constructed to be invariant under \sl $\lambda$-dependent 
$3$-diffeomorphisms\normalfont.  From the 3 + 1 perspective, the hidden foliation invariance\fn{Whereas at 
first glance, one would expect the BSW action to be invariant only with respect to the global 
reparametrization $\lambda \longrightarrow \lambda^{\prime}(\lambda)$ for $\lambda^{\prime}$ 
a monotonic arbitrary function of $\lambda$ (in accord with Noether's theorem), in fact the 
action is invariant under the far more general local transformation: foliation invariance (equivalent to 
the embeddability notion mentioned in App C): 
\be
\lambda \longrightarrow \lambda^{\prime}(\lambda),\mbox{  }
h_{ij}(x,\lambda) \longrightarrow h_{ij}(x,\lambda^{\prime}),
\mbox{ } \xi_i(x,\lambda) \longrightarrow
\frac{d\lambda^{\prime}}{d\lambda}\xi_i(x,\lambda) 
\mbox{ } . 
\ee} 
of GR is associated with the propagation of ${\cal H}$.  
Finally, I note that in contrast with Sec 3-4, the Hamiltonian constraint arises in place 
of time-label conditions that ensure the recovery of the standard form of the equations of physics.  
There are to be no such privileged time-labels in GR!

\section{Example 4: the 3-Space Approach to Relativity}

\subsection{Postulates and Main Working}

Rather than presupposing GR's ADM split or embeddability into spacetime, the idea is to 
take as general an ansatz as possible built by the {\bf RI} and Diff-{\bf AF} implementations.  
We are to see how well our principles do `by themselves' (simplicities will also be listed).

\mbox{ }

\noindent  
{\bf RI[TR]}: a {\sl local square root reparametrization-invariant} implementation is used.  
The pure gravity actions considered are of Baierlein--Sharp--Wheeler (BSW) type 
with kinetic term \sffamily T \normalfont homogeneously quadratic in the velocities and 
ultralocal in the metric.  

\mbox{ }

\noindent{\bf AF[GSR]}: the action is to be constructed in an arbitrary Diff-frame, i.e 
using an arbitrary coordinate grid with coordinates $s^i$.  From this one can deduce 
BF\'{O}'s postulate that

\mbox{ }

\noindent{\bf BM[GSR]}: the {\sl best matching} rule is used to implement the {\sl{3-d}} 
diffeomorphism invariance by correcting the bare metric velocities 
$\dot{h}_{ij} \longrightarrow \mbox{\ss}_{\xi}h_{ij} \equiv 
\dot{h}_{ij} - \pounds_{\xi}h_{ij}$, for $\xi^i = \dot{s}^i$. 

\mbox{ }  

\noindent My analysis below however differs from BF\'{O}'s since, as doccumented below, 
they missed out a number of possibilities.  My trial BSW-type action is 

\noindent
\be
\mbox{\sffamily I\normalfont}_{\mbox{\scriptsize BSW-type \normalsize}}
= \int \textrm{d}\lambda \int \textrm{d}^3x \sqrt{h} \sqrt{\sigma R + \Lambda}
\sqrt{\mbox{\sffamily T\normalfont}^{\mbox{\scriptsize g\normalsize}}_{\mbox{\scriptsize WY\normalsize}}} 
\mbox{ } , \mbox{ } \mbox{ } 
\mbox{\sffamily T\normalfont}^{\mbox{\scriptsize g\normalsize}}_{\mbox{\scriptsize WY\normalsize}} 
= \frac{1}{\sqrt{h}Y}G_{\mbox{\scriptsize W\normalsize}}^{abcd}
{\&}_{\dot{\mbox{\scriptsize s\normalsize}}}{h}_{ab}
{\&}_{\dot{\mbox{\scriptsize s\normalsize}}}{h}_{cd} 
\mbox{ } ,
\label{VASBSW}
\ee
where
\be
G^{ijkl}_{\mbox{\scriptsize W\normalsize}} \equiv \sqrt{h}(h^{ik}h^{jl} - Wh^{ij}h^{kl}) 
\mbox{ } , \mbox{ } \mbox{ }
W \neq \frac{1}{3} \mbox{ } , 
\ee
is the inverse of the most general (invertible) ultralocal supermetric
\be
G_{abcd} = h_{ac}h_{bd} - \frac{X}{2}h_{ab}h_{cd} \mbox{ } , \mbox{ } X = \frac{2W}{3W - 1} 
\ee
and w.l.o.g $\sigma \in \{-1, 0, 1\}$.  More general potentials are discussed in Sec 5.4.       
Setting $2N \equiv \sqrt{          
\frac{        \mbox{\sffamily\scriptsize T\normalsize\normalfont}^{\mbox{\tiny g\normalsize}}_{\mbox{\tiny WY\normalsize}}        }
{\sigma R + \Lambda}        }$, the gravitational momenta are 
\be
p^{ij} \equiv \frac{\partial\mbox{\sffamily{L}\normalfont} }{ \partial\dot{h}_{ij}} =
\frac{\sqrt{h}Y}{2N}G^{ijcd}_{\mbox{\scriptsize W\normalsize}}
{\&}_{\dot{\mbox{\scriptsize s\normalsize}}}{h}_{cd} 
\mbox{ } .
\label{wmom}
\ee
The primary constraint\fn{$A \circ B$ denotes $A_{ij}B^{ij}$.} 
\be
{\cal H } \equiv \frac{Y}{\sqrt{h}}
\left( 
p \circ p - \frac{X}{2}p^2
\right) - \sqrt{h}(\sigma R + \Lambda)  = 0 
\label{VGRHam}
\ee
then follows merely from the local square-root form of the Lagrangian.  
In addition, FEP variation w.r.t $s^i$ (or standard variation w.r.t $\xi^i$) 
leads to a secondary constraint which is the usual 
momentum constraint (\ref{mommy}). 

The ELE's are 
$$
\dot{p}^{ij} = \frac{\delta\mbox{\sffamily L\normalfont}}{\delta h_{ij}} =  \sqrt{h}Nh^{ij}(\sigma R + \Lambda) -\sqrt{h}\sigma NR^{ij}
- \frac{2NY}{\sqrt{h}}
\left(
p^{im}{p_m}^j -\frac{X}{2}p^{ij}p
\right)
$$
\be
\mbox{\hspace{0.5in}}
+ \sqrt{h}\sigma (D^iD^j N - h^{ij}D^2 N) + \pounds_{\dot{\mbox{\scriptsize s\normalsize}}}p^{ij} 
\mbox{ } . 
\label{wel}
\ee
The propagation of ${\cal H}$ then gives \cite{Sanderson} 
\be
\dot{{\cal H}} = \frac{Y\sigma}{N}D^i(N^2{{\cal H}_i})
+ \frac{(3X - 2)NpY}{2\sqrt{h}}{\cal H}
+ \pounds_{\dot{\mbox{\scriptsize s\normalsize}}}{\cal H} 
+ \frac{2}{N}(1 - X)Y\sigma D_i\left(N^2 D^ip\right).
\label{mastereq}
\ee
The first three terms of this are functionals of existing constraints and thus vanish weakly in 
the sense of Dirac.  However note that the last term is not related to the existing 
constraints. It has 4 factors which could conceivably be zero:
\be
(1 - X)Y\sigma D_i(N^2D^ip) 
\mbox{ } .
\label{keyterm}
\ee
Any of the first three factors being zero would be strong equations restricting the form of 
the ansatz.  The fourth factor might however lead to new constraints and thus vanish weakly.

\subsection{Interpretation of the Consistency Condition}

The above `Relativity without Relativity' (RWR) argument \cite{BOF} succeeds in demonstrating 
that GR can be derived solely from spatial arguments, that is, without resort to any 
arguments involving $4$-d general covariance (spacetime structure).  This success stems from 
the combined restrictions of the propagation of ${\cal H }$ and the Diff-invariance which 
ensures the propagation of ${\cal H}_i$, which mean that one is left with at most 
2 d.o.f's per space point. 
It turns out that foliation invariance 
does not usually hold for the generalization (\ref{VASBSW}) of the BSW 
action. Rather, further constraints arise, which lead exhaustively to inconsistencies.  
To have a consistent theory, it is required that the term (\ref{keyterm}) vanishes.  
The expression I provide generalizes BF\'{O}'s result. 

The first factor enforces the ($W = X = 1$) DeWitt supermetric of GR, which is 
the basis of BF\'{O}'s RWR result.  Note that the Lorentzian 
signature ($\sigma = 1$) of GR does not arise alone; one can just as well obtain 
Euclidean GR ($\sigma = -1$) in this way.  Earlier work of Giulini already noted that the 
$W = 1$ supermetric has special properties \cite{SGGiulini}.  

Moreover, GR is not entirely uniquely picked out, because the second, third and fourth 
factors give alternatives.  The mere fact that these arise from an exhaustive route 
to GR motivates their study.  Can they be overruled to provide a unique derivation of GR, 
or can any of them seriously rival GR?  In addition, study of these alternative theories 
can be motivated from the interesting theoretical properties that they possess 
\cite{Sanderson, ABFO}.   

The second factor arises from my explicit inclusion of $Y$, which is scaled to 1 in BF\'{O}.  
My approach makes clear how a `Galilean' alternative arises for $Y = 0$ 
i.e in the degenerate case in which the gravitational momenta completely vanish in the Hamiltonian 
constraint.  That this possibility arises together with the Lorentzian signature Relativity 
possibility makes it clear that the condition that (\ref{keyterm}) vanishes is closely related to the choice of postulates that 
Einstein faced in setting up Relativity (App A).  This point is further discussed in Sec 7, 
since it requires first the introduction of matter (Sec 6).    

The third factor gives strong gravity theories \cite{sg, Sanderson}.    
$\sigma = 0$, $W = 1$ is the conventional strong gravity, i.e the strong-coupled limit of GR,  
of relevance near singularities, while I showed $\sigma = 0$, $W \neq 1$ gives analogous regimes in 
scalar-tensor theories \cite{Sanderson}.  
   
The fourth factor leads to conformal theories with privileged slicing 
\cite{conformal, ABFO, Kelleher, ABFKO}.  One can try avoiding 
privileged slicing by 
trying to ensure $N$ remains freely specifiable.  Thus $D_i p = 0$, which implies the 
{\it constant mean curvature (CMC) condition}
$\frac{p}{\sqrt{h}} = C(\lambda)$.  But this new constraint must also propagate.  This leads 
to a nontrivial {\it lapse-fixing equation (LFE)} which (if soluble) gives a 
{\it CMC foliation}.  The LFE is 
\be
\frac{\pa}{\pa\lambda}\left(\frac{p}{\sqrt{h}}\right) = B(\lambda) =  3\Lambda N + 2\sigma (NR - D^2N) 
+ \frac{(3X - 2)NYp^2}{2h} \mbox{ } ,
\label{SGslicingeq}
\ee
which is a nontrivial equation for the lapse $N$ for $\sigma \neq 0$, 
and is the standard CMC LFE (\ref{CMCLFE}) in the GR case ($\sigma = Y = W = 1$).   

These alternatives are conformal in that in addition to quotienting out Diff, they involve 
quotienting out some conformal group, corresponding to implementing $\frac{p}{\sqrt{h}} = C(\lambda)$ 
or the subsidiary {\it maximal condition} $p = 0$.  This sort of mathematics is that used in the GR IVF (see App D).  
As explained in Sec 14, these alternatives include both new formulations of GR and a number of privileged 
slicing alternative theories. 

To mathematically distinguish GR from these other theories, I use the 

\mbox{ }

\noindent
{\bf GR-specifying TSA postulate 3}: the theory does not rely on privileged foliations 
and has Lorentzian signature.

\mbox{ }

\noindent I am currently seeking to overrule the alternative conformal theories 
fundamental grounds, by thought experiments or by use of current astronomical data, 
which would tighten the uniqueness of GR as a viable 3-space theory \sl on physical 
grounds\normalfont.  If such attempts persistently fail, these theories will become 
established as serious alternatives to GR.

\subsection{Deducing BM from the RI BSW Form Alone}

If one considers starting off with \cite{Sanderson, SGNiall}

\mbox{ }

\noindent \bf Integrability[TR] \normalfont: the use of 
`bare velocities' rather than Diff-{\bf AF} or Diff-{\bf BM} ones,  

\mbox{ } 

\noindent 
one nevertheless discovers the momentum constraint as an integrability of ${\cal H}$: 
\be
\dot{{\cal H}} = \frac{Y\sigma}{N}D^i(N^2{{\cal H}_i})
+ \frac{(3X - 2)NpY}{2\sqrt{h}}{\cal H} + \frac{2}{N}(1 - X)Y\sigma D_i\left(N^2 D^ip\right).
\ee   
One then argues that this `discovered' constraint may be `encoded' into the bare action by 
the introduction of an auxiliary variable $\xi^i$.  It is then this encoding that may be 
thought of as the content of BM.  One then goes back and re-evaluates the momenta and ELE's, 
obtaining now the full GR ones (\ref{GRmom}) and (\ref{GReleq}). Thus the guess that G = id 
is a suitable choice turns out here to be untenable.  In other words, in such an approach 
spatial relationalism is {\sl emergent} rather than imposed.   However, for $\sigma = 0$ or 
$Y = 0$ no additional constraints arise.  Thus constraints may be `missed out' rather than 
`discovered' if one relies on integrability.  This example also illustrates that 
non-spatially relational metrodynamical theories sometimes exist alongside spatially 
relational geometrodynamical ones.

\subsection{Higher Derivative Potentials}

Using the potential \sffamily V \normalfont $= \sigma R + \Lambda$ assumed in 1.2.3 
amounts to applying a temporary 

\mbox{ }

\noindent{\bf TSA gravity simplicity 4}: the pure gravity action is constructed with at most 
second-order derivatives in the potential, and with a homogeneously quadratic kinetic term. 

\mbox{ }

\noindent Furthermore, BF\'{O} considered potentials that are more complicated scalar concomitants 
of the 3-metric $h_{ij}$ than the above:  \sffamily V\normalfont $ = R^n$ and 
\sffamily V\normalfont $= C_1R^2 + C_2R\circ R + C_3D^2R$ (the most general fourth-order 
curvature correction in 3-d because of the Gauss--Bonnet theorem).  Among these the 
potential of GR alone permits the Hamiltonian constraint to propagate.  Also, recently, 
\'{O} Murchadha \cite{Niall03} considered actions based on matrices $M_{ab}(x_i, \lambda)$ 
and their conjugates $P^{ab}(x_i, \lambda)$ which contain all terms in the former with the use 
of up to two derivatives and are ultralocal in the latter.  In this bare approach, he recovers 
the combination $R(M_{ab})$ for the form of the potential (along with the strong and conformal 
options), as singled out by the propagation of the local square root constraint.

\section{Example 5: TSA to Relativity Coupled to Fundamental Matter Fields}  

The capacity to include matter would strengthen the TSA as a viable ontology.  The first TSA 
works \cite{BOF, AB} futhermore appear to give some striking derivations of the classical 
laws of bosonic physics.  Rather than being presupposed, both the null cone structure shared 
between gravitation and classical bosonic matter theories, and Gauge Theory, are enforced and 
share a common origin in the propagation of the Hamiltonian constraint.  Gauge Theory arises 
through the discovery of gauge-theoretic Gauss constraints as secondary constraints from the 
propagation of the Hamiltonian constraint.  This is then encoded by an auxiliary field which 
occurs as gauge-theoretic {\bf BM} corrections to the velocities.  Thus electromagnetism and 
Yang--Mills theory {\sl emerge} in this work.  

The matter considered was subject to the 

\mbox{ } 

\noindent{\bf TSA matter simplicity 5}: the matter potential has at most first-order derivatives and the kinetic 
term is  ultralocal and homogeneous quadratic in the velocities.  


\mbox{ }

\noindent As explained in \cite{Vanderson} there is a further tacit simplicity hidden in the 
`adding on' of matter.  This is linked to how including general matter can alter the 
gravitational part of the theory.  This issue is tied both to the relationship between 
the TSA and the {\bf Principle of Equivalence (POE)} (see Sec 12), and 
in my contesting BF\'{O}'s speculation that the matter results ``hint at partial unification"  
(see Sec 11).  

\mbox{ }

The action is to be {\bf RI}.  And written in the arbitrary Diff-frame, 
from which it 
follows that, alongside the gravitational velocities, {\sl all} the velocities of the 
matter fields $\Psi_{\mbox{\sffamily\scriptsize A\normalsize\normalfont}}$ pick up Diff-{\bf AF} 
Lie derivative corrections: 
$\dot{\Psi}_{\mbox{\sffamily\scriptsize A\normalsize\normalfont}} \longrightarrow 
\&_{\dot{\mbox{\scriptsize s\normalsize}}}
{\Psi}_{\mbox{\sffamily\scriptsize A\normalsize\normalfont}} \equiv 
\dot{\Psi}_{\mbox{\sffamily\scriptsize A\normalsize\normalfont}} - 
\pounds_{\dot{\mbox{\scriptsize s\normalsize}}}
\Psi_{\mbox{\sffamily\scriptsize A\normalsize\normalfont}}$.  
Alternatively the matter field velocities pick up Diff-{\bf BM} corrections 
$\dot{\Psi}_{\mbox{\sffamily\scriptsize A\normalsize\normalfont}} \longrightarrow 
\mbox{\ss}_{\xi}{\Psi}_{\mbox{\sffamily\scriptsize A\normalsize\normalfont}} \equiv 
\dot{\Psi}_{\mbox{\sffamily\scriptsize A\normalsize\normalfont}} 
- \pounds_{\xi}\Psi_{\mbox{\sffamily\scriptsize A\normalsize\normalfont}}$. 
Thus gravitational {\bf AF} or {\bf BM} kinematics takes a {\sl universal} form: it is the 
same within each rank of tensor.  Each of these forms may be built up as explained in Sec 4.

\subsection{TSA Ansatz for a Single 1-Form Field}

To include a single $1$-form field $A_a$, BF\'{O} considered the {\bf RI} action

\noindent
\be 
\mbox{\sffamily I\normalfont}^{\mbox{\scriptsize A\normalsize}}_{\mbox{\scriptsize BSW\normalsize}}
= \int \textrm{d}\lambda \int \textrm{d}^3x \sqrt{h} 
\sqrt{R + \mbox{\sffamily U\normalfont}^{\mbox{\scriptsize A\normalsize}}} 
\sqrt{\mbox{\sffamily T\normalfont}^{\mbox{\scriptsize g\normalsize}} 
+ \mbox{\sffamily T\normalfont}^{\mbox{\scriptsize A\normalsize}}}  
\ee 
for
$\mbox{\sffamily T\normalfont}^{\mbox{\scriptsize A\normalsize}} = h^{ab}
\&_{\dot{\mbox{\scriptsize s\normalsize}}}{A}_a
\&_{\dot{\mbox{\scriptsize s\normalsize}}}{A}_b$ 
the quadratic Diff-{\bf AF} kinetic term of $A_a$, and the potential ansatz 
$\mbox{\sffamily U\normalfont}^{\mbox{\scriptsize A\normalsize}} =
C_1D_bA_{a}D^bA^{a} + C_2D_bA_{a}D^aA^{b} + C_3D^a{A_a}D^b{A_b} + 
\sum_{(k)} B_{(k)}(A_aA^a)^{k} $.  The first part of this can be expressed
more conveniently for some purposes by using a generalized supermetric 
$C^{abcd} = C_1h^{ab}h^{cd} + C_2h^{ad}h^{bc} + C_3h^{ac}h^{bd}$.  

Defining $2N \equiv \sqrt{     \frac{    \mbox{\sffamily\scriptsize T\normalsize\normalfont}^{\mbox{\tiny g\normalsize}} 
+ \mbox{\sffamily\scriptsize T\normalsize\normalfont}^{\mbox{\tiny A\normalsize}}    }
{    R + \mbox{\sffamily\scriptsize U\normalsize\normalfont}^{\mbox{\tiny A\normalsize}}    }     }$, the conjugate momenta are given by 
(\ref{GRmom}) and 
\be 
\pi^i \equiv \frac   {\partial\mbox{\sffamily{L}\normalfont}}
{\partial \dot{A_i}}  = \frac{\sqrt{h}}{2N} \&_{\dot{\mbox{\scriptsize s\normalsize}}} A^i 
\mbox{ } .
\label{1formmom}
\ee 
Then, the local square root gives as a primary constraint a Hamiltonian-type constraint 
\be 
^{\mbox{\scriptsize A\normalsize}}{\cal H } \equiv  
\frac{1}{\sqrt{h}}
\left(
p \circ p - \frac{1}{2}p^2 + \pi^a\pi_a
\right) 
- 
\sqrt{h}(R + \mbox{\sffamily U\normalfont}^{\mbox{\scriptsize A\normalsize}}) 
= 0 \mbox{ } .
\ee 
FEP variation w.r.t $\dot{s}^i$ gives as a secondary constraint the momentum constraint 
\be
^{\mbox{\scriptsize A\normalsize}}{\cal H}_i \equiv 
-2D_j{p^{j}}_{i} + \pi^{c}(D^j{A_{i}} - D_i{{A}^j}) - D_c{\pi^{c}}A_{i} = 0 \mbox{ } . 
\label{emmomcon}
\ee 

Then, propagating $^{\mbox{\scriptsize A\normalsize}}{\cal H }$ 
gives
$$
^{\mbox{\scriptsize A\normalsize}}\dot{{\cal H}} = 
- \frac{    D^i(N^2 \mbox{ }^{A}{\cal H}_i)    }{    N    }  
+ \frac{   Np \mbox{ }^{A}{\cal H}   }{   2   } 
+ \pounds_{\dot{\mbox{\scriptsize s\normalsize}}} \mbox{ }^{A}{\cal H} 
$$
$$
+ \frac{1}{N} 
\left[
(4C_1 + 1)D_b(N^2\pi^aD^b{A_{a}}) + (4C_2 - 1)D_b(N^2\pi^aD_a{A^b}) + 4C_3D_a(N^2\pi^aD_b{A^b}) 
\right]
$$
$$
- \frac{1}{N} D_b(N^2D_a{\pi^a}A^b) 
- \frac{1}{N} 
\left[
N^2 D_a
\left(
p_{ij} - \frac{p}{2} h_{ij}
\right) 
D_dA_b 
(2A^iC^{ajbd} - A^aC^{ijbd}) 
\right]  
\mbox{ } .
$$
Now, the system has a priori 5 d.o.f's per space point, that is 2 geometric d.o.f's  
and the 3 d.o.f's of the 1-form field itself.  The constraints cannot include $N$, so the 
penultimate line includes a 3-vector of constraints multiplied by $\pa^aN$, which would take 
away all the 1-form d.o.f's, thus rendering a trivial theory, unless the  cofactor of 
$\pa^aN$ vanishes strongly.  This gives a nontrivial theory only for $C_1 = -C_2 = -\frac{1}{4}$,
$C_3 = 0$ and if there is a secondary constraint 
\be
{\cal G} \equiv D_a{\pi^a} = 0 \mbox{ } .
\label{emergau}
\ee 
The fixed value that $C_1$ and $C_2$ take means that the 1-form field shares the null cone of 
gravity.  Furthermore, that $C_1 = - C_2$ and $C_3 = 0$ mean that the derivative terms in 
$\mbox{\sffamily U\normalfont}_{\mbox{\scriptsize A\normalsize}}$ are the `Maxwellian curl combination' 
$-\frac{1}{4}|\mbox{\b{$\pa$}} \mbox{ \scriptsize $\times$ \normalsize} \mbox{\b{A}\normalfont}|^2$.  
${\cal G}$ may be identified as the Gauss constraint of electromagnetism.   

The propagation of ${\cal G}$ gives
$$ 
\frac{\partial}{\partial \lambda}(D_i{\pi^i}) =  2\sqrt{h}D_a
\left[
N\sum_{(k)}kB_{(k)}(A^aA_a)^{k - 1} A^i
\right] 
+ \pounds_{\dot{\mbox{\scriptsize s\normalsize}}}(D_i{\pi^i}) \mbox{ } .
$$
Again, one can argue that constraints should not depend on $N$, and then that the only way of 
avoiding triviality of the 1-form field due to the terms in  $\pa_aN$ is to have all the 
$B_{(k)}$ be zero. In particular, $B_{(1)} = 0$ means that this working leads to massless 1-forms.  

Now, the allowed form  $\mbox{\sffamily U\normalfont}^{\mbox{\scriptsize A\normalsize}} = 
-\frac{1}{4}D_bA_b(D^bA^a - D^aA^b)$, is invariant under the gauge transformation 
$A_a \longrightarrow A_a - \pa_a\Lambda$, so one is dealing with a Gauge Theory.  Note first 
how the Gauge Theory and the fixing of the light-cone to be equal to the gravity-cone arise 
together in the same part of the above calculation.  These are two aspects of the same 
consistency condition arising from the role of the momentum constraint in the propagation of 
the Hamiltonian constraint.  Second, because we have a gauge symmetry, if we introduce an 
auxiliary variable $\Phi$ ( = $\dot{\Lambda}$) into 
$\mbox{\sffamily T\normalfont}^{\mbox{\scriptsize A\normalsize}}$ such that variation w.r.t  
it encodes ${\cal G}$, then we should do so according to U(1)-{\bf BM}.   This uniquely fixes 
the form of $\mbox{\sffamily T\normalfont}^{\mbox{\scriptsize A\normalsize}}(A, \Phi)$ to be 
$\mbox{\sffamily T\normalfont}^{\mbox{\scriptsize A\normalsize}} = h^{ab} 
(\dot{A}_a - \pounds_{\dot{\mbox{\scriptsize s\normalsize}}}A_a - \pa_a\Phi) 
(\dot{A}_b - \pounds_{\dot{\mbox{\scriptsize s\normalsize}}}A_b - \pa_b\Phi)$.  Thus, if one 
identifies $\Phi$ as $A_{0}$, this derivation forces the 4-d 1-form $A_A = [\Phi, A_i]$ 
to obey Maxwell's equations minimally-coupled to gravity. Moreover, \cite{giulini, BOF}, the 
massive (Proca) 1-form field does not fit into this TSA formulation despite being a 
perfectly good generally covariant theory. BF\'{O} originally took this to be evidence that 
the TSA does not yield all generally covariant theories.

\subsection{More General Matter Treated within the TSA}

A similar treatment to the above has been carried out for many interacting 1-forms \cite{AB}, 
leading to Yang--Mills theory, and for a single 1-form interacting with scalar fields 
\cite{BOF2}, leading to U(1)-scalar Gauge Theory.  In each of these cases, Gauss-type 
constraints arise as integrabilities of ${\cal H}$.  I presented a systematic approach to do 
such calculations in \cite{Thanderson}, rather than the aforementioned case-by-case analyses.  
The Gauss-type constraints are present from the start if one furthermore adopts not the 
`discover and encode' {\bf integrability} implementation but rather the {\bf AF} one.  This 
last approach is more accommodating of further theories (see \cite{Thanderson} and Sec 11).

\section{Emergence of the Relativity Principles in the TSA}

I investigate the emergence of the {\bf Relativity Principles} in the TSA.  The conventional treatment of these 
is provided for comparison in App A.  This investigation requires having treated several 
sorts of matter beforehand.  Whereas {\bf RP1} is about one transformation law for all of nature, 
{\bf RP2} specifies which.  Three possibilities are investigated below:  Galilean, Lorentzian and Carrollian (this last one 
corresponds to strong gravity).    
  
Starting from the relational 3-space ontology, the TSA gives Hamiltonian-type constraints 
\be
{\cal H}^{\mbox{\scriptsize trial\normalsize}} \equiv 
\sqrt{h}(\sigma R + \Lambda + \mbox{\sffamily U\normalfont}_{\Psi})  
- \frac{ 1 }{  \sqrt{h}  }
\left\{
Y
\left(
p \circ p - \frac{X}{2}p^2
\right)
+ G_{\mbox{\sffamily\scriptsize AB\normalfont\normalsize}}
\Pi^{\mbox{\sffamily\scriptsize A\normalfont\normalsize}}\Pi^{\mbox{\sffamily\scriptsize B\normalfont\normalsize}}
\right\} = 0 
\ee
as identities from {\bf RI}.    
Consistency alone then dictates what options are available for 
${\cal H}^{\mbox{\scriptsize trial\normalsize}}$ -- the Dirac approach.  
I have included matter fields $\Psi_{\mbox{\sffamily\scriptsize A\normalfont\normalsize}}$ 
since I have found that conclusions are best made only once this is done.  
$\Psi_{\mbox{\sffamily\scriptsize A\normalfont\normalsize}}$ is s.t  
$\mbox{\sffamily T\normalfont}^{\Psi}$ is homogeneous quadratic in its velocities and 
$\mbox{\sffamily U\normalfont}^{\Psi}$ at worst depends on connections (rather than their 
derivatives). I then get the following master equation for the propagation of the 
Hamiltonian-type constraint: 
$$
\dot{{\cal H}}^{\mbox{\scriptsize trial\normalsize}} \approx  \frac{2}{N}D^a
\left\{
N^2
\left(
Y
\left\{
\sigma
\left(
D^bp_{ab} + \{X - 1\}D_ap
\right)
+ 
\right.
\right.
\right.
\mbox{ } \mbox{ } \mbox{ } \mbox{ } \mbox{ } \mbox{ } \mbox{ } \mbox{ } \mbox{ } \mbox{ } \mbox{ } 
\mbox{ } \mbox{ } \mbox{ } \mbox{ } \mbox{ } \mbox{ } \mbox{ } \mbox{ } \mbox{ } \mbox{ } \mbox{ } 
\mbox{ } \mbox{ } \mbox{ } \mbox{ } \mbox{ } \mbox{ } \mbox{ } \mbox{ } \mbox{ } \mbox{ } \mbox{ }
$$
\be
\left.
\left.
\left.
\left(
p_{ij} - \frac{X}{2}ph_{ij}
\right)
\left(
\frac{\pa \mbox{\sffamily U\normalfont}^{\Psi}}{\pa {\Gamma^c}_{ia}}h^{cj} -  
\frac{1}{2}\frac{\pa \mbox{\sffamily U\normalfont}^{\Psi}}{\pa{\Gamma^c}_{ij}}h^{ac}
\right)
\right\}
+ G_{\mbox{\sffamily\scriptsize AB\normalfont\normalsize}}\Pi^{\mbox{\sffamily\scriptsize A\normalfont\normalsize}}
\frac{\pa \mbox{\sffamily U\normalfont}^{\Psi}}{\pa(\pa_a\Psi_{\mbox{\sffamily\scriptsize B\normalfont\normalsize}})}
\right)
\right\}
\mbox{ } .
\label{sku}
\ee

\mbox{ }

The {\bf Galilean RP2} arises if one declares that $Y = 0$.  This kills 
all but the last factor.  It would then seem natural to take 
$\Pi^{\mbox{\sffamily\scriptsize A\normalfont\normalsize}} = 0$, whereupon the fields are not 
dynamical.  They are however \sl not \normalfont trivial: they include fields obeying 
analogues of Poisson's law, or Amp\`{e}re's, which are capable of governing a wide variety of 
complicated patterns. One would then have an entirely nondynamical `Galilean' world.  Although 
this possibility cannot be obtained from a BSW-type Lagrangian (the \sffamily T \normalfont 
factor is badly behaved), this limit is unproblematic in the Hamiltonian description.  Of course, 
the Hamiltonian-type constraint ceases to be quadratic: 
\be
{\cal H}_{(\mbox{\scriptsize Y = 0\normalsize})} = \sigma R + \Lambda + \mbox{\sffamily U\normalfont}_{\Psi} = 0 
\mbox{ } .
\ee
Now one might still vary w.r.t the metric, obtaining a multiplier equation in place of the ADM 
(or BSW) evolution equation, $N(h^{ij}R - R^{ij}) = h^{ij}D^2N - D^iD^jN$.  In vacuum the 
trace of this and ${\cal H} = R = 0$ leads to $D^2N = 0$ which in the absence of privileged 
vectors implies that $N$ is independent of position so that clocks everywhere march in step.  
Then also $R_{ij} = 0$.  The cosmological constant alone cannot exist in an unfrozen CWB 
world.  But the inclusion of matter generally breaks these results.  One might well however 
not vary w.r.t the metric and consider the worlds with a fixed spatial background metric.  
This includes as a particular case the Hamiltonian study of the flat spatial background world 
in the local square root version of App II.B, but permits generalization to curved backgrounds. 

\mbox{ }
  
The {\bf Carrollian RP2} arises if one declares that $\sigma = 0$.  One still has 
the penultimate term so presumably one further declares that 
$\mbox{\sffamily U\normalfont}^{\Psi}$  contains no connections (the possibility of 
connections is investigated more fully in Sec 10).  It is `natural' then to take the second 
factor of the last term to be 0 thus obtaining a world governed by Carrollian Relativity.  

\mbox{ }

The {\bf Lorentzian RP2} is somewhat more colourful.  $\sigma = 1$ will be required.  Take (\ref{sku}), use 
$0 = - 1 + 1$, reorder and invent a momentum constraint: 
$$
\dot{{\cal H}}^{\mbox{\scriptsize trial\normalsize}}\mbox{$\approx$}\frac{2D^a}{N} 
\left(
N^2
      \left\{
Y
             \left(
\sigma
                   \left\{
                          \left(
D^bp_{ab}\mbox{--}\frac{1}{2}
\left[
\Pi^{\mbox{\sffamily\scriptsize A\normalfont\normalsize}}
\frac{\delta\pounds_{\dot{\mbox{\scriptsize s\normalsize}}}\Psi_{\mbox{\sffamily\scriptsize A\normalfont\normalsize}}}
{\delta\dot{s}^a}
\right]
                          \right)
\mbox{+}\frac{1}{2}
\left[
\Pi^{\mbox{\sffamily\scriptsize A\normalfont\normalsize}}
\frac{\delta\pounds_{\dot{\mbox{\scriptsize s\normalsize}}}\Psi^{\mbox{\sffamily\scriptsize A\normalfont\normalsize}}}{\delta\dot{s}^a}
\right]
                   \right)
            \right\}
\mbox{+}G_{\mbox{\sffamily\scriptsize AB\normalfont\normalsize}}\Pi^{\mbox{\sffamily\scriptsize A\normalfont\normalsize}}
\frac{\pa \mbox{\sffamily U\normalfont}^{\Psi}}{\pa(\pa_a\Psi_{\mbox{\sffamily\scriptsize B\normalfont\normalsize}})}
      \right.
\right.
$$
\be
\mbox{ } \mbox{ } \mbox{ } \mbox{ } \mbox{ } \mbox{ } \mbox{ } \mbox{ } \mbox{ } \mbox{ } \mbox{ } \mbox{ } \mbox{ } \mbox{ } \mbox{ } \mbox{ } \mbox{ } \mbox{ }
\left.
       \left.
            + Y\sigma(X - 1)D_ap
            + Y
            \left(
p_{ij} - \frac{X}{2}ph_{ij}
            \right)
            \left(
            \frac{\pa \mbox{\sffamily U\normalfont}^{\Psi}}{\pa {\Gamma^c}_{ia}}h^{cj} 
          - \frac{1}{2}\frac{\pa \mbox{\sffamily U\normalfont}^{\Psi}}{\pa{\Gamma^c}_{ij}}h^{ac}
            \right)
      \right\}
\right) 
\mbox{ } . 
\ee 
Here, $\frac{\delta}{\delta}$ denotes the functional derivative, and the square bracket 
delineates the factors over which its implied integration by parts is applicable.  
Now go for the orthodox general covariance option: that the third and fourth terms cancel, 
enforcing the null cone.  This needs to be accompanied by doing something about the fifth term.  
One can furthermore {\sl opt} for the orthodox $X = 1$: the recovery of embeddability into 
spacetime corresponding to GR (RWR result), or for the preferred-slicing but GR IVF-like 
worlds of $D_ap = 0$.  Either will do: the recovery of locally-Lorentzian physics does not 
happen for generally-covariant theories alone!  One requires also to get rid of the connection 
terms but the Dirac procedure happens to do this automatically for our big ans\"{a}tze.  Thus 
GR spacetime arises alongside preferred slicing, Carrollian and Galilean worlds, in which 
aspects of GR-like spacetime structure are not recovered.  

\mbox{ }  

With the above in mind, a clarification is required as regards the previous use of exhaustive proofs.  
The ultralocal and nondynamical strategies for dealing with the last term in (\ref{sku}) are available in {\sl all} 
the above options.  It may not shock the reader that degenerate and dual-degenerate possibilities might 
coexist.  Indeed Carrollian matter in the Galilean option permits a BSW Lagrangian to exist... But in the Lorentzian 
case this means \bf RP1 \normalfont is not fully replaced! At the moment, we do derive that gravitation 
enforces a unique finite propagation speed, but the possibility of fields with infinite and 
zero propagation speeds is not precluded.  Thus the objection 
that Newtonian Mechanics and Maxwellian Electromagnetism have different Relativities is precisely not 
being countered!  So in this approach, if one were to observe an analogue of electrostatics 
(a Poisson law), or of magnetostatics, one could not infer that there is a missing displacement 
current (or any other  appropriate individual `Lorentzifications' of electrostatics and magnetostatics in the 
absence of a good reason such as Faraday's Law to believe in unifying these two analogue 
theories).  One would suspect that formulating physics in this way would open the door to analogue Aethers 
coexisting in a universe with Einstein's equations.  

In more detail, BF\'{O} dismissed this possibility as trivial from counting arguments.  
But these are generally misleading, since they do not take into account the geometry of the 
restrictions on the solution space.  It is true that if there are more conditions than degrees 
of freedom then there is typically no solution, but some such systems will 
nevertheless have {\sl undersized} and not empty solution spaces.  

As a first example of this, consider the flat spacetime single 1-form case of Sec 3.  
The crucial term is then $(1 - C)\pi_iF^{ij}$.  The $C = 1$ option gives the 
universal light-cone, but the other factors could be zero in a variety of situations: 
they mean a vanishing Poynting vector: $(E \mbox{ \scriptsize $\times$ \normalsize} B)_i = 0$.  
This includes $E_i = 0$ (magnetostatics), $B_i$ = 0 (electrostatics) and 
$E_i \mbox{ } || \mbox{ } B_i$.  Each of these cases 
admits a number of solutions.  These include complicated patterns analogous to those which 
can occur in electrostatics and magnetostatics, which could not be described as trivial.  

As a second example, consider the single 1-form in homogeneous curved spacetimes.  
$\pi_i = E_i = 0$ imposes a severe but not total restriction \cite{Hughston} on the Minisuperspace of homogeneous spaces.  
The Bianchi types {\sl IV}, {\sl V}, {\sl VI} (h $\neq -1$), {\sl VII} 
(h $= 0$), {\sl VIII}, {\sl IX} are banned outright, whereas the fields in Bianchi types 
{\sl II}, {\sl VI} (h = $-1$), {\sl VII} (h $\neq 0$) have less degrees of freedom than 
expected pointwise in Einstein--Maxwell theory.\fn{h is an invariant which exists for 
Bianchi types {\sl VI} and {\sl VII}.  It is given by $(1 - \mbox{h})
{L^{\mbox{\tiny\bf A\normalfont\normalsize}}}_{\mbox{\tiny\bf BA\normalfont\normalsize}}
{L^{\mbox{\tiny\bf D\normalfont\normalsize}}}_{\mbox{\tiny\bf CD\normalfont\normalsize}} 
= - 2\mbox{h}
{L^{\mbox{\tiny\bf A\normalfont\normalsize}}}_{\mbox{\tiny\bf DB\normalfont\normalsize}}
{L^{\mbox{\tiny\bf D\normalfont\normalsize}}}_{\mbox{\tiny\bf AC\normalfont\normalsize}}$ 
for 
${L^{\mbox{\tiny\bf A\normalfont\normalsize}}}_{\mbox{\tiny\bf BC\normalfont\normalsize}}$ 
the structure constants of each Bianchi model's associated Lie Algebra.}  Nevertheless, solutions exist 
(see p 202 of \cite{MC}).  
The treatment of $B_i = 0$ is identical to that of $E_i = 0$ by dual rotation.  
$E_i \mbox{ } || \mbox{ } B_i$ also admits nontrivial solutions such as the charged Taub 
metric, or its generalization on p 195 of \cite{MC}.  These are not trivial models.  Thus one has indeed an 
undersized but still interesting solution space.  Now, these solutions could all be 
interpreted as belonging not just to Einstein--Maxwell theory, but also to a theory T with 
Einstein cones and distinct (even degenerate) cones belonging to some exotic 1-form theory.  

However, despite these examples illustrating non-triviality, {\bf RP1} is safe.  
For, the theory T permits no macroscopic 1-form propagations, since $E_i = 0$ means no 
momentum, $B_i = 0$ means the theory is ultralocal so 1-form information does not propagate 
away from any point, and $E_i \mbox{ } || \mbox{ } B_i$ means that there is none of the 
mutual orthogonality that ensures the continued propagation of light in electromagnetism.  
In the absence of such propagation, the concept of a 1-form particle moving in a background 
solution of theory T makes no sense (since this is but an approximation to the field equations 
of theory T, which permit no 1-form propagation).  Thus such a 1-form is causally irrelevant, 
so the recovery of {\bf RP1 } from the TSA is not affected.   

Moreover, one does have a source of potentially nontrivial scenarios from this insight: 
such nonpropagating Carrollian or non-($c=1$) Lorentzian or Galilean fields could nevertheless be coupled 
via potential terms to propagating fields, leading to scattering of the propagating fields.  
Whether such unusual fields are capable of producing interesting 
theoretical cosmology results may deserve further investigation.

\section{Relation of Space and Spacetime Points of View} 

\subsection{Motivation}

In \cite{Vanderson}, I showed that there are two sorts of difficulty with BF\'{O}'s suggested use of 
BSW-type actions.  
First, even in Minisuperspace, the associated geometry is plagued with zeros from the 
potential. Furthermore, if one attempts more generally to use the BSW action as a 
{\sl metric function} ${\cal F}$, 
one finds that the associated configuration space metric ${\cal G}_{ijkl} \equiv \frac{\pa^2 {\cal F}}{\pa h_{ij}\pa h_{kl}}$ 
is infinite-dimensional, velocity dependent 
(so the geometry is not Riemannian), degenerate (so the geometry is not even Finslerian 
\cite{giusan})
and containing non-cancelling delta functions and integrals 
(so the metric is not even a function).  
Second, {\bf RI} actions are far more general.  They only look like the BSW action 
if they correspond to theories whose Lagrangians are homogenous quadratic in their 
velocities.  Whereas this case covers all bosonic fields, it does not cover fermionic fields, 
nor phenomenological matter nor charged particles \cite{Thanderson}

I prefer to consider those actions which may be cast into a suitably general BSW form, i.e 
{\bf RI-castable} actions rather the specific implementation by {\bf RI} actions.  I formulate 
this notion more precisely in Sec 9.  This viewpoint allows for fruitful comparison 
with the `split spacetime framework' (SSF) of Kucha\v{r} \cite{KucharI, KucharII, KucharIII, KucharIV}, 
where canonical formulations for very general consistent matter theories are constructed by 
presupposing spacetime and correctly implementing the resulting kinematics.  This leads to 
existence being established for the usual matter fields 
in terms of which nature is described; albeit these are not tightly picked out, there is some degree of picking out 
involved which is perhaps related to the {\bf POE}.    The presupposition of spacetime 
leads to 3 sorts of kinematics that are universal per rank of tensor.

\subsection{Spacetime kinematics}  

First, there is {\it derivative-coupling kinematics}, i.e metric-matter cross-terms in {\sffamily T}, 
which lead to momenta that are less straightforward to invert, and the gravitational part of the 
Hamiltonian constraint being altered.  Thus nonderivative-coupled fields are a lot simpler to deal 
with than derivative-coupled ones.  Furthermore, I realize that this 
is a \sl tacit assumption \normalfont in almost all of BF\'{O}'s work.

\mbox{ }

\noindent \bf Gravity--Matter Simplicity 0 \normalfont : the implementation of `adding on' matter is for 
matter contributions that do not interfere with the structure of the gravitational theory.

\mbox{ }

\noindent This amounts to the absence of Christoffel symbols in the matter Lagrangians, 
which is true of minimally-coupled scalar fields ($D_a\varsigma = \pa_a\varsigma$) and of Maxwell and 
Yang--Mills theories and their massive counterparts (since $D_aA_b - D_bA_a = \pa_aA_b - \pa_bA_a$).  
Thus it suffices to start off by considering the nonderivative-coupled case on the grounds that it 
includes all the fields hitherto thought to fit in with the BF\'{O} scheme, and also the interesting 
example of massive 1-form fields.   

\mbox{ }

Second, there is {\it tilt kinematics}.  This concerns spatial derivatives of $\alpha$.  Thus it 
potentially concerns barriers to having a {\bf RI-castable} action arising from an algebraic BSW 
procedure.  
However, as I demonstrate in Secs 9 and 11, in some cases tilt can be removed by parts, 
by $\alpha$-dependent change of variables or by `accident'.  
  
\mbox{ }

Third, there is {\it shift kinematics}, which concerns coordinate changes on the 3-space itself.  This takes 
a universal Lie derivative form.  

\mbox{ } 

If there is no derivative coupling and if one can arrange for the tilt to play no part in a 
formulation of a matter theory, then all that is left of the hypersurface kinematics is the 
shift kinematics, which may be identified with the {\bf AF} or {\bf BM} implementations of 
{\bf GSR}.  But complying with hypersurface kinematics is a guarantee for consistency for 
established spacetime theories so in these cases {\bf BM} suffices for consistency.  So, 
within a GR spacetime ontology for which it is available, the SSF is a powerful and 
advantageous tool for writing down consistent TSA theories of GR coupled to matter.  Note 
furthermore that the less structure is assumed in theoretical physics, the more room is left 
for predictability.  Could it really be that nature has less kinematics than the GR SSF might 
have us believe?

\section{Exhaustive Inclusion: Bosons}\normalsize 

It strengthens the case for the TSA that it can naturally accommodate the standard set  
of classical fundamental fields matter fields on which can be based the simplest explanation 
of all we know about nature.  The earlier papers \cite{BOF, AB} constructively picked out 
electromagnetism and Yang--Mills theory, all coupled to GR.  I then switched \cite{Vanderson} 
to inspecting the SSF and deducing whether the full standard set of classical fundamental 
fields belongs to its {\bf TSA-castable} fragment.  This is how I here approach scalar-1-form 
Gauge Theory (e.g describing the interactions between gauge fields and hypothetical but 
strongly favoured Higgs fields).  In Sec 10, I likewise approach spin-$\frac{1}{2}$ fermions 
and fermion-1-form Gauge Theory (e.g the classical actions underlying QED, QCD and the 
Weinberg--Salam electroweak theory, as well as the Yukawa coupling by which Higgs scalars 
can interact with fermions).

\subsection{Universal Kinematics of 1-Forms in Split Spacetime}

The SSF treatment of minimally-coupled scalars is trivial so I do not provide it.  
For 1-forms, given the second-order action  
$$
\mbox{\sffamily I\normalfont}_{\mbox{\scriptsize A\normalsize}}
= \int \textrm{d}^4x \sqrt{|g|} \mbox{\sffamily L\normalfont}(A_{A}, \nabla_{B}A_{C}, g_{DE}) 
\mbox{ } , 
$$
introduce $\lambda^{AB} = \frac{\pa L}{\pa (\nabla_{B}A_{A})}$, and use the Legendre 
transformation 

\noindent$(A_{A}, \nabla_{B}A_{C}, \mbox{\sffamily L\normalfont}) 
\longrightarrow (A_{A}, \lambda_{CB}, L),$
where $L$ is the {\it Lagrangian potential}
$
L = [\lambda^{AB}\nabla_{B}A_{A} - \mbox{\sffamily L\normalfont}](A_{A}, \lambda_{BC}, g_{DE}).
$
The SSF decompositions of $A_A$ and $\lambda_{AB}$ are 
$$
A_A = n_AA_{\perp} + {e_A}^aA_a \mbox{ } ,
$$
$$
\lambda^{AB} = \lambda^{\perp\perp}n^{A}n^{B} + 
\lambda^{a\perp}e^{A}_a n^{B} 
+ \lambda^{\perp b}n^{A}e^{B}_b + 
\lambda^{ab}e^{A}_ae^{B}_b \mbox{ }.  
$$
Furthermore, the definition of canonical momentum means that 
$
\lambda^{a\perp} = \pi^a, \mbox{ } \lambda^{\perp\perp} = \pi^{\perp}.
$

Then the `hypersurface Lagrangian' for the 1-form is
\be
\mbox{\sffamily L\normalfont}^{\mbox{\scriptsize hyp\normalsize}}_{\mbox{\scriptsize A\normalsize}}
= \int_{\Sigma}\textrm{d}^3x(\pi^{\perp}\delta_{\beta}A_{\perp}
+ \pi^a\delta_{\check{\beta}}A_a -
\alpha \mbox{ }_{\mbox{\scriptsize A\normalsize}}{\cal H}^{\mbox{\scriptsize o\normalsize}}
- \beta^a { }_{\mbox{\scriptsize A\normalsize}}{\cal H}^{\mbox{\scriptsize o\normalsize}}_a) 
\mbox{ } ,   
\label{Vcovectoraction}
\ee
where $\delta_{\beta}$ is explained in Fig 4.
\begin{figure}[h]
\centerline{\def\epsfsize#1#2{0.6#1}\epsffile{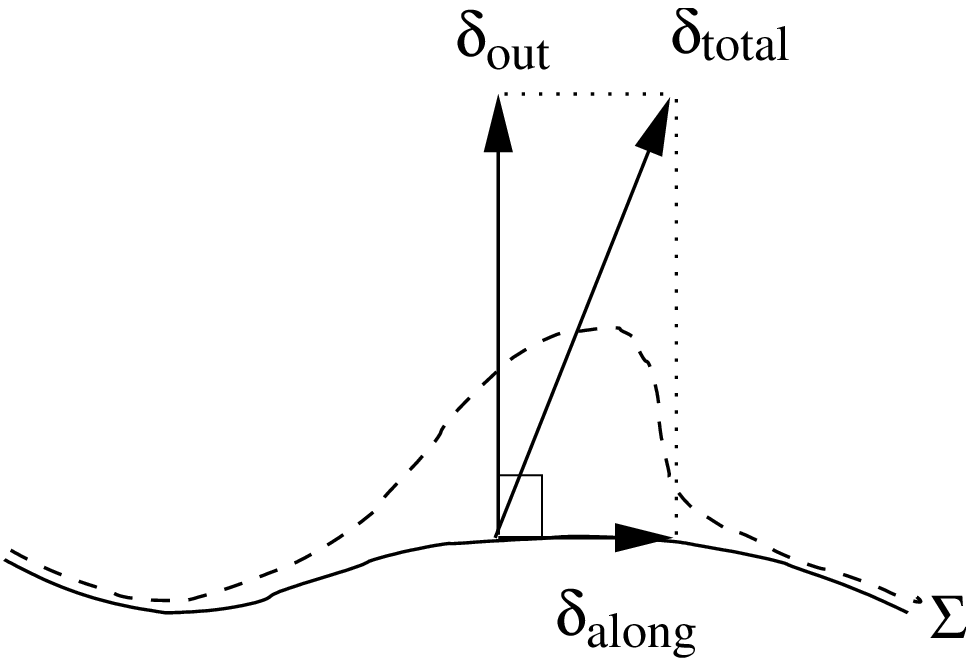}}
\caption[]{\label{TO8.ps}
\footnotesize The change along an arbitrary deformation of the hypersurface $\Sigma$ 
is split according to $\delta_{\mbox{\scriptsize total\normalsize}} = 
\delta_{\mbox{\scriptsize out\normalsize}} + 
\delta_{\mbox{\scriptsize along\normalsize}}$.  Then \cite{KucharI, KucharIII}, 
the {\it hypersurface derivative} $\delta_{\beta} 
\equiv \delta_{\mbox{\scriptsize out\normalsize}} = 
\delta_{\mbox{\scriptsize total\normalsize}} - \delta_{\mbox{\scriptsize along\normalsize}} = 
\frac{\pa}{\pa\lambda} - \pounds_{\beta}$.\normalsize}
\end{figure}
$_{\mbox{\scriptsize A\normalsize}}{\cal H}^{\mbox{\scriptsize o\normalsize}}$ is the 
$A_i$-contribution to the Hamiltonian constraint on a fixed background.  Such contributions are 
decomposed as follows. First, they are decomposed into translation and tilt parts of the deformation
(Fig 5). 
\begin{figure}[h]
\centerline{\def\epsfsize#1#2{0.4#1}\epsffile{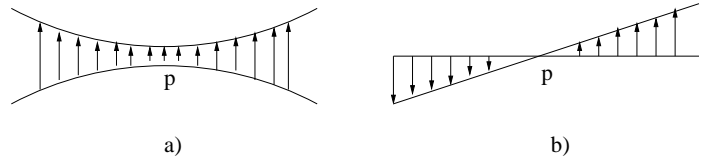}}
\caption[]{\label{TO7.ps}
\scriptsize a)  The \it translation \normalfont part is such that $\alpha(p) \neq 0$, $[\pa_a\alpha](p) = 0$.  
 b) The \it tilt \normalfont part is such that $\alpha(p) = 0$, $[\pa_a\alpha](p) \neq 0$.
\normalsize}
\end{figure}

\be
_{\mbox{\scriptsize A\normalsize}}{\cal H}^{\mbox{\scriptsize o\normalsize}}  =
{}_{\mbox{\scriptsize A\normalsize}}{\cal H}^{\mbox{\scriptsize o\normalsize}}_{\mbox{\scriptsize t\normalsize}}  
+ {_{\mbox{\scriptsize A\normalsize}}{\cal H}^{\mbox{\scriptsize o\normalsize}}_{\not-}} 
\mbox{ } .
\label{Vhamdec}
\ee
The translational part 
$_{\mbox{\scriptsize A\normalsize}}{\cal H}^{\mbox{\scriptsize o\normalsize}}_{\mbox{\scriptsize t\normalsize}}$ 
may contain a term $2 {}_{\mbox{\scriptsize A\normalsize}}P^{ab}K_{ab}$ due to the possibility of 
\it derivative coupling \normalfont of the metric to the 1-form, whilst the remainder of 
$_{\mbox{\scriptsize A\normalsize}}{\cal H}^{\mbox{\scriptsize o\normalsize}}_{\mbox{\scriptsize t\normalsize}}$ 
is denoted by $_{\mbox{\scriptsize A\normalsize}}{\cal H}_{\mbox{\scriptsize t\normalsize}}$: 
\be
_{\mbox{\scriptsize A\normalsize}}{\cal H}^{\mbox{\scriptsize o\normalsize}}_{\mbox{\scriptsize t\normalsize}}  =
{}_{\mbox{\scriptsize A\normalsize}}{\cal H}_{\mbox{\scriptsize t\normalsize}} + 2 {}_{\mbox{\scriptsize 
A\normalsize}}P^{ab}K_{ab} 
\mbox{ } .
\label{Vtransdec}
\ee 

For the 1-form, the tilt part is (up to a divergence)
\be
_{\mbox{\scriptsize (A)\normalsize}}{\cal H}^{\mbox{\scriptsize o\normalsize}}_{\not-} = A^aD_a\pi^{\perp} 
+ A_{\perp} D_a\pi^a 
\mbox{ } .
\label{univtilt}
\ee
The translational part is
$$
_{\mbox{\scriptsize A\normalsize}}{\cal H}_{\mbox{\scriptsize t\normalsize}} = L + \sqrt{h}
(\lambda^{\perp a}D_a A_{\perp} - \lambda^{ab}D_aA_b) 
\mbox{ } .
$$
which is non-universal since $L$ appears.
The derivative-coupling part is
\be
_{\mbox{\scriptsize A\normalsize}}P^{ab} = \frac{\sqrt{h}}{2}( - 
A^{(a|}\lambda^{\perp|b)}
+ A_{\perp}\lambda^{(ab)} - A^{(a}\pi^{b)}) 
\mbox{ } . 
\label{V1formderiv}
\ee

The $A_i$-contribution to the momentum constraint 
$_{\mbox{\scriptsize A\normalsize}}{\cal H}^{\mbox{\scriptsize o\normalsize}}_a$ 
is obtained as in Fig 5 and integrating by parts where necessary.  This `shift kinematics' follows immediately 
from the form of the 1-form Lie derivative.    

For the 1-form, $\lambda^{\perp a}$ and $\lambda^{ab}$ play the role of Lagrange multipliers;
one would then use the corresponding multiplier equations to attempt to eliminate the 
multipliers from (\ref{Vcovectoraction}).  In my examples below, $A_{\perp}$ will also occur 
as a multiplier, but this is generally not the case.

\subsection{$A_{\perp}$ Formulation of Proca Theory}

Consider first the $A_{\perp}$ formulation for the Proca 1-form theory 
(massive analogue of Maxwell theory).  Unlike for Maxwell theory, I show that this 
formulation cannot be cast as a TSA theory.   For the rest of this section, consideration 
of massless cases suffices.  The Proca Lagrangian is
\be
\mbox{\sffamily L\normalfont}^{\mbox{\scriptsize A\normalsize}}_{\mbox{\scriptsize Proca\normalsize}}
= - \nabla_{[A}A_{B]}\nabla^{[A}A^{B]} - 
\frac{m^2}{2}A_{A}A^{A} 
\mbox{ } ,
\label{Vlagproca}
\ee
with corresponding Lagrangian potential
\be
L = -\frac{1}{4}\lambda^{[AB]}\lambda_{[AB]} + 
\frac{m^2}{2}A_{A}A^{A} 
\mbox{ } .
\ee  
The first term in the tilt (\ref{univtilt}) vanishes since $\pi^{\perp} = 0$ by 
antisymmetry for the 1-forms described by (\ref{Vlagproca}).
Also $_{\mbox{\scriptsize A\normalsize}}P^{ab} = 0$ by antisymmetry so
\be
_{\mbox{\scriptsize A\normalsize}}{\cal H}^{\mbox{\scriptsize o\normalsize}} =
\sqrt{h}\left[ -\frac{1}{4}\lambda_{ab}\lambda^{ab} + \frac{1}{2h}\pi_a\pi^a
+ \frac{m^2}{2}(A_aA^a - A_{\perp}^2) - \lambda^{ab}A_{[a,b]} \right] + 
A_{\perp}D_a\pi^a 
\mbox{ } .
\label{VProcaham}
\ee
by (\ref{Vhamdec}, \ref{Vtransdec}).  
The multiplier equation for $\lambda_{ab}$ gives 
\be 
\lambda_{ab} = -2 D_{[b}A_{a]} \equiv B_{ab} 
\mbox{ } .
\label{Vlme}
\ee
For $m \neq 0$, the multiplier equation for $A_{\perp}$ gives
\be
A_{\perp} = - \frac{1}{m^2\sqrt{h}} D_a\pi^a 
\mbox{ } , 
\label{Vlightening}
\ee
and elimination of the multipliers in (\ref{VProcaham}) using (\ref{Vlme}, \ref{Vlightening}) gives
\be
_{\mbox{\scriptsize A\normalsize}}{\cal H}^{\mbox{\scriptsize o\normalsize}} = \frac{1}{2\sqrt{h}}\pi_a\pi^a 
+ \frac{\sqrt{h}}{4}B_{ab}B^{ab}
+ \frac{m^2\sqrt{h}}{2}A_aA^a + \frac{1}{2m^2\sqrt{h}}(D_a\pi^a)^2 
\mbox{ } ,
\label{Vunel}
\ee
which is non-ultralocal in the momenta.  I note that this does nothing to eliminate 
the remaining term in the tilt: the Proca field has nonzero tilt.  

But, for $m = 0$, the $A_{\perp}$ multiplier equation gives instead the Gauss constraint of electromagnetism 
\be  
{\cal G} \equiv D_a\pi^a \approx 0 
\mbox{ } .  
\label{VGaussweak}
\ee
This would not usually permit $A_{\perp}$ to be eliminated from (\ref{Vunel}) but the final 
form of $_{\mbox{\scriptsize A\normalsize}}{\cal H}^{\mbox{\scriptsize o\normalsize}}$ for $m = 0$ is 
\be
_{\mbox{\scriptsize A\normalsize}}{\cal H}^{\mbox{\scriptsize o\normalsize}} = \frac{\sqrt{h}}{4}B^{ab}B_{ab} + 
\frac{1}{2\sqrt{h}}\pi_a\pi^a + A_{\perp}(D_a\pi^a \approx 0) 
\mbox{ } , 
\label{cthulu}
\ee
so the cofactor of $A_{\perp}$ in (\ref{cthulu}) weakly vanishes by (\ref{VGaussweak}), 
so $A_{\perp}$ may be taken to `accidentally' drop out.
This means that the tilt of the Maxwell field may be taken to be zero.

\subsection{Inclusion of Yang--Mills Theory}

The SSF Yang--Mills Lagrangian is\fn{By $\mbox{\sffamily g\normalfont}
{f^{\mbox{\bf\tiny A\normalsize\normalfont}}}_{\mbox{\bf\tiny BC\normalsize\normalfont}}$ 
I strictly mean $\mbox{\sffamily g\normalfont}^{\cal A}
f^{\mbox{\bf\tiny A\normalsize\normalfont}}_{{\cal A}\mbox{\bf\tiny BC\normalsize\normalfont}}$ 
where ${\cal A}$ indexes each gauge subgroup in a direct product.  Then each such gauge 
subgroup can be associated with a distinct coupling constant 
$\mbox{\sffamily g\normalfont}^{\cal A}$.}  
\be
\mbox{\sffamily L\normalfont} = -
\left(
\nabla_{[A}A^{\mbox{\bf\scriptsize A\normalsize\normalfont}}_{B]}
+ \frac{\mbox{\sffamily g\normalfont}}{2}
{f^{\mbox{\bf\scriptsize A\normalsize\normalfont}}}_{\mbox{\bf\scriptsize BC\normalsize\normalfont}}
A^{\mbox{\bf\scriptsize B\normalsize\normalfont}}_{B}
A^{\mbox{\bf\scriptsize C\normalsize\normalfont}}_{A}
\right)
\left(
\nabla^{[A}{A_{\mbox{\bf\scriptsize A\normalsize\normalfont}}}^{B]}
+ \frac{\mbox{\sffamily g\normalfont}}{2}
{f_{\mbox{\bf\scriptsize ADE\normalsize\normalfont}}}A^{\mbox{\bf\scriptsize D\normalsize\normalfont}B}
A^{\mbox{\bf\scriptsize E\normalsize\normalfont}A}
\right)
\mbox{ } .
\label{VlagYMmass}
\ee
I define 
$\lambda^{AB}_{\mbox{\bf\scriptsize M\normalsize\normalfont}}
= \frac{\pa \mbox{\sffamily \scriptsize L\normalsize\normalfont}}
{\pa (\nabla_{B}A^{\mbox{\bf\tiny M\normalsize\normalfont}}_A)}$.   
The Lagrangian potential is then 
\be
L = -\frac{1}{4}\lambda^{[AB]}_{\mbox{\bf\scriptsize M\normalsize\normalfont}}
\lambda^{\mbox{\bf\scriptsize M\normalsize\normalfont}}_{[AB]}
- \frac{\mbox{\sffamily g\normalfont}}{2}
f_{\mbox{\bf\scriptsize BDE\normalsize\normalfont}}
A_{B}^{\mbox{\bf\scriptsize D\normalsize\normalfont}}
A_{A}^{\mbox{\bf\scriptsize E\normalsize\normalfont}}
\lambda^{AB{\mbox{\bf\scriptsize B\normalsize\normalfont}}}
+ \frac{m^2}{2}A_{B{\mbox{\bf\scriptsize M\normalsize\normalfont}}}
A^{A{\mbox{\bf\scriptsize M\normalsize\normalfont}}} 
\mbox{ } .
\ee
The overall tilt contribution is now the sum of the tilt contributions of the individual fields, so  
$_{(\mbox{\scriptsize A\normalsize}_{\mbox{\bf\tiny M\normalsize\normalfont}})}{\cal H}^{\mbox{\scriptsize o\normalsize}}_{\not-}
= A^{\mbox{\bf\scriptsize M\normalsize\normalfont}}_{\perp}D_a
\pi^a_{\mbox{\bf\scriptsize M\normalsize\normalfont}}$
suffices to
generate the
tilt change.
Again,
$_{{\mbox{\scriptsize A\normalsize}}_{\mbox{\bf\tiny M\normalsize\normalfont}}}P^{ab} = 0$
by antisymmetry so 
$$
_{\mbox{\scriptsize A\normalsize}_{\mbox{\bf\tiny M\normalsize\normalfont}}}
{\cal H}^{\mbox{\scriptsize o\normalsize}}
=  \sqrt{h}\left[ -\frac{1}{4}\lambda^{\mbox{\bf\scriptsize 
M\normalsize\normalfont}}_{ab}
\lambda_{\mbox{\bf\scriptsize M\normalsize\normalfont}}^{ab}
+ \frac{1}{2h}\pi^{\mbox{\bf\scriptsize M\normalsize\normalfont}}_a
\pi_{\mbox{\bf\scriptsize M\normalsize\normalfont}}^a
-\lambda_{\mbox{\bf\scriptsize M\normalsize\normalfont}}^{ab}
A^{\mbox{\bf\scriptsize M\normalsize\normalfont}}_{[a,b]}\right]  
$$
\be
+ A^{\mbox{\bf\scriptsize M\normalsize\normalfont}}_{\perp}{\mbox{\bf 
D\normalfont}}_a
\pi_{\mbox{\bf\scriptsize M\normalsize\normalfont}}^a -
\frac{\mbox{\sffamily g\normalfont}}{2}
f_{\mbox{\bf\scriptsize MPQ\normalsize\normalfont}}
(\sqrt{h}\lambda^{ab{\mbox{\bf\scriptsize M\normalsize\normalfont}}}
A^{\mbox{\bf\scriptsize P\normalsize\normalfont}}_bA^{\mbox{\bf\scriptsize 
Q\normalsize\normalfont}}_a +
2 \pi^{\mbox{\bf\scriptsize M\normalsize\normalfont}}
A^{\mbox{\bf\scriptsize P\normalsize\normalfont}}_{\perp}
A^{\mbox{\bf\scriptsize Q\normalsize\normalfont}}_a) 
\ee
by (\ref{Vhamdec}, \ref{Vtransdec}).  
The multipliers are
$\lambda^{ab}_{\mbox{\bf\scriptsize M\normalsize\normalfont}}$
and $A_\perp^{\mbox{\bf\scriptsize M\normalsize\normalfont}}$,
with corresponding
multiplier equations
\be
\lambda^{\mbox{\bf\scriptsize M\normalsize\normalfont}}_{ab}
= -2 D_{[b}A^{\mbox{\bf\scriptsize M\normalsize\normalfont}}_{a]}
\equiv B^{\mbox{\bf\scriptsize M\normalsize\normalfont}}_{ab} 
\mbox{ } ,
\ee
the second multiplier equation gives instead the Yang--Mills Gauss constraint
\be
{\cal G}^{\mbox{\bf\scriptsize M\normalsize\normalfont}}
\equiv {\mbox{\bf D\normalfont}}_a\pi^{\mbox{\bf\scriptsize 
M\normalsize\normalfont}a} \approx 0 
\mbox{ } . 
\label{VYMGaussweak}
\ee
In this case, the tilt is nonzero, but the Yang--Mills Gauss constraint 
`accidentally' enables the derivative part of the tilt to be converted into 
an algebraic expression, which then happens to cancel with part of the Lagrangian potential: 
\be
_{{\mbox{\scriptsize A\normalsize}}_{\mbox{\bf\tiny M\normalsize\normalfont}}}{\cal H}^{\mbox{\scriptsize o\normalsize}} 
= \frac{\sqrt{h}}{4}
B^{ab}_{\mbox{\bf\scriptsize M\normalsize\normalfont}}
B_{ab}^{\mbox{\bf\scriptsize M\normalsize\normalfont}} 
+ \frac{1}{2\sqrt{h}}
\pi^{\mbox{\bf\scriptsize M\normalsize\normalfont}}_a
\pi^a_{\mbox{\bf\scriptsize M\normalsize\normalfont}} 
+ A^{\mbox{\bf\scriptsize M\normalsize\normalfont}}_{\perp}
(D_a\pi^a_{\mbox{\bf\scriptsize M\normalsize\normalfont}} 
+ \mbox{\sffamily g\normalfont}
f_{\mbox{\bf\scriptsize LMP\normalsize\normalfont}}
\pi^{\mbox{\bf\scriptsize L\normalsize\normalfont}a}
A_a^{\mbox{\bf\scriptsize P\normalsize\normalfont}} \approx 0) 
\mbox{ } .
\ee

\subsection{Inclusion of Scalar--1-Form Gauge Theory}

For U(1) 1-form--scalar Gauge Theory the Lagrangian is 
\be
\mbox{\sffamily L\normalfont}
= - \nabla_{[A}A_{B]}\nabla^{[A}A^{B]} +
(\pa_{A}\varsigma - ieA_{A}\chi)(\pa^{A}\varsigma^* + ieA^{A}\varsigma^*) - 
\frac{m_{\varsigma}^2}{2}\varsigma^*\varsigma 
\mbox{ } .
\label{VlagU(1)}
\ee
Now, in addition to $\lambda^{AB}$, I define 
$
\mu^{A} = \frac{\pa\mbox{\sffamily\scriptsize L\normalsize\normalfont}}{\pa (\nabla_{A}\varsigma)} 
\mbox{ and }
\nu^{A} = \frac{\pa\mbox{\sffamily\scriptsize L\normalsize\normalfont}}
{\pa (\nabla_{A}\varsigma^*)} \mbox{ } , 
$
so the Lagrangian potential is
\be
L = -\frac{1}{4}\lambda^{[AB]}\lambda_{[AB]} + 
\frac{m^2}{2}A_{A}A^{A}
+ \mu^{A}\nu_{A} - ieA_{A}(\varsigma^*\nu^{A} - 
\varsigma\mu^{A}) + \frac{m_{\varsigma}^2}{2}\varsigma^*\varsigma 
\mbox{ } .
\mbox{\hspace{0.5in}}
\ee
$_{\mbox{\scriptsize (A)\normalsize}}{\cal H}^{\mbox{\scriptsize o\normalsize}}_{\not-} =   A_{\perp} 
D_a\pi^a$ still
suffices to generate the tilt (as scalars contribute no tilt), 
$$
\mbox{$_{\varsigma, \varsigma^*, \mbox{\scriptsize A\normalsize}}P^{ab} = 0$, and }
_{\varsigma, \varsigma^*, \mbox{\scriptsize A\normalsize}}{\cal H}^{\mbox{\scriptsize o\normalsize}}_{\mbox{\scriptsize t\normalsize}} =  \sqrt{h}
\left[ 
- \frac{1}{4}\lambda_{ab}\lambda^{ab} + \mu_a\nu^a 
+ \frac{1}{h}
\left(
\frac{1}{2}\pi_a\pi^a 
- \pi_{\varsigma}\pi_{\varsigma^*}
\right) 
+ \frac{m_{\varsigma}^2}{2}\varsigma^*\varsigma 
\right.
\mbox{\hspace{2in}}
$$
\be
\mbox{\hspace{2in}}
\left.
- ie\left( A_a[\varsigma^*\nu^a - \varsigma\mu^a] 
- \frac{A_{\perp}}{\sqrt{h}}[\varsigma^*\pi_{\varsigma^*} - \varsigma\pi_{\varsigma}]\right)   
\right] 
\mbox{ } . 
\mbox{\hspace{2in}}
\label{VunelUH}
\ee
The $\lambda_{ab}$ multiplier equation is (\ref{Vlme}) again, whilst the 
$A_{\perp}$ multiplier equation is now 
\be
{\cal G}_{\mbox{\scriptsize U\normalsize\normalfont}(1)}
\equiv D_a\pi^a + ie(\varsigma^*\pi_{\varsigma^*} - \varsigma\pi_{\varsigma}) = 0 
\mbox{ } ,
\label{VsourceGauss}
\ee
which can be explained in terms of electromagnetism now having a fundamental source.  
In constructing $_{\varsigma, \varsigma^*, \mbox{\scriptsize A\normalsize}}{\cal H}^{\mbox{\scriptsize o\normalsize}} $ 
from (\ref{Vhamdec}, \ref{Vtransdec}, \ref{VunelUH}), I can convert the tilt to an algebraic expression by the sourced Gauss law (\ref{VsourceGauss}) 
which again `accidentally' happens to cancel with a contribution from the Lagrangian potential: 
$$
_{\varsigma, \varsigma^*, \mbox{\scriptsize A\normalsize}}{\cal H}^{\mbox{\scriptsize o\normalsize}} 
= -\lambda^{ab}A_{[a,b]} - (\mu^a + \nu^a)\varsigma_{,a} + { }_{(A)}{\cal H}^{\mbox{\scriptsize o\normalsize}}_{\not-} 
+ _{A,\varsigma,\varsigma^*}{\cal H}^{\mbox{\scriptsize o\normalsize}}_{\mbox{\scriptsize t\normalsize}} 
\mbox{\hspace{0.4in}}
$$
$$ 
\mbox{\hspace{2in}}
= 
\left[  
\frac{1}{4}B_{ab}B^{ab} - \mu_a\nu^a 
+ \frac{1}{h}
\left(
\frac{1}{2}\pi_a\pi^a 
- \pi_{\varsigma}\pi_{\varsigma^*}
\right) 
+ \frac{m_{\varsigma}^2}{2}\varsigma^*\varsigma
\right]
\mbox{\hspace{2in}}
$$
\be
\mbox{\hspace{2in}}
+ A_{\perp}[D_a\pi^a + ie(\varsigma^*\pi_{\varsigma^*} - \varsigma\pi_{\varsigma}) \approx 0] 
\mbox{ } . 
\mbox{\hspace{2in}}
\ee  
It is not too hard to show that the last two accidents also accidentally conspire together to 
wipe out the tilt contribution in Yang--Mills 1-form--scalar Gauge Theory used to obtain broken 
SU(2) $\times$ U(1) bosons for the electroweak force.  This theory is also obviously 
nonderivative-coupled.

\subsection{Concluding Remarks about the Inclusion of Bosonic Theories}

The above `accidents' are all of the following form.  
They arise from eliminating $A_{\perp}$ from its multiplier equation.  For this to make sense, $A_{\perp}$ must 
{\sl be} a multiplier, thus $\pi^{\perp} = 0$.  Then for general $L$, the multiplier equation is
\be
\frac{\pa L}{\pa A_{\perp}} +D_a\pi^a = 0.
\label{multipleq}
\ee
Then the requirement that $A_{\perp}D_a\pi^a + L$ be independent of $A_{\perp}$ on using 
(\ref{multipleq}) means that $- A_{\perp}\frac{\pa L}{\pa A_{\perp}} + L$ is independent of 
$A_{\perp}$.  Thus the `accidents' occur whenever the Lagrangian potential is linear in $A_{\perp}$.
This is a particular instance of {\bf RI-castability}.  Other strategies considered elsewhere include 
changes of variables and integration by parts.  So Einstein--Maxwell theory, Einstein--Yang--Mills theory, 
and their corresponding scalar Gauge Theories have `shift kinematics' alone.  The removability of their tilts 
ensures that these coupled to GR may be cast as TSA theories.  Such workings begin to illustrate what sorts of 
obstacles within the SSF ontology might be regarded as responsible for the uniqueness 
results for bosonic matter within BF\'{O}'s TSA ontology.

\section{Inclusion of Spin-1/2 Fermions}

One also needs to be able to account for nature's spin-$\frac{1}{2}$ fields.  I use same idea 
as in the preceding section to demonstrate that the TSA can accommodate these.  I do so by 
considering the following 4-component spacetime spinor formalism.  First, I use 4-spinors 
$\psi^{\rho}$, usually suppressing their indices.  In flat spacetime, these solve the Dirac 
equation $i\gamma^{\bar{A}}\pa_{\bar{A}}\psi - m\psi = 0$ where $\gamma^{\bar{A}}$ are Dirac 
matrices obeying the Dirac algebra 
\be
\gamma^{\bar{A}}\gamma^{\bar{B}} + \gamma^{\bar{B}}\gamma^{\bar{A}} = 2\eta^{\bar{A}\bar{B}} 
\mbox{ } .  
\ee
For convenience below, I adopt the chiral representation for these: 
\be
\gamma^{\bar{0}} = 
\left(
\begin{array}{ll}
0 & \mbox{ } 1| \\ 1| & \mbox{ } 0
\end{array}
\right)
\mbox{ } , 
\mbox{ } 
\gamma^{\bar{a}} = \left(
\begin{array}{ll}
\mbox{ } 0 & \sigma^{\bar{a}} \\ -\sigma^{\bar{a}} & 0
\end{array}
\right)
\label{chiralgamma}
\ee
where $1|$ is the unit 2 $\times$ 2 matrix and 
\be
\sigma^{\bar{1}} =
\left(
\begin{array}{ll}
0 & \mbox{ } 1 \\ 1 & \mbox{ } 0
\end{array}
\right)
\mbox{ } \mbox{ } , 
\mbox{ } \mbox{ }
\sigma^{\bar{2}} =
\left(
\begin{array}{ll}
0 & -i \\ i & \mbox{ } 0
\end{array}
\right)
\mbox{ } \mbox{ } , 
\mbox{ } \mbox{ }
\sigma^{\bar{3}} =
\left(
\begin{array}{ll}
1 & \mbox{ } 0 \\ 0 & -1
\end{array}
\right)
\ee 
are the Pauli matrices.  
Next, introduce Dirac's suited triads \cite{VDiracrec} 
${\mbox{\sc e}_{\bar{A}}}^{B}$, where the barred indices are flat spacetime indices.  
These obey 
$\mbox{\sc e}_{\bar{0}a} = 0$, 
$\mbox{\sc e}_{\bar{0}0} = - \alpha$, 
${\mbox{\sc e}_{\bar{A}}}^{B} \mbox{\sc e}_{\bar{C}B} = \eta_{\bar{A}\bar{C}}$ and 
$\mbox{\sc e}_{\bar{A}B} {\mbox{\sc e}^{\bar{A}}}_{C} = g_{BC}$.
The spacetime {\it spinorial covariant derivative} is 
\be
\nabla^{\mbox{\scriptsize S\normalsize}}_{\bar{A}}\psi 
= \pa_{\bar{A}}\psi - \frac{1}{4}\Omega_{\bar{R}\bar{S}\bar{A}}\gamma^{\bar{R}}\gamma^{\bar{S}}\psi 
\mbox{ } , 
\ee
where
\be
\Omega_{\bar{R}\bar{S}\bar{P}} = (\nabla_{B}\mbox{\sc e}_{\bar{R}A})
                                            {\mbox{\sc e}^{A}}_{\bar{S}} {\mbox{\sc e}^{B}}_{\bar{P}} 
\label{4scdef}
\ee
is the spacetime {\it spin connection}.  The form of the spinorial covariant derivative 
can be justified from the point of view of fibre bundles as follows \cite{Kosmann}.  The spin bundle on a curved manifold 
has that manifold as its base space, $C^2$ fibres and structure group SL(2, C), which is 
the universal covering group of SO(3, 1) (which is relevant because at each point of the 
manifold, the manifold is locally Minkowskian).  Consider 2 maps from SL(2, C).  The first is 
the adjoint action $\Lambda$ mapping to SO(3, 1) and the other is the representation 
$\Gamma$, thus mapping to GL(4, C).
$$
\begin{array}{llllll}
\underline{\mbox{space}}
&
SO(3, 1) 
& 
\mbox{ } \stackrel{\Lambda}{\longleftarrow} \mbox{ }
& 
SL(2, C) 
& 
\mbox{ } \stackrel{\Gamma}{\longrightarrow} \mbox{ } 
&  
GL(4, C) 
\\
\underline{\mbox{elements}}
&
\Lambda(A) 
&
&
A
&
&
\left(
\begin{array}{cc}
A
&
0
\\
0
&
A^{\dagger - 1}
\end{array}
\right)
\\
\underline{\mbox{connections}}
&
w = \Lambda_*W
&
&
W
&
&
\Omega = \Gamma_*W
\end{array}
$$
The idea is to use knowledge of SO(3, 1) to deduce the form of the GL(4, C), 
by composition of the inverse of $\Lambda$ and $\Gamma$. Now, the connection 
$W = \sum_{\bar{\i}<\bar{\j}}E_{{\bar{\i}}{\bar{\j}}}{w^{\bar{\i}}}_{\bar{\j}} 
= \frac{1}{2}\sum_{\bar{\i}<\bar{\j}}
\sigma^{\bar{\i}}\sigma^{\bar{\j}}w^{{\bar{\i}}{\bar{\j}}}$ 
\cite{Frankel}
where $E_{{\bar{\i}}{\bar{\j}}}$ are the basis 
$$
E_{{\bar{1}}{\bar{2}}} = 
\left(
\begin{array}{cccc}
0 & \mbox{ } 0 & 0 & \mbox{ } 0 \\ 
0 & \mbox{ } 0 & 1 & \mbox{ } 0 \\
0 &     -1     & 0 & \mbox{ } 0 \\
0 & \mbox{ } 0 & 0 & \mbox{ } 0 \\
\end{array}
\right) \mbox{ } , \mbox{ } 
E_{{\bar{2}}{\bar{3}}} = 
\left(
\begin{array}{cccc}
0 & \mbox{ } 0 & \mbox{ } 0 & 0 \\ 
0 & \mbox{ } 0 & \mbox{ } 0 & 0 \\
0 & \mbox{ } 0 & \mbox{ } 0 & 1 \\
0 & \mbox{ } 0 &    -1      & 0 \\
\end{array}
\right) \mbox{ } , \mbox{ }
E_{{\bar{3}}{\bar{1}}} = 
\left(
\begin{array}{cccc}
0 & \mbox{ } 0 & \mbox{ } 0 & \mbox{ } 0 \\ 
0 & \mbox{ } 0 & \mbox{ } 0 &     -1     \\
0 & \mbox{ } 0 & \mbox{ } 0 & \mbox{ } 0 \\
0 & \mbox{ } 1 & \mbox{ } 0 & \mbox{ } 0 \\
\end{array}
\right) \mbox{ } , \mbox{ }
$$
$$
E_{{\bar{0}}{\bar{1}}} = 
\left(
\begin{array}{cccc}
0 & \mbox{ }1 & \mbox{ }0 & \mbox{ } 0 \\ 
1 & \mbox{ }0 & \mbox{ }0 & \mbox{ } 0 \\
0 & \mbox{ }0 & \mbox{ }0 & \mbox{ } 0 \\
0 & \mbox{ }0 & \mbox{ }0 & \mbox{ } 0 \\
\end{array}
\right) \mbox{ } , \mbox{ } 
E_{{\bar{0}}{\bar{2}}} = 
\left(
\begin{array}{cccc}
0 & \mbox{ } 0 & \mbox{ } 1 & \mbox{ } 0 \\ 
0 & \mbox{ } 0 & \mbox{ } 0 & \mbox{ } 0 \\
1 & \mbox{ } 0 & \mbox{ } 0 & \mbox{ } 0 \\
0 & \mbox{ } 0 & \mbox{ } 0 & \mbox{ } 0 \\
\end{array}
\right) \mbox{ } , \mbox{ }
E_{{\bar{0}}{\bar{3}}} = 
\left(
\begin{array}{cccc}
0 & \mbox{ } 0 & \mbox{ } 0 & \mbox{ } 1 \\ 
0 & \mbox{ } 0 & \mbox{ } 0 & \mbox{ } 0 \\
0 & \mbox{ } 0 & \mbox{ } 0 & \mbox{ } 0 \\
1 & \mbox{ } 0 & \mbox{ } 0 & \mbox{ } 0 \\
\end{array}
\right) \mbox{ } , \mbox{ }
$$
of SO(3, 1) matrices, the first step is just an expansion i.t.o a basis and the second step 
uses $\Lambda_{*}(\sigma_{\bar{\i}}\sigma_{\bar{\j}}) = 2E_{{\bar{\i}}{\bar{\j}}}$.  
Finally, $\Omega = \frac{1}{4}w_{{\bar{\i}}{\bar{\j}}}\gamma^{\bar{\i}}\gamma^{\bar{\j}}$ 
by $\Omega = \Gamma_*W$ and the identities 
$$
\gamma_{\bar{0}}\gamma_{\bar{b}} = \left(\begin{array}{cc}\sigma_{\bar{b}} & 0 \\ 0 & -\sigma_{\bar{b}}\end{array}\right), 
\gamma_{\bar{a}}\gamma_{\bar{b}} = \left(\begin{array}{cc}\sigma_{\bar{a}}\sigma_{\bar{b}} & 0 \\ 0 & \sigma_{\bar{a}}\sigma_{\bar{b}}\end{array}\right).
$$   

I next decompose this, keeping track of the geometrical significance of the various pieces, 
in the style of Henneaux \cite{Hen}.  I supply each piece with contracting gamma matrices as 
suits its later application.  As
\be
\Omega_{(\bar{P}\bar{Q})\bar{R}} = 0
\mbox{ } , 
\label{spantisym}
\ee
there are 4 components in its decomposition. $\omega_{\bar{p}\bar{q}\bar{r}}$ may be used as 
it is, to form the 3-d spinorial covariant derivative: 
\be
D^{\mbox{\scriptsize S\normalsize}}_{\bar{p}}\psi = \pa_{\bar{p}}\psi - \frac{1}{4}\omega_{\bar{r}\bar{s}\bar{p}}\gamma^{\bar{r}}\gamma^{\bar{s}}\psi 
\mbox{ } , 
\ee
where
\be
\omega_{\bar{r}\bar{s}\bar{p}} = (D_{b}\mbox{\sc e}_{\bar{r}a})
                                 {\mbox{\sc e}^{a}}_{\bar{s}}{\mbox{\sc e}^{b}}_{\bar{p}}
\label{3scdef}
\ee
is the spatial spin connection.  

By (\ref{3scdef}), suited tetrad properties, the Dirac algebra 
and the well-known properties of the extrinsic curvature 
$
K_{ab} = K_{ba} \mbox{ } , \mbox{ } K_{a\perp} = 0 \mbox{ } , 
$ 
one arrives at 
\be  
\gamma^{\bar{d}}\gamma^{\bar{b}}\omega_{\bar{0}\bar{d}\bar{b}} = - K 
\mbox{ } . 
\label{womble}
\ee
By (\ref{3scdef}), the form of the metric connection and use of suited tetrad properties, 
\be
\omega_{\bar{b}\bar{0}\bar{0}} = - \frac{\pa_{\bar{b}}\alpha}{\alpha} 
\mbox{ } . 
\ee
By (\ref{3scdef}), 
\be
\omega_{\bar{b}\bar{d}\bar{0}} = \pa_{\bar{0}}\mbox{\sc e}_{[\bar{b}|l}{\mbox{\sc e}^l}_{\bar{d}]} 
- {\alpha}(\pounds_{\beta}\mbox{\sc e}_{\bar{d}a}){\mbox{\sc e}^a}_{\bar{b}} 
\mbox{ } . 
\label{ursa}
\ee

Then, using (\ref{4scdef}), (\ref{spantisym}) and (\ref{3scdef}), 
$$
\gamma^{\bar{A}}\nabla^{\mbox{\scriptsize S\normalsize}}_{\bar{A}}\psi 
= \gamma^{\bar{0}}\nabla^{\mbox{\scriptsize S\normalsize}}_{\bar{0}}\psi 
+ \gamma^{\bar{l}}\nabla^{\mbox{\scriptsize S\normalsize}}_{\bar{l}}\psi 
$$                     
$$
= \gamma^{\bar{0}}\nabla^{\mbox{\scriptsize S\normalsize}}_{\bar{0}}\psi + 
\left(
\gamma^{\bar{l}}\pa_{\bar{l}}\psi 
- \frac{1}{4}\gamma^{\bar{l}}\omega_{\bar{b}\bar{d}\bar{l}}\gamma^{\bar{b}}\gamma^{\bar{d}}\psi 
- \frac{1}{4}\gamma^{\bar{l}}\omega_{\bar{b}\bar{0}\bar{l}}\gamma^{\bar{b}}\gamma^{\bar{0}}\psi 
- \frac{1}{4}\gamma^{\bar{l}}\omega_{\bar{0}\bar{d}\bar{l}}\gamma^{\bar{0}}\gamma^{\bar{d}}\psi
- \frac{1}{4}\gamma^{\bar{l}}\omega_{\bar{0}\bar{0}\bar{l}}\gamma^{\bar{0}}\gamma^{\bar{0}}\psi
\right)
$$
\be
= \gamma^{\bar{0}}\nabla^{\mbox{\scriptsize S\normalsize}}_{\bar{0}}\psi 
+ \gamma^{\bar{l}}D^{\mbox{\scriptsize S\normalsize}}_{\bar{l}}\psi 
+ \frac{1}{2}\gamma^{\bar{0}}\gamma^{\bar{l}}\gamma^{\bar{b}}\omega_{\bar{0}\bar{b}\bar{l}}\psi 
\label{threeterms}
\ee
Then the first term may be replaced by 
\be
\alpha\nabla^{\mbox{\scriptsize S\normalsize}}_{\bar{0}}\psi 
= \dot{\psi} 
- \pounds^{\mbox{\scriptsize S\normalsize}}_{\beta}\psi 
- \pa_{\mbox{\scriptsize R\normalsize}}\psi 
+ \frac{1}{2}\pa_{\bar{b}}\alpha\gamma^{\bar{0}}\gamma^{\bar{b}} 
\ee
by splitting (\ref{ursa}) into two pieces; the first of these is directly geometrically 
meaningful, whereas the second is geometrically meaningful when combined with 
$\pounds_{\beta}$:
\be
\pa_{\mbox{\scriptsize R\normalsize}}\psi 
\equiv \frac{1}{4}{\mbox{\sc e}^i}_{[\bar{r}}\dot{\mbox{\sc e}}_{\bar{s}]i}\gamma^{\bar{r}}\gamma^{\bar{s}}\psi    
\mbox{\hspace{1in}}
\mbox{ (triad rotation correction) ,}
\label{VRottcod}
\ee 
and
\be
\pounds^{\mbox{\scriptsize S\normalsize}}_{\beta}\psi \equiv \beta^i\pa_i\psi 
- \frac{1}{4}{\mbox{\sc e}^i}_{[\bar{r}|}\pounds_{\beta}
             \mbox{\sc e}_{|\bar{s}]i}\gamma^{\bar{s}}\gamma^{\bar{r}}\psi 
\mbox{\hspace{0.7in}}
\mbox{ (Lie derivative) .}
\label{Vsplied}
\ee 
Thus the tensorial Lie derivative $\pounds_{\beta}\psi = \beta^i\pa_i\psi$ is but a piece of 
the spinorial Lie derivative \cite{GH, Kosmann}.  

Here I consider the origin of this expression for the spinorial Lie derivative.  
While the dragging first principles formulation of Lie derivatives cannot be generalized to 
spinors, it is the covariant derivative reformulation of Lie derivatives available on affine manifolds 
that is to be extended to spinors.  This requires mapping from the 
group on the tangent space GL(4, R) to SO(3, 1) in addition to the maps mentioned 
in the definition of the spinorial covariant derivative.  
The new map is just the antisymmetrization map a  
$$
\begin{array}{lll}
GL(4, R)
&
\stackrel{\mbox{a}}{\longrightarrow}
&
SO(3, 1)  
\\
M_{ij}
&
&
M_{[ij]} 
\end{array}
\mbox.
$$

The second term in (\ref{threeterms}) is already in 
clear-cut spatial form, while the last term is just $-\frac{\gamma^{\bar{0}}K}{2}$, by (\ref{womble}).  Thus
\be
\sqrt{|g|}\bar{\psi}\gamma^{\bar{A}}
\nabla^{\mbox{\scriptsize S\normalsize}}_{\bar{A}}\psi 
= i\sqrt{h}{\psi}^{\dagger}
\left[
\alpha\gamma^{\bar{0}}\gamma^{\bar{l}}D^{\mbox{\scriptsize S\normalsize}}_{\bar{l}}\psi 
+ \frac{\alpha K}{2}\psi + \pa_{\bar{l}}\alpha\gamma^{\bar{0}}\gamma^{\bar{l}}\psi  
- (\dot{\psi} - \pounds^{\mbox{\scriptsize S\normalsize}}_{\beta}\psi 
- \pa_{\mbox{\scriptsize R\normalsize}}\psi)
\right].  
\label{Vspinsplit}
\ee

Next, although derivative coupling (second term) and tilt (third term) appear to be present 
in (\ref{Vspinsplit}), G\'{e}h\'{e}niau and Henneaux \cite{GH} observed that these simply 
cancel out in the Dirac field contribution to the Lagrangian density, 
$$
\sqrt{|g|}\mbox{\sffamily L\normalfont}_{\mbox{\scriptsize D\normalsize}} = 
\sqrt{|g|}
\left[
\frac{1}{2}(\bar{\psi}\gamma^{\bar{A}}\nabla^{\mbox{\scriptsize S\normalsize}}_{\bar{A}}\psi
- \nabla^{\mbox{\scriptsize S\normalsize}}_{\bar{A}}\bar{\psi}\gamma^{\bar{A}}\psi) 
- m_{\psi}\bar{\psi}\psi
\right]
$$
$$
= \frac{i\sqrt{h}}{2}
\left(
\psi^{\dagger}
\left[
\alpha\gamma^{\bar{0}}\gamma^{\bar{l}}D^{\mbox{\scriptsize S\normalsize}}_{\bar{l}}\psi 
+ \frac{\alpha K}{2}\psi + \pa_{\bar{l}}\alpha\gamma^{\bar{0}}\gamma^{\bar{l}}\psi  
- (\dot{\psi} - \pounds^{\mbox{\scriptsize S\normalsize}}_{\beta}\psi 
- \pa_{\mbox{\scriptsize R\normalsize}}\psi)
\right]
- 
\right.
$$
$$
\left.
\left[
\alpha\gamma^{\bar{0}}\gamma^{\bar{l}}D^{\mbox{\scriptsize S\normalsize}}_{\bar{l}}\psi^{\dagger} 
+ \frac{\alpha K}{2}\psi^{\dagger} 
+ \pa_{\bar{l}}\alpha\gamma^{\bar{0}}\gamma^{\bar{l}}\psi^{\dagger}  
- (\dot{\psi}^{\dagger} - \pounds^{\mbox{\scriptsize S\normalsize}}_{\beta}\psi^{\dagger} 
- \pa_{\mbox{\scriptsize R\normalsize}}\psi^{\dagger})
\right] 
\right)
- \sqrt{h}\alpha m_{\psi}\bar{\psi}\psi
$$
$$
= 
\frac{i\sqrt{h}}{2}
\left[
\psi^{\dagger}\alpha\gamma^{\bar{0}}\gamma^{\bar{l}}D^{\mbox{\scriptsize S\normalsize}}_{\bar{l}}\psi  
- \psi^{\dagger}(\dot{\psi} - \pounds^{\mbox{\scriptsize S\normalsize}}_{\beta}\psi 
- \pa_{\mbox{\scriptsize R\normalsize}}\psi)
- \alpha\gamma^{\bar{0}}\gamma^{\bar{l}}(D^{\mbox{\scriptsize S\normalsize}}_{\bar{l}}\psi^{\dagger})\psi  
- (\dot{\psi}^{\dagger} - \pounds^{\mbox{\scriptsize S\normalsize}}_{\beta}\psi^{\dagger} 
- \pa_{\mbox{\scriptsize R\normalsize}}\psi^{\dagger})\psi 
\right]
$$
\be
- \sqrt{h}\alpha m_{\psi}\bar{\psi}\psi 
\mbox{ } .
\label{Vfermilag}
\ee

The  
G\'{e}h\'{e}niau--Henneaux formulation is clearly encouraging for the TSA.  For, using the BSW 
procedure on the combined split Einstein--Dirac action, 
one immediately obtains a {\bf RI} action:  
$$
\mbox{\sffamily I\normalfont} = \int \textrm{d}\lambda
\int \textrm{d}^3x \sqrt{h}
\left[
\sqrt{      \Lambda + \sigma R + 
\psi^{\dagger}\gamma^{\bar{0}}\gamma^{\bar{l}}D^{\mbox{\scriptsize S\normalsize}}_{\bar{l}}\psi  
- \gamma^{\bar{0}}\gamma^{\bar{l}}(D^{\mbox{\scriptsize S\normalsize}}_{\bar{l}}\psi^{\dagger})\psi  
- m_{\psi}\bar{\psi}\psi       }
\sqrt{\mbox{\sffamily T\normalfont}^{\mbox{\scriptsize g\normalsize}}} 
\right.
$$
\be
\left.
- \psi^{\dagger}(\dot{\psi} - \pounds^{\mbox{\scriptsize S\normalsize}}_{\beta}\psi 
- \pa_{\mbox{\scriptsize R\normalsize}}\psi)
+ (\dot{\psi}^{\dagger} - \pounds^{\mbox{\scriptsize S\normalsize}}_{\beta}\psi^{\dagger} 
- \pa_{\mbox{\scriptsize R\normalsize}}\psi^{\dagger})\psi 
\right] 
\mbox{ }  
\label{specificfermi}
\ee  

Thus one knows that 
$$
\mbox{\sffamily I\normalfont}  
= \int \textrm{d}\lambda\int \textrm{d}^3x \sqrt{h}
\left[
\sqrt{      \Lambda + \sigma R + 
\psi^{\dagger}\gamma^{\bar{0}}\gamma^{\bar{l}}D^{\mbox{\scriptsize S\normalsize}}_{\bar{l}}\psi  
- \gamma^{\bar{0}}\gamma^{\bar{l}}(D^{\mbox{\scriptsize S\normalsize}}_{\bar{l}}\psi^{\dagger})\psi  
- m_{\psi}\bar{\psi}\psi       }
\sqrt{\mbox{\sffamily T\normalfont}^{\mbox{\scriptsize g\normalsize}}(\&_{\dot{\mbox{\scriptsize s\normalsize}}}h_{ij})} 
\right.
$$
\be
\left.
- \psi^{\dagger}\&_{\dot{\mbox{\scriptsize s\normalsize}}, \mbox{\scriptsize r\normalsize}}{\psi} 
+ \&_{\dot{\mbox{\scriptsize s\normalsize}}, \mbox{\scriptsize r\normalsize}}{\psi}^{\dagger}\psi 
\right]
\mbox{ }  
\label{specificfermi3}
\ee  
will work as a spatial ontology starting-point for Einstein--Dirac theory. Note that this 
is {\bf RI} but not a homogeneous quadratic {\bf RI} form such as the BSW action of GR.  Note 
that in addition to the more complicated form of the Diff-{\bf BM} correction to the Dirac 
velocities, there is also a triad rotation correction (\ref{VRottcod}) (which requires a new 
rotation auxiliary $r^i$).  Thus the {\bf BM[GSR]} implementation should be generalized to 
accommodate this additional, natural geometric correction: given two spinor-bundle 
3-geometries $\Sigma_1$, $\Sigma_2$, the (full spinorial) drag shufflings of $\Sigma_2$ 
(keeping $\Sigma_1$ fixed) are accompanied by the rotation shufflings of the triads glued to 
it (Fig 6).   The triad rotation correction then leads to a further `locally Lorentz' 
constraint \cite{VDiracrec}.  
\begin{figure}[h]
\centerline{\def\epsfsize#1#2{0.4#1}\epsffile{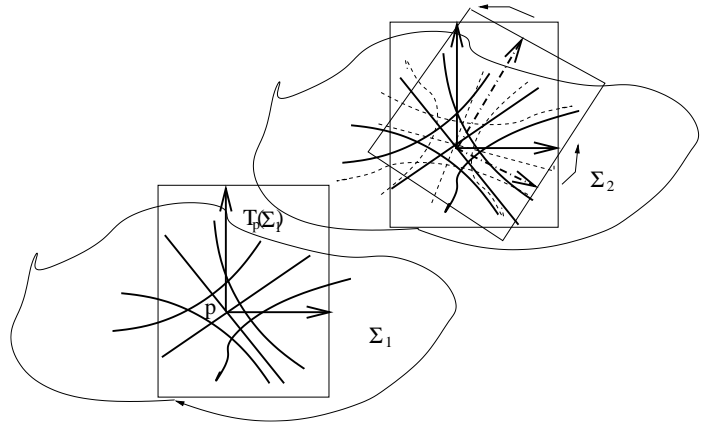}}
\caption[]{\label{TO7.ps}
\scriptsize Best matching of coodinate grids and of the triads used to define spatial spinors.  
One dimension is suppressed.  I view the triads as basis vectors for the tangent planes 
$\mbox{T}_p(\Sigma)$ at each $p$.      
\normalsize}
\end{figure}


Also note I succeed in including the 1-form--fermion interaction terms of the Einstein--Standard 
Model theory: 
\be
\mbox{\sffamily g\normalfont}^{\cal A}\tau_{{\cal A}\mbox{\scriptsize\bf I\normalfont\normalsize}}
\bar{\psi}\gamma^{\bar{B}}
{\mbox{\sc e}^{A}}_{\bar{B}}A^{\mbox{\scriptsize\bf I\normalfont\normalsize}}_{A}\psi
\label{Vstar}
\ee
where ${\cal A}$ takes the values U(1), SU(2) and SU(3).  The decomposition of these into 
spatial quantities causes no difficulties.  In particular, they clearly 
contribute linearly in $A_{\perp}$ to the Lagrangian potential, so by the argument at the end 
of VI.2.2, an accident occurs ensuring that tilt kinematics is not necessary.  Also, clearly 
the use of the form (\ref{Vfermilag}) is compatible with the inclusion of the interactions 
(\ref{Vstar}) since, acting on $\bar{\psi}$ the gauge correction is the opposite sign.  So 
our proposed formulation's combined Standard Model matter Lagrangian is\fn{Here 
$\mbox{\sffamily L\normalfont}_{\mbox{\tiny YM\normalsize}}$
is given by the SU(3) $\times$ SU(2) $\times$ U(1) version of (\ref{VlagYMmass}) and 
one would need to sum the square bracket over all the known fundamental fermionic species.}    
\be
\mbox{\sffamily L\normalfont}_{\mbox{\scriptsize SM\normalsize}} = 
\frac{1}{2}
\left[
\bar{\psi}\gamma^{\bar{A}}
(\nabla^{\mbox{\scriptsize S\normalsize}}_{\bar{A}} 
- \mbox{\sffamily g\normalfont}_{\cal A}\tau^{\cal A}_{\mbox{\scriptsize\bf I\normalfont\normalsize}}
{\mbox{\sc e}^{B}}_{\bar{A}} A^{\mbox{\scriptsize\bf I\normalfont\normalsize}}_{B})\psi
- (\nabla^{\mbox{\scriptsize S\normalsize}}_{\bar{A}} 
+ \mbox{\sffamily g\normalfont}_{\cal A}\tau^{\cal A}_{\mbox{\scriptsize\bf I\normalfont\normalsize}}
{\mbox{\sc e}^{B}}_{\bar{A}}A^{\mbox{\scriptsize\bf I\normalfont\normalsize}}_{B}     
\bar{\psi}\gamma^{\bar{A}}\psi) 
\right]
- m_{\psi}\bar{\psi}\psi
+ \mbox{\sffamily L\normalfont}_{\mbox{\scriptsize YM\normalsize}} 
\mbox{ } .  
\ee
which can be adjoined to the split GR Lagrangian, 
and successfully subjected to the BSW procedure, so an enlarged version of 
(\ref{specificfermi3}) will serve as a spatial ontology starting-point.  
  
There is also no trouble with the incorporation of the Yukawa interaction term 
$\varsigma\bar{\psi}\psi$ which could be required for some fermions to pick up mass from a 
Higgs scalar.  Thus the Lagrangian for all the known fundamental matter fields can be built 
from the TSA principles.   

In starting from consideration of 3-space, the natural 3-d kinematics embodied in 
(\ref{VRottcod}) and (\ref{Vsplied}) happen to suffice throughout for difference-type 
Lagrangians.  The action is not yet in entirely spatial terms because it is i.t.o gamma 
matrices and 4-component spinors which are natural to spacetime.  But these features can be 
ammended.  For, one can choose to work in the chiral representation and use how the Dirac 
matrices are built out of the Pauli matrices associated with SO(3),\fn{This is standard use 
of representation theory, based on the accidental Lie algebra relation 
SO(4)$\cong$ SO(3) $\bigoplus$ SO(3), which depends on space being 3-d.  By SO(3, 1) and SO(3) 
spinors, I strictly mean spinors corresponding to their universal covering groups, SL(2, C) 
and SU(2) respectively.} and use how the 4-spinor $\psi^{\rho}$ may be rewritten as 

\noindent
$$ 
\psi^{\rho} = 
\left[
\begin{array}{l}
\psi_{\mbox{\scriptsize D\normalsize}}^{\mbox{\tt A\normalfont}} \\
\psi_{\mbox{\scriptsize L\normalsize}}^{\mbox{\tt A\normalfont}} \\
\end{array}
\right] 
\mbox{ } , 
\label{chiralspinor}
$$
where D (`dextero') and L (`laevo') stand for right- and left-handed SO(3) 2-spinors.  Of 
course, SO(3) is the 3-d spatial rotation group, so the action is now in entirely spatial 
terms.  

To accommodate neutrino (Weyl) fields, one would consider a single 
$\psi_{\mbox{\scriptsize L\normalsize}}$ SO(3) spinor, i.e set its pair 
$\psi_{\mbox{\scriptsize D\normalfont}}$ and its mass to zero before the variation is carried 
out.  Whilst we are free to accommodate all the known fundamental fermionic fields in the 
TSA, one cannot predict the number of Dirac and Weyl fields present in nature nor their 
masses nor the nongravitational forces felt by each field.

\section{The TSA Does Not Exhaustively Pick Out the `Fields of Nature'}

Does the TSA pick out the matter fields standardly used to describe nature, or does it also 
permit fields offering other, more complicated explanations, or fields which have no currently 
observed consequences?  I first stress that the TSA has no control over how many fields there 
are in nature, nor any control of their masses, interaction strengths or gauge groups.  I next 
explored whether more exotic departures from the 
standard fields are possible in the TSA.  In \cite{Vanderson}, I showed that the massless 2-form, 
for which there is no current evidence in physical observations,  is also possible in the TSA.  Here I show how 
to include various other sorts of fields.  This provides evidence against the TSA picking out the matter fields 
standardly used to describe nature.  The earlier papers' matter results in fact rely heavily on simplicities and not 
on the TSA's relational principles.  This evidence also counts firmly against BF\'{O}'s speculation that the TSA 
``hints at partial unification'' of gravity and electromagnetism.

\subsection{A Means of Including Proca Theory}

I can begin to relate this occurrence to the BSW or generalized BSW implementation of 
\bf GSR\normalfont.  For, suppose an action has a piece depending on 
$\pa_a\alpha$ in it.  Then the immediate elimination of $\alpha$ from it {\sl is not 
algebraic}, i.e the BSW procedure is not possible.  By definition, the tilt part of 
the Hamiltonian constraint is built from the $\pa_a\alpha$ contribution using integration by 
parts.  But, for the $A_{\perp}$-eliminated Proca Lagrangian, this integration by parts 
gives a term that is non-ultralocal in the momenta, $(D_a\pi^a)^2$, which again contains 
$\pa_a\alpha$ within.  Thus, for this formulation of Proca theory, within the SSF one cannot build a 
TSA Einstein--Proca action to start off with.   Of importance, this difficulty with spatial 
derivatives was not foreseen in the simple analogy with the Jacobi principle in mechanics,
where there is only one independent variable.  

\noindent  
Note that whether a theory can be cast into TSA form can only be treated formalism by formalism.    
I explain here how to obtain a formalism in which Proca theory is allowed.  
The tricks used {\sl do not} suffice to put all other theories I considered into 
TSA form.  Thus the TSA retains some selectivity, albeit less than previously assumed.  
I furthermore tie this selectivity to the {\bf Principle of Equivalence (POE)} in Sec 12.  

Note that Proca theory does have its uses, so its inclusion should be viewed as a favourable 
result.  Proca theory appears phenomenologically e.g in superconductivity.  Also, having included 
Proca theory, it is then easy to see how to include massive Yang--Mills theory as a phenomenological theory 
of what the weak bosons look like today.  Whereas these applications are quite peripheral, it is 
nevertheless reassuring that one need not abandon the TSA to do phenomenology.  

I begin with another way of looking at electromagnetism.  The `accident' method of 
Sec 9 `lets go' of the constraint; fortunately it is `caught again' because 
it arises as an integrability, but one would not generally expect this to be 
the case.  One could rather avoid the tilt by {\sl redefining variables} 
according to $A_{\perp} \longrightarrow A_0 = - \alpha A_{\perp}$.  Then one never 
`lets go' of the constraint.  

The above approach then generalizes to Proca theory, leading to the action
\be
\mbox{\sffamily I\normalfont}
= \int\textrm{d}\lambda \int\textrm{d}^3x\sqrt{h}
\sqrt{       D_{[a}A_{b]}D^{[a}A^{b]}\mbox{+}m^2 A_aA^a\mbox{+}R    }
\sqrt{       [\dot{A}_a\mbox{--}\pounds_{\dot{\mbox{\scriptsize s\normalsize}}}A_a\mbox{--}\pa_a\dot{\Xi}]^2 + m^2 
\dot{\Xi}^2
\mbox{+}\mbox{\sffamily T\normalfont}_{\mbox{\scriptsize g\normalfont}}   }    
\ee
where $\dot{\Xi}$ may be identified with $A_0$.  Moreover, objections to this approach on 
grounds of it giving an action that is not best matched are not valid.  One can 
gravitationally best match the auxiliary velocity $\dot{\Xi}$  if one wishes, since this 
only disturbs the equations of motion weakly.  

Also note that unlike Gauge Theories' constraints \cite{HTbook}, the Proca constraint is 
second-class \cite{Dirac}. It then makes no sense by definition to work immediately with 
constraint propagation.  Rather, the way to proceed is to obtain the Proca constraint, use it 
on the other constraints to eliminate $A_0$ and then obtain constraints which close.  The 
earlier TSA papers however proceeded via contstraint propagation, and missed out Proca theory. 
To include Proca, I require rather the more broad-minded approach in which all the auxiliaries 
are treated on the same footing by being present in the action from the outset.  This works 
here by starting with actions for 3-geometries together with one scalar and one 1-form matter 
fields.  To have Proca theory, that scalar then turns out to be the above auxiliary.

\subsection{Use of Split Spacetime Ans\"{a}tze}

Here I investigate what fragment of theories included in the general SSF ansatz for a single 
1-form give TSA theories.  My use of `the general' is subject to the assumption that there is 
no fundamental underlying theory so that  local flat spacetime na\"{\i}ve renormalizability 
makes sense.  This puts a stringent bound on how many terms can be in the 1-form ansatz, by 
forbidding products of more than four 1-form fields (non-renormalizable interactions).  
I also restrict attention to first-order Lagrangians.    The ansatz is
\be
\mbox{\sffamily L\normalfont}^{\mbox{\scriptsize A\normalsize}} = C^{ABCD}\nabla_B A_A\nabla_DA_C 
 + \bar{C}^{ABCD}\nabla_B A_A A_DA_C + mA^2 + qA^4
\ee
(the other possible contributions are total derivatives or zero by symmetry--antisymmetry).  

Using the SSF derivative decomposition formulae \cite{KucharII} 
\be
\nabla_bA_{\perp} = D_bA_{\perp} - K_{bc}A^c 
\mbox{ } ,
\label{Vderivproj1}
\ee
\be
\alpha\nabla_{\perp} A_{a} = - \delta_{\check{\beta}} A_a - \alpha K_{ab}A^b - A_{\perp}\pa_a\alpha  
\mbox{ } ,
\label{Vderivproj2}
\ee
\be
\nabla_b A_a = D_bA_a - A_{\perp}K_{ab} 
\mbox{ } ,
\label{Vderivproj3}
\ee
\be
\alpha\nabla_{\perp}A_{\perp} = - \delta_{\check{\beta}}A_{\perp} - A^a\pa_a\alpha  
\mbox{ } , 
\label{Vderivproj4}
\ee
I obtained the split of the first term in \cite{Vanderson}.  If this term contributes, 
TSA-castability enforces the `Maxwellian curl combination' $C_1 = - C_2$, $C_3 = 0$, since 
otherwise there are tilt terms preventing algebraic elimination of $\alpha$.  Note that 
whereas one might have suspected the `simplicity' absence of kinetic cross-terms, of terms 
linear in the velocities and of 1-form dependence in the kinetic metric (which are all symptoms 
of derivative coupling) these effects are invariably partnered by tilt terms in the quadratic part 
of this ansatz, and are thus not directly responsible for the picking out `Maxwellian curl 
combination'.  This tilt explanation accounts within the SSF's spacetime ontology for how 
BF\'{O}'s 3-space assumptions result in `Maxwellian curl combination' 
theories being picked out.  Note also that whereas `Maxwellian curl combination' theories 
can be `added on' to the Dirac--ADM--DeWitt canonical study of pure GR, the other combinations 
of derivatives would seriously complicate the canonical structure, influencing 
and thus invalidating the canonical study of pure GR.  See \cite{KucharIV, IN, Vanderson} for a discussion of this 
and of other undesirable features of these theories. Thus it is 
fortunate that the TSA excludes such fields.    

The second term of the ansatz has w.l.o.g just two pieces 

\noindent $\bar{C}_1g^{AB}\nabla_AA_B A^2 +  \bar{C}_2g^{AC}g^{BD}A_CA_D\nabla_AA_B$, which may 
similarly be decomposed.  The last two terms of the ansatz are trivial to decompose.  Now, 
using (\ref{Vderivproj1}--\ref{Vderivproj4}), the $\bar{C_2}$ contribution i.t.o $A_{\perp}$ 
has no tilt, whereas the $\bar{C}_2$ contribution has no tilt by parts.  

One then either requires the $A_{\perp}$ multiplier equation `accident' or the $A_0$ 
formulation to write a TSA form.  If the second term is also present, it so happens that 
$A_{\perp}$ is not a multiplier, and passing to the $A_0$ formulation shifts tilt from the 
first term into the second term.  Thus the first and second terms look mutually incompatible 
in the TSA.  However, either of these terms could be present, and both are compatible with 
the third and fourth terms.  

Thus I have found two classes of single 1-form theories which I can cast into TSA form:  
\be
\mbox{\sffamily L\normalfont}^{\mbox{\scriptsize A\normalsize}}_1 = a\nabla_{[A}A_{B]}\nabla^{[A}A^{B]} +   mA^2 + qA^4 
\mbox{ } , 
\ee
\be
\mbox{\sffamily L\normalfont}^{\mbox{\scriptsize A\normalsize}}_2 = bA^2\nabla_AA^A + cA^AA^B\nabla_B A_A  + mA^2 + qA^4 
\mbox{ } .  
\label{T2}
\ee 

The first are the `Maxwellian curl combination' theories.  These are: Proca theory if $q = 0$, or if   
$q \neq 0$, its $A_0$ formulation has a Lagrangian of type 
$$
\sqrt{g}\mbox{\sffamily L\normalfont}^{\mbox{\scriptsize A\normalsize}} = \alpha
\left(
A + \frac{B}{\alpha^2}  +  \frac{C}{\alpha^4}
\right) 
\mbox{ } ,
$$
so the $\alpha$-multiplier equation is $A\alpha^4 - B\alpha^2 - 3C = 0$, so by the quadratic formula, 
\be 
\mbox{\sffamily L\normalfont}^{\mbox{\scriptsize A\normalsize}}_{\mbox{\scriptsize TSA form of 1\normalsize}} = 
\left(
\frac{2A}{B \pm \sqrt{B^2 + 12AC}}
\right)^{\frac{3}{2}}
\left[
\frac{      B(B \pm \sqrt{B^2 + 12AC})}{A} + 4C
\right] 
\mbox{ } .
\label{N4}
\ee
Compared to $\mbox{\sffamily L\normalfont}^{\mbox{\scriptsize A\normalsize}} = \alpha
\left(
A + \frac{B}{\alpha}
\right)
$ which gives $\mbox{\sffamily L\normalfont} = 2\sqrt{AB}$, the above is far more 
complicated but is nevertheless a valid TSA 
presentation.  Thus  $A^4$-theory was excluded by BF\'{O} on simplicity grounds rather 
than for fundamental reasons.   

The second coupled to GR is linear as regards $N$ (because it is i.t.o $A_{\perp}$):
 and thus gives a TSA theory with Lagrangian of form 
\be
\mbox{\sffamily L\normalfont} = 2\sqrt{AB} + D 
\ee
where $D$ is the linear kinetic term.  This is similar in layout to the TSA formulation 
with spin-$\frac{1}{2}$ fermions, except that here the derivative coupling terms are 
unavoidably manifest.  Moreover, at least for a single 1-form, these theories do not produce dynamical equations.  
But note that the many 1-form case is suggestive and deserves further study.  

Distinct further examples are easy to come by if 
na\"{\i}ve renormalizability is rendered irrelevant by some underlying fundamental theory.  
Clearly adjoining any polynomial in 
$A^2$ to the spacetime Maxwell Lagrangian will do, and the TSA forms just get nastier.  A messy Cardano 
formula is required if an $A^6$ term is present.  If terms as high as  $A^{10}$ are present, 
despite being algebraic the $\alpha$-multiplier equation is not generally explicitly exactly 
soluble for $\alpha$ by Galois' well-known result.  I also considered Born—-Infeld theory.  
The easiest non-Maxwell Born—-Infeld theory [{\sffamily L} = $(F \circ F)^2$] can be cast 
into the TSA form (\ref{N4}), while the string-inspired Born—-Infeld theory 
[{\sffamily L} = det($g_{ab} + F_{ab}$)] gives a $\alpha$-multiplier equation not generally 
exactly soluble for $\alpha$.  

Thus the TSA in fact admits a broad range of single 1-form theories.

\section{Does the TSA Implement the Principle of Equivalence?}

Curved spacetime matter field equations are locally Lorentzian if they contain no worse than 
Christoffel symbols (by applying the Christoffel symbols' transformation law).  
The gravitational field equations are given a special separate status in the {\bf POE} 
(`all the laws of physics bar gravity').   
However, derivatives of Christoffel symbols, be they from double derivatives or 
straightforward curvature terms muliplied by matter factors, cannot be eliminated 
likewise and are thus {\bf POE}-violating terms.  

Let me translate this to the level of the Lagrangians I am working with.  If the 
Lagrangian may be cast as  functionally-independent of Christoffel symbols, its 
field equations clearly will not inherit any, so the {\bf POE} is satisfied.  If the 
Lagrangian is a {\sl function} of the Christoffel symbols, then by the use of 
integration by parts in each Christoffel symbol's variation, generally derivatives 
will appear acting on the cofactor Christoffel symbols, leading to {\bf POE}-violating 
field equations.  A clear exception is when the Lagrangian is a {\sl linear} 
function of the Christoffel symbols.  Lagrangians unavoidably already containing 
matter-coupled Christoffel symbol derivatives lead to {\bf POE}-violating field equations.  

The `Maxwellian curl combination' 1-form Lagrangian above contains no Christoffel symbols by antisymmetry 
(as do Yang--Mills theory and the various bosonic Gauge Theories).   Linearity 
gives a guarantee of protection to Dirac theory, in a different way from the 
fortunate rearrangement in Sec 9: despite being derivative-coupled this behaves according 
to the {\bf POE } by this linearity.  This means also holds for my second 1-form theory above.  
On the other hand, the excluded 1-form theories have Lagrangians nonlinear in the 
Christoffel symbols.  Thus, in addition to these theories being undesirably complicated and damaging of 
the canonical study of pure GR, they also correspond to {\bf POE} violation.  
So the TSA and the {\bf POE} are acting in a similar way as regards the selection of admissible theories.  
This leads me to tentativiely conjecture that (possibly subjected to some restrictions) 
the TSA leads to the {\bf POE}.   

Crudely, I) tilt and derivative coupling come together in the spacetime split, and tilt 
tends to prevent TSA  formulability. II) Tilt and derivative coupling originate 
in spacetime Christoffel symbol terms, which are {\bf POE}-violating.   At a finer level, 
1) I know tilt and derivative coupling need not always arise together by judicious 
construction otherwise [theories along the lines of (\ref{T2})].  This could potentially 
cause discrepancies between TSA formulability and obedience of the {\bf POE} from derivative-coupled 
but untilted examples. 2) Christoffel-linear actions are not {\bf POE}-violating; moreover in 
the examples considered [Dirac theory, the standard interacting theories related to 
Dirac theory, and theories along the lines of (\ref{T2})] this coincides with unexpected TSA 
formulability.  2) also overrules the given example of 1) from becoming a counterexample 
to the conjecture.      

There is one limitation I am aware of within the examples I considered.  I have shown 
\cite{Thanderson} that 
Brans--Dicke (BD) theory, whose potential contains Christoffel symbol derivatives 
multiplied by matter terms (i.e a $e^{-\chi}\check{R}$ term for $\chi$ the BD field), happens to have derivative 
coupling but no tilt.  Hence this is another example of 1), but now BD theory is TSA 
formulable but {\bf POE}-violating.  Thus one should first classify {\bf POE}-violating Lagrangians 
into e.g ones which merely containing covariant derivatives and ones which contain matter coupled 
to the curvature scalar, and then have a conjecture only about the former.  
I suggest a further systematic 
search for (counter)examples should be carried out: including many 1-forms, higher derivative 
gravity terms, torsion.  …

\section{On Alternative Foundations for Gauge Theory }

In contrast to the traditional point of view on unbroken and broken Gauge Theories in Sec 3.3, 
I emphasize instead that gravitation may alternatively be viewed as leading to all these 
theories without resort to flat spacetime symmetry arguments.  Non-Maxwellian combinations of 
derivative terms in flat spacetime are just as Lorentz-invariant as the Maxwellian curl 
combination, but accepting that we live in a curved GR spacetime,\fn{This includes the 
{\bf POE} holding perfectly; the argument is unaffected by replacing this part of GR by the 
{\bf POE} holding up to somewhat past our current stringent observational limit on {\bf POE} 
violation.} the former {\sl could not locally arise in the first place since they are POE 
violators}.  Thus the TSA leads to both broken and unbroken Gauge Theory.  Similarly, 
Teitelboim \cite{Teitelthesis, Teitelboim} obtained Gauge Theory in the context explained in 
App C.   

One issue is whether new physics could be inspired by this point of view.  For example, 
might theories along the lines of (\ref{T2}), which obey the {\bf POE} via 
Christoffel-linearity rather than through the Maxwellian curl combination's absence of Christoffel symbols, 
also be present in nature? 
This might lead to interesting cosmology or particle physics.  
N.B whether such additional theories could be 
overruled on further grounds does not affect the viability of the above viewpoint on the 
origin of Gauge Theory.  

\vspace{2in}

\section{Example 6: Conformal Alternatives}

Consider again the configuration space Riem, but now let the redundant motions be both 
3-coordinate transformations and conformal transformations.  These do not quite form 
Diff $\times$ Conf as Diff $\bigcap$ Conf $\neq$ 0.  Nevertheless one can take the RCS to be
\be
\mbox{Conformal Superspace CS } = \frac{\mbox{Riem}}{\mbox{Diff} \times \mbox{Conf}}
\ee
because it does not affect the orbit structure if the effect of certain transformations is 
quotiented out twice.  This may also be established by showing the equivalence of 
implementing the non-conformal diffeomorphisms by use of
\be
h_{ij} \longrightarrow h_{ij} - 2
\left(
D_{(i}s_{j)} - \frac{D_ks^k}{3}h_{ij}
\right)
\ee 
(conformal Killing form correction rather than a Killing form correction).  This example is 
related to the attempted isolation of the true d.o.f's of CWB GR and to the GR IVF (see App D).  
CS is the space of shapes (excluding scale), which each have 2 d.o.f's per space point.  

What should be noted is that the conformal transformations are different to all the other 
transformations considered so far in this chapter.  The extent to which this difference is 
manifested depends on how they are treated.  

In the `York style' (parallelling the GR IVF of App D), one writes the action in the 
arbitrary conformal frame, building it out of good conformally covariant objects.  One must 
bear in mind that this involves treating what might have been regarded as different pieces of 
the same tensor as distinct objects which scale differently (see below).  One must also bear 
in mind that formulations in this style {\sl only has a temporary technically-convenient 
conformal gauge symmetry}, since the Lichnerowicz--York equation (conformally-transformed 
${\cal H}$, see App D) then gauge-fixes the conformal factor $\psi$ by specifically mapping 
to a particular point on the Conf 
orbit.\fn{Thus this use of gauge theory is different from the U(1), Yang--Mills and Diff uses.  
There, the choice of gauge is unphysical, whereas here it is physical because it is 
prescribed by an additional condition required for passage from the initially prescribed 
unphysical metric $h_{ij}$ to the physical 3-metric $\psi^4h_{ij}$.}  Consequently there is 
no unphysical dragging along conformal orbits, so no conformal best matching of velocities 
arises in this style.  Rather, conformally-bare velocities are to be regarded as already 
conformally-covariant.  

But there is also a distinct `Barbour style' in which one considers actions with true 
conformal symmetry, with scale factor $\omega$.  The conformal transformations act by 
transforming both the metric and a conformal auxiliary $\phi$ in a compensatory fashion (see 
below).  Then unphysical dragging is permitted along the conformal orbits, so conformal best 
matching of velocities is indeed required.  Moreover, York's distinct scaling laws are no 
longer a natural assumption from Barbour's first principles.  They will rather eventually 
{\sl emerge}.

The above choice of style reflects whether the choice of adjoining Conf to Diff is to be a 
permanent feature of the alternative. 

Now, note that unlike for the other transformations covered so far in this chapter, in the 
`Barbour style', the auxiliary conformal factor $\phi$ occurs alongside $\dot{\phi}$ in the 
action.  Thus this auxiliary is not cyclic.  Nevertheless it is an auxiliary, so FEP 
variation is still to be used, and corresponds to a new type of gauge theory 
\cite{PD, ABFO, ABFKO}.  The auxiliary canonical coordinate $\phi$ appears in the actions 
because the geometrical objects being employed in building these actions are not by 
themselves conformally covariant tensors and $\phi$ occurs so as to compensate for this 
lack.  In the `York style', $\phi$ alone occurs; whereas one may still formally carry out 
the FEP part of the variation, it no longer yields any information.  

Conf is associated with $p = 0$, which in GR is the condition for a slice to be 
{\it maximal}.  One then requires the maximal lapse-fixing equation (LFE)
$$
\triangle \alpha = \alpha R
$$  
to be soluble for the lapse $\alpha$ if $p = 0$ is to be maintained on a series of slices.  
But this is well-known to be an insoluble equation for CWB GR, by the following 
`integral inconsistency' argument.  
\be
0 = \oint_{\pa\Omega}D_i\alpha\textrm{d}S^i =\int\textrm{d}\Omega \triangle \alpha =
\int\textrm{d}\Omega \alpha R \Rightarrow \not{\exists} \mbox{ } \alpha
\mbox{ if } \alpha R  \mbox{ is of fixed sign.}
\ee
This last step follows from supposing there is some 
point $x_0$ at which the integrand $I(x_0) = \epsilon > 0$.  Then for $I$ 
continuous, $|I(x) - I(x_0)| < \frac{\epsilon}{2} \mbox{ } \forall \mbox{ } |x - x_0| 
< \delta$, so $I(x) > \frac{\epsilon}{2} \mbox{ } \forall \mbox{ } |x - x_0| < \delta$, so 
$\int \textrm{d}^3xI(x) > K\delta^3\epsilon >0$, where $K$ is some positive constant, which is 
a contradiction.  
This last step is indeed applicable here because $R$ is positive-definite from ${\cal H}$ and 
the lapse $\alpha$ is strictly positive by definition, so the integrand is positive and hence 
cannot vanish.

\subsection{Alternative theory of gravity on CS}

The first conformal alternative is `conformal gravity' \cite{conformal, ABFO}, which uses the 
`Barbour style' together with 
a `solution by quotienting' of the integral inconsistency.  
One works in the arbitrary Conf-frame by use of 
\be
\bar{h}_{ij}=\phi^4h_{ij} \mbox{ } .  
\label{CGCoCo} 
\ee 
which has conformal invariance under  
\be
h_{ab} \longrightarrow \omega^4 h_{ab} \mbox{ } , 
\label{splitban}
\ee
\be
\phi\longrightarrow {\frac{\phi}{\omega}} \mbox{ } .
\label{CGConBanal}
\ee
The potential is then
\be
\bar{R} = \phi^{-4}\left(R - \frac{8D^2\phi}{\phi}\right) \mbox{ } , 
\label{Rphi}
\ee
while the kinetic term contains the arbitrary Conf-frame velocities
\be
{\&}_{\dot{\scriptsize s\normalsize}}\bar{h}_{ab} = 
{\&}_{\dot{\scriptsize s\normalsize}}(\phi^4h_{ab}) \equiv \phi^4
\left(
{\&}_{\dot{\scriptsize s\normalsize}}h_{ab} + 
{4}\frac{{\&}_{\dot{\scriptsize s\normalsize}}\phi}{\phi}h_{ab}
\right) \mbox{ } .
\label{CGBabel}
\ee
The action is additionally to be made homogeneous of degree 0 in $\phi$ so as to be invariant 
under constant rescalings.  This is implemented by division by the suitable power of the 
conformalized volume of the universe 
$$
V = \int \textrm{d}^3x \sqrt{h}\phi^6 
\mbox{ } ,  
$$
so the action is 
$$
\mbox{\sffamily I\normalfont}_{\mbox{\scriptsize \normalsize}}\mbox{=}\int\textrm{d}\lambda\int\textrm{d}^3x
\left(
\frac{\sqrt{h}\phi^6}{V}
\right)
\sqrt{
\left(
\frac{\vc}{\phi^4}
\right)
\left( 
R\mbox{--}\frac{8D^2\phi}{\phi}
\right)
                             } 
\sqrt{\mbox{\sffamily T\normalfont}^{\mbox{\scriptsize g\normalsize}}_{\mbox{\scriptsize C\normalsize}\mbox{\scriptsize W\normalsize}} }
\mbox{=}\int \textrm{d}\lambda \frac{\int\textrm{d}^3x\sqrt{h}\phi^4 \sqrt{R\mbox{--}\frac{8D^2\phi}{\phi}}
\sqrt{\mbox{\sffamily T\normalfont}^{\mbox{\scriptsize g\normalsize}}_{\mbox{\scriptsize C\normalsize}\mbox{\scriptsize W\normalsize}}  }    } {\vc }. 
\label{CGSCGR} 
$$ 

Now evaluating $p^{ij}$ and $p_{\phi}$ reveals the primary constraint 
\be
p = \frac{\phi}{4}p_{\phi} 
\mbox{ } . 
\label{CGprimconstr} 
\ee 
[a direct consequence of the invariance (\ref{splitban}, \ref{CGConBanal})], but also $p_{\phi} = 0$ by FEP variation.  
Thus one obtains the maximal condition 
$$
p = 0 \mbox{ } .
$$  
Consequently w.l.o.g $W = 0$.  Furthermore, a close analogue of the Lichnerowicz equation 
(see App D) arises as a square root primary constraint:
$$
{\cal H}^{\mbox{\scriptsize C\normalsize}} \equiv 
\frac{\vc}{\sqrt{h}\phi^4}p\circ p 
- \frac{\sqrt{h}\phi^4}{V^{\frac{2}{3}}}
\left( 
R - \frac{8D^2\phi}{\phi} 
\right)
= 0 
\mbox{ } .
$$
$s^i$ and $\phi$ variation respectively give as secondary constraints the usual momentum constraint 
$$
D_jp^{ij} = 0
$$
and the conformal gravity LFE
$$
\triangle N = NR - <NR>
$$ 
in the distinguished representation ($\phi = 1$ gauge \cite{ABFO}), where 
$$
<A> \equiv \frac{\int \textrm{d}^3x \sqrt{h} A} {\int \textrm{d}^3x\sqrt{h}} 
$$
is the usual notion of global average.  
This LFE indeed avoids the integral inconsistency, thanks to its new term 
which in turn arises from the quotient implementation of homogeneity of degree 0 in $\phi$.   
This works simply because $\int\textrm{d}\Omega (NR - <NR>)$ is trivially 0, so the integral 
inconsistency is rendered irrelevant by construction. 

Conformal gravity is an as-yet largely unexplored alternative theory of gravity 
see \cite{conformal, ABFO, Thanderson} 
for what is currently known).

\subsection{New GR formulation and alternative theories on CS+V}

The other conformal alternatives make use of York's generalization from maximal to 
CMC surfaces \cite{York, CBY}.  
In this case, the relevant LFE is
$$
\triangle N = N\left(R + \frac{p^2}{4h}\right) + C(\lambda) \mbox{ }
$$
which avoids the integral inconsistency because $C(\lambda)$ may be taken to be negative. 

One is now adjoining the volume of universe to the two conformal geometry d.o.f's, i.e working on a CS + V 
relative configuration space.  In York's original IVF work \cite{CS}, this entailed having one copy of each 
conformal orbit per value of the volume.   My collaborators and I \cite{ABFKO} rather quotient out VPConf - 
the volume-preserving conformal transformations.  This splits up each conformal orbit into finer orbits, one per volume.  

We have considered two implementations of VPConf.  There is the Laplacian implementation  
$$
\bar{h}_{ab} = \left(1 + \frac{2}{3}\triangle\zeta\right)h_{ab}
$$
so that 
$$
\bar{V} = \int\textrm{d}\lambda \tilde{h} = \int\textrm{d}\lambda \sqrt{h}(1 + \triangle\zeta) = 
\int\textrm{d}\lambda \sqrt{h} = V \mbox{ } ,
$$
by use of the divergence theorem and CWB.
This is the infinitesimal version of the 
implementation used in \cite{ABFO}; the 
finite version's transformations do not 
close as a group.  There is also Foster's 
implementation that we use in \cite{ABFKO}:
$$
\bar{h}_{ij} = \hat{\phi}^4h_{ij} \mbox{ } , \mbox{ } \hat{\phi} = \frac{\phi}{<\phi^6>^{\frac{1}{6}}} 
$$
so that 
$$
\bar{V} = \int\textrm{d}\lambda \bar{h} = \int\textrm{d}\lambda \sqrt{h}\hat{\phi}^6 
= \frac{\int\textrm{d}\lambda \sqrt{h}\phi^6}{\frac{\int\textrm{d}\lambda \sqrt{h}\phi^6}{\int\textrm{d}\lambda \sqrt{h}}} = 
\int\textrm{d}\lambda \sqrt{h} = V \mbox{ } .
$$
These are permitted to be finite.  Here I rather present `York-style' and 
`Barbour-style' actions built using the Laplacian implementation of VPConf.  

\mbox{ }

In the `York style', what might have been regarded as the trace and tracefree 
parts of the momentum are now rather regarded as distinct objects which are allotted distinct 
conformal rank as befits the formation of conformally-covariant derivatives.  In particular, 
a relative scaling between\fn{I use 
$u \equiv tr({\&}_{\dot{\scriptsize s\normalsize}}h_{ij})$ and 
$u^{\mbox{\tiny T\normalsize}}_{ij} 
\equiv ({\&}_{\dot{\scriptsize s\normalsize}}h_{ij})^{\mbox{\tiny T\normalsize}}$, where 
$A_{ij}^{\mbox{\tiny T\normalsize}} \equiv A_{ij} -\frac{A}{3}h_{ij}$} 
$u^{\mbox{\scriptsize T\normalsize}}\circ u^{\mbox{\scriptsize T\normalsize}}$ and $u^2$ of 
$\phi^{12}$ arises.  This corrects the na\"{\i}ve mismatch in amount of $Np^2$ between the 
auxiliary variation LFE and the LFE required to propagate the CMC condition.  
In this approach, there 
are two distinct auxiliaries, and the one encoding the CMC condition ($\eta$) 
should probably be regarded as a multiplier, not a best matching.  The action is 
\be
\mbox{\sffamily I\normalsize} = 
\int d\lambda\int d^3x\sqrt{h}
\left(
1 + 2\frac{\triangle\zeta}{3} 
\right)
\sqrt{
                                                \left( 
                                                R - \frac{4\triangle^2\zeta}{3}
                                                \right)                                       }
                                    \sqrt{      u^{\mbox{\scriptsize T\normalsize}}\circ u^{\mbox{\scriptsize T\normalsize}} 
                                                - \frac{2}{3}(1 - 2\triangle\zeta)
                                                \left(
                                                u + \triangle\eta
                                                \right)^2                                       }  
\mbox{ } .
\ee

\mbox{ } 

In the `Barbour style', VPConf-{\bf BM} is indeed required\fn{It is in fact only a 
nontrivial correction to $u$ since $h^{\mbox{\tiny T\normalsize}}_{ij} = 0$.} to make 
$\dot{h}_{ij}$ into a good VPConf object.  The natural action is then 
$$
\mbox{\sffamily I\normalfont} = \int\textrm{d}\lambda \int\textrm{d}^3x\sqrt{h}\phi^6
                                \sqrt{\phi^{-4}
                                \left(
                                R - \frac{8\triangle \phi}{\phi}
                                \right)} 
                                \sqrt{G^{ijkl}
                                \left(
                                \&_{\dot{\scriptsize s\normalsize}}{h}_{ij} + 
                                \frac{4\&_{\dot{\scriptsize s\normalsize}}{\phi}}{\phi}h_{ij}
                                \right)
                                \left(
                                \&_{\dot{\scriptsize s\normalsize}}{h}_{kl} + 
                                \frac{4\&_{\dot{\scriptsize s\normalsize}}{\phi}}{\phi}h_{kl}
                                \right)}
$$
with $\phi = 1 + \frac{\triangle\zeta}{6}$, $\zeta$ infinitesimal, so that 
\be
\mbox{\sffamily I\normalsize} = 
\int d\lambda\int d^3x\sqrt{h}
\left(
1 + \frac{2\triangle\zeta}{3} 
\right)
\sqrt{
                                                \left( 
                                                R - \frac{4\triangle^2\zeta}{3}
                                                \right)                                       }
                                    \sqrt{      u^{\mbox{\scriptsize T\normalsize}}\circ u^{\mbox{\scriptsize T\normalsize}} - \frac{2}{3}
                                                \left(
                                                u + 2\frac{\pa (\triangle\zeta)}{\pa \lambda}
                                                \right)^2                                       }  
\mbox{ } .
\label{Barac}
\ee

The momenta are 
\be
\mbox{\tt p}^{ij} = p^{ij} + D^k(pD_k\zeta)\frac{h^{ij}}{3}
\label{CS+Vmom}
\ee
for $p^{ij}$ the GR expression for momentum, and 
\be
p^{\zeta} = \frac{2}{3}\triangle p \mbox{ } .  
\ee
Then FEP variation and the trace of (\ref{CS+Vmom}) give 
$$
\frac{p}{\sqrt{h}} = C(\lambda) \mbox{ } , \mbox{ } \mbox{\tt p} = p(1 + \triangle\zeta) \mbox{ i.e. }
\mbox{\tt p} = p\phi^6
$$
so that the na\"{\i}ve $p$ is replaced by the `scaled-up' {\tt p}. 

Note that (\ref{Barac}) does not have a relative scaling of $\phi^{12}$ 
between $u^{\mbox{\scriptsize T\normalsize}}\circ u^{\mbox{\scriptsize T\normalsize}}$ and $u^2$.  
Na\"{\i}vely, this would lead to a mismatch in the $Np^2$ terms between the $\zeta$-variation 
and $\frac{p}{\sqrt{h}} = C(\lambda)$ propagation LFE's, so one might expect (\ref{Barac}) to be 
inconsistent.  However, it turns out that $\zeta$-variation of (\ref{Barac}) has a new source of 
$Np^{2}$ terms.  
\be
\frac{\delta}{\delta\zeta}
\left[
\frac{\pa(\triangle\zeta)}{\pa\lambda}
\right] = 
\triangle
\left(
\frac{Np}{2\sqrt{h}}
\right)
\ee
occurs with a cofactor proportional to 
$\frac{p}{\sqrt{h}}$, but as this is a spatial constant, 
it can be taken inside $\triangle$, and 
this happens to contribute just the right amount of $Np^2$ to the 
$\zeta$-variation LFE to obtain the CMC LFE.

Note also that a relative scaling of $\phi^{12}$ emerges in the primary constraint arising from squaring 
the momenta (taking into account the cross-terms arising from the nonminimal coupling):  
$$
G_{ijkl}\mbox{\tt p}^{ij}\mbox{\tt p}^{kl} = 
\mbox{\tt p}^{\mbox{\scriptsize T\normalsize}} \circ \mbox{\tt p}^{\mbox{\scriptsize T\normalsize}} 
- \frac{\mbox{\tt p}^2}{6} = 
p^{\mbox{\scriptsize T\normalsize}} \circ p^{\mbox{\scriptsize T\normalsize}} - \frac{p^2}{6}(1 + 2\triangle{\zeta})
$$
i.e
$$
\phi^4\left(R - \frac{8\triangle\phi}{\phi}\right) = p^{\mbox{\scriptsize T\normalsize}} \circ p^{\mbox{\scriptsize T\normalsize}} - \phi^{12}\frac{p^2}{6} 
\mbox{ } .
$$
This may be identified with the Lichnerowicz--York equation (see App D), and constitutes an 
alternative derivation of it.  The scaling up of $p$, and emergence of the Lichnerowicz--York 
equation occur similarly in \cite{ABFKO}.  In that work, unexpected extra terms (distinct 
from those here) also ensure that a consistent LFE emerges.  

\mbox{ }

Note that the above actions may be taken to originate from the possibility of setting 
the $D^a(N^2D_ap)$ factor in (\ref{keyterm}) to 0.  But the theories so far considered in this 
subsection are also resliceable since the $X - 1$ factor is also zero.  Thus their constraint 
algebras close just as well if CMC slices {\sl are not} chosen.  Thus 
I identify this subsection's actions as corresponding not to alternative theories but to new formulation of CWB GR in 
the CMC gauge.  However there is no longer any consistency reason for setting $W = 1$.  
So I can consider the arbitrary-$W$ versions of the two approaches above in order to 
obtain alternative theories of gravity which are about truly non-resliceable 
stacks of maximal or CMC hypersurfaces. These follow 
from replacing the $-\frac{2}{3}$'s in the actions by $\frac{1- 3W}{2}$.  Thus, if one takes 
York's IVF mathematics more seriously than GR itself, then one is entitled to consider a range 
of privileged-slicing theories in addition to resliceable GR.

For the $W = 1$ CS+V theory, the use of a CMC stack of hypersurfaces is thus ultimately a 
{\sl gauge choice}, which is available provided that the LFE is soluble.  It is a {\sl partial} 
gauge choice since the point-identification (shift) between hypersurfaces in the stack is still 
unspecified.  That the LFE enodes this gauge choice means that one is automatically provided 
with a partially {\sl gauge-fixed} action.  Any pathology in this CMC gauge might then go away 
under the valid procedure of reslicing so as to be in another gauge.  

But $W \neq 1$ CS+V theories are not just written to favour a particular slicing or possess 
a privileged slicing.  They are not generally resliceable because this leads to inconsistency.  
Thus these describe stacks of CMC slices and not pieces of GR-like spacetime.  As a result of 
this, pathologies of the stack of CMC slices become real effects since reslicing to avoid these 
is not possible.  Thus while $W = 1$ CS+V theory is just (a restriction of) CMC-sliced GR, $W \neq 1$ 
CS+V theories are quite distinct at a conceptual level.    
To complete the picture, for conformal gravity the choice of $W$ does not affect the 
equations of motion.  However, setting $W = 1$ permits reslicing.  
But unlike in CS+V, these new slicings {\sl remember} the privileged $p = 0$ slicing, since the volume of these slices gets 
incorporated into the field equations.

Here is a theoretical consequence of not being able to reslice.  Suppose one has access to a 
compact object whose curvature profile permits (GR-inspired) collapse of the lapse ($\alpha \longrightarrow 0$) to occur 
well outside its horizon.  Then in a GR world, one could send an observer past where the lapse 
collapses, and as nothing physical occurs there and the observer is still safely away from the 
horizon, the observer can `return to Earth' and report that $W \neq 1$ CS+V theory has been 
falsified.  But in a $W \neq 1$ CS+V theory world, the observer would have become frozen forever 
where the lapse collapses and thus would not be able to return.  Note that this is somewhat 
similar to the frozen star concept which predated the GR notion of black holes, except that 
the freezing could be occurring {\sl outside} the horizon.  Although $W \neq 1$ CS+V theory 
could therefore be an improvement as regards strong cosmic censorship (the occurrence of 
singularities {\it at all}), there is also the GR-inspired possibility in a  
Tolman--Bondi example of Eardley and Smarr \cite{ES}:  that sufficiently steep curvature profiles generate too slow a collapse of the lapse 
to avoid singularities.   Also, the collapse of the lapse would not save one from other 
non-curvature blowup pathologies usually regarded as singularities.  

$W \neq 1$ CS+V theory ought to also be testable much as Brans--Dicke theory is, by solar 
system tests.   While plain `arbitrary-$W$ GR' was suggested as another useful testbed for GR 
\cite{KiefGiu} (c.f the use of Brans--Dicke theory), this is not of any direct use because it's 
inconsistent \cite{SGGiulini} (also the RWR result).  What I have demonstrated however is that 
{\sl the idea of this `arbitrary-$W$ GR' can be salvaged because it is a consistent theory 
provided that it is treated as a (non-resliceable) stack of CMC hypersurfaces, in which case 
it becomes CS+V theory}.  Thus I provide a 1-parameter family of theories to test against (not 
just extreme but also) everyday GR.  Moreover, by the nature of the conformal mathematics in 
which they are so naturally expressed, they should be easily useable as testbeds for 
Theoretical Numerical Relativity.  This should require but minor modifications of existing 
codes.  Conformal gravity could also be used/tested in this way.  {\sl Numerical Relativity 
uses conformal mathematics, not necessarily any notion of embeddability into GR-like spacetime.  
The alternative theories of this section may be seen as arising from taking this conformal 
mathematics in its own right as possibly a serious alternative to GR itself.} Finally, $W = 1$ 
CS+V theory may be seen as a formulation of GR proper, and thus still be directly useful (both 
conceptually and as a tool) in Theoretical Numerical Relativity even if the suggested 
alternatives to GR are dismissed or heavily bounded by future compact-object observations and 
analysis. 

\vspace{3in}

\section{Conclusion}  

A Leibniz--Mach approach to physics has been laid out.  
Much standard physics is recovered, and is moreover accompanied by new alternative possibilities.  
Point particle mechanics for the whole universe is distinct, but the usual theory is easily 
recoverable for subsystems.  Standard Gauge Theory is natural in the context of this approach.  
So is CWB GR, which is thus, in a precise sense, Machian.  The TSA derivation of 
GR from among more general 3-space possibilities is furthermore an answer to Wheeler's question 
(in App C) about first principles for geometrodynamics, and shows that GR is rather special, 
even fragile, from a canonical perspective that does not presuppose spacetime.  
In the TSA, the structure of spacetime is {\sl emergent} for CWB GR.   
Furthermore, spacetime is {\sl not} emergent in all TSA theories. 
New gravitational theories arise that are based instead on conformal mathematics related to 
that used in GR IVF.  New derivations of the GR IVF itself accompanies these theories.  
Whereas GR admits both spacetime and conformal IVF formulations, the conformal formulation is 
found to be able to exist independently of whether there is spacetime structure.    

The {\bf Relativity Principles} are emergent in the TSA with matter included.  The TSA admits 
all the fundamental classical fields required to describe nature.  Moreover, it is not as 
strongly selective of these matter fields a previously thought, while retaining a certain 
amount of selectiveness.  It is interesting that this selectiveness may be tied to the 
{\bf Principle of Equivalence (POE)} being emergent in the TSA.  The TSA's selectiveness over the 
types of matter allowed may also lead to new ideas for particle physics and cosmology.  
One possible route to this is via what unusual theories are included in adopting the suggested 
alternative `curved space and {\bf POE}' foundations for Gauge Theory.  

The TSA study of GR assigns prominence to dynamics on the configuration space of 3-spaces, 
rather than assigning the usual prominence to the spacetime arena.  This difference of 
perspective is furthermore suggestive of differences in how Quantum Gravity should be approached, 
in particular of differences in how to attempt to resolve its Problem of Time 
\cite{POT1, POT2}.  The TSA is suggestive of timeless resolutions such as the Na\"{\i}ve 
Schr\"{o}dinger Interpretation \cite{NSI}, Barbour's variation thereof 
(see \cite{B94II, EOT} and the critical response in \cite{Butterfield, Smolin}), 
or the Conditional Probabilities Interpretation \cite{CPI}.  
On the other hand, spacetime is suggestive of internal time approaches \cite{POT1, POT2}, 
in which one seeks a change of variables from the 6 $h_{ij}$ to a clean split of 2 true 
dynamical d.o.f's and 4 embedding (time) variables.  

\mbox{ }

\noindent\large{\bf Acknowledgements}\normalsize 

\mbox{ }

I would like to thank Julian Barbour, Harvey Brown, Brendan Foster, Bryan Kelleher, 
Malcolm MacCallum, James Nester and Niall \'{O} Murchadha for discussions, and 
Jeremy Butterfield and Oliver Pooley for inviting me to speak at the Oxford 
conference on Spacetime.  I acknowledge Peterhouse College Cambridge and the Izaak 
Walton Killam Memorial Foundation for making the continuation of my career 
possible, and thank Malcolm MacCallum, Reza Tavakol, Julian Barbour, Don Page and 
Gonville and Caius College Cambridge for their support of me.  I thank Claire 
Bordenave, Ed Nokes, Bryony Baines, Yves Jacques, Alex Churchill, Suzy Tucker, 
Rebecca Heath, Charlie Baylis, Martin Lester and Rosemary Warner for support in 
difficult times.  

Future Addresses: Peterhouse College Cambridge CB2 1RD as of October 2004, 
and also Physics Department, Avadh Bhatia Physics Laboratory, University of Alberta, 
Edmonton, Alberta, Canada T6G 2J1 as of January 2005.  

\mbox{ }

\vspace{3in}

\noindent\Large{\bf Appendix A Standard Foundations of Relativity}\normalsize

\mbox{ }

\noindent Newtonian Mechanics posseses Galilean invariance, while Maxwell's Electromagnetic 
Unification possesses Lorentz invariance. Moreover, Electromagnetic Unification unexpectedly 
included light, now identified as electromagnetic waves.  Now, contemporary knowledge about 
other sorts of waves strongly suggested there should be a medium associated with 
electromagnetism, the excitations of which would be light.  Given such an {\it Aether}, its 
rest frame would be expected to be be privileged by Maxwell's equations, so the lack of 
Galilean invariance was not perceived as an immediate impasse.  It rather led to the proposal 
{\bf P} that as electromagnetism does not possess Galilean invariance, experiments involving 
electromagnetic phenomena  could be used to determine motion w.r.t the Aether rest frame.  
There were furthermore speculation that this Aether rest-frame might coincide with Newton's 
absolute space.  

However, in the Michelson--Morley experiment a null result was obtained for the velocity of 
the Earth relative to the Aether.  Furthermore, within the framework of Aether theory, this 
could not be made consistent with Bradley's observation of stellar aberration.  Although 
Fitzgerald and Lorentz \cite{Lorentz} attempted to explain these observations  
{\it constructively} in terms of somehow the inter-particle distances of particles travelling 
parallel to the Aether flow being contracted, Einstein had a different, {\it axiomatic }
strategy akin \cite{CGconstructive, CGspacetime} to how thermodynamics is based on the 
non-existence of perpetual motion machines.  He elevated the outcome of the Michelson--Morley 
experiment from a null result about motion and electromagnetism to a {\it universal} 
postulate.  Rather than mechanics possessing Galilean invariance, electomagnetism possessing 
Lorentz invariance and other branches of physics possessing goodness knows what invariance, 
he postulated that \cite{E05}

\mbox{ }

\noindent{\bf RP1 (Relativity Principle)}: 
all inertial frames are equivalent for the formulation of all physical laws.

\mbox{ }

From this it follows that the laws of nature share a universal transformation group under 
which they are invariant.  There is then the issue of which transformation group this should be.  
{\bf RP1} narrows this down to two obvious physical possibilities, distinguished by whether the laws of 
nature contain a finite or infinite propagation speed 
$v_{\mbox{\scriptsize prop\normalsize}}$.  
If one adopts absolute time as a second postulate the (\bf Galilean RP2\normalfont), the infinite is 
selected, and one has universally Galileo-invariant physics.  
The finite is selected if one adopts instead a constant velocity postulate such as 

\mbox{ }

\noindent{\bf Lorentzian RP2}: light signals in vacuo are propagated rectilinearly with the same velocity 
at all times, in all directions, in all inertial frames.

\mbox{ }

The chosen velocity serves universally [and so is unique so taking 
$v_{\mbox{\scriptsize prop\normalsize}}$ =  (the speed of light $c$) is w.l.o.g].  One has 
then a universally Lorentz-invariant physics.  In the former case, electromagnetism must be 
corrected, whereas in the latter case Newtonian Mechanics must be corrected.  
Einstein chose the latter.  ({\bf RP1} and {\bf Lorentzian RP2} constitute Special Relativity).  
Notice that this is the option given by a law of nature and not 
some postulated absolute structure; also whereas there was ample experimental evidence 
for Maxwellian electromagnetism, existing experimental evidence for Newtonian Mechanics 
was confined to the low velocity ($v \ll c$) regime for which Galilean transformations are an excellent 
approximation to Lorentz transformations.  Indeed the investigation of the high velocity 
regime promptly verified Einstein's corrections to Newtonian Mechanics.  This example of the 
great predictive power of SR is compounded by the universality:  for each 
branch of physics, one obtains specific corrections by requiring the corresponding laws 
to be Lorentz-invariant.  The concept of non-materially substantiated media and the 
proposal {\bf P} were thus destroyed, and physics was rebuilt on the premises that there 
was no room in any of its branches for analogous concepts and proposals.  

Minkowski pointed out that whilst Newton's notions of absolute space and time are also 
destroyed because privileged surfaces of simultaneity cease to exist, one could 
geometrize space and time together as a flat manifold equipped with a $-+++$ 
indefinite signature metric $\eta_{AB}$: {\it spacetime}.  Now it is the null cones permitted by the 
indefinite signature which play the role of privileged surfaces.  
These correspond to the 
surfaces on which the free motion of light occurs (and of all other massless particles, by 
Einstein's postulates: one has a \it universal null cone structure \normalfont in classical 
physics).  And massive particles are permitted only to travel from an event 
(spacetime point) into the interior of the future null cone of that event.  Of particular significance, in free `inertial' 
motion all massive particles follow timelike straight lines whereas all massless particles 
follow null straight lines.  Following from such a geometrization, it makes sense to implement 
the laws of physics in terms of the 4-tensors corresponding to Minkowski's 4-d spacetime.  
However, Einstein found that attempting to accommodate gravity in this scheme 
presented significant difficulties.  

Nearby freely-falling particles in a (non-uniform) gravitational field experience a relative 
acceleration.  Thus gravitation requires the replacement of the inertial frames of 
Newtonian Mechanics (which are supposedly of infinite extent) by local inertial frames.  In  
order to be able to define these it is crucial that inertial mass be identically proportional 
to gravitational mass for all materials, for else each material would require its own 
definition of local inertial frame.  This is the \bf Principle of Equivalence \normalfont 
({\bf POE}).   Einstein \cite{poe} then adopted the somewhat stronger supposition that 
gravitation is not locally distinguishable from acceleration by physical experiments anywhere 
in the universe, and can thus be transformed away by passing to the suitable local inertial 
frame.  He then guessed that the inertial frames of SR were to be identified with the local 
inertial frames of freely-falling massive particles.  To Einstein the {\bf POE} strongly 
suggested \cite{EG13} that gravitation could be included within Relativity by the bold  
postulate that spacetime with gravitation would not be flat Minkowski 
spacetime but rather a spacetime curved by the sources of gravitation so that the straight 
timelike lines followed by free massive particles in Minkowski spacetime are bent into the curves 
followed by relatively-accelerated freely-falling massive particles.  The straight null lines 
constituting the lightcones of Minkowski spacetime would then likewise be bent by the sources 
of gravitation.

The mathematics of the connection permits the incorporation of the above features of the 
gravitational field.  The coordinates in which the connection may be set to zero at each 
particular point are to correspond to the freely-falling frame at that point.  
The privileged curves followed by freely falling particles and by light rays are to be the 
timelike and null affine geodesics of the geometry; at any point in the freely-falling frame 
these reduce to the straight lines of Minkowski spacetime.\fn{Thus this implementation of the 
{\bf POE } tacitly includes {\bf Lorentzian RP2 }, the assumption of Lorentz signature.}  
The geodesic 
equation 
\be
\frac{\pa x^A}{\pa\lambda^2} 
+ {\Gamma^A}_{BC}\frac{\pa x^B}{\pa\lambda}\frac{\pa x^C}{\pa\lambda} 
= 0
\ee 
is of the form of the combination of Newton's second law  
and Newton's law of gravitation\fn{$\phi$ is the Newtonian gravitational potential.}
\be
\frac{\pa x^a}{\pa t^2} + \pa^a\phi = 0 
\ee for a 4-d connection whose only nonzero components are  
${{\Gamma}^i}_{00} = \pa_i\phi$; from this it follows that the only nonzero Riemann 
tensor components are 
\be
{\mbox{\tt{R}\normalfont}^i}_{0j0} = \pa_{i}\pa_{j}\phi \mbox{ } ,
\label{NewtR}
\ee
so that one obtains agreement between the Newtonian tidal equation 
\be
\frac{\pa^2 \mbox{\scriptsize $\Delta$\normalsize}{x}_i}{\pa t^2} = 
- \pa_i\pa_j \mbox{\scriptsize $\Delta$\normalsize} x_j
\ee
and a piece of the geodesic deviation equation 
\be
\frac{D^2z^A}{D\lambda^2} = - {\mbox{\tt R\normalfont}^A}_{BCD}
\frac{\pa x^B}{\pa\lambda}\frac{\pa x^C}{\pa \lambda}z^D
\mbox{ } 
\ee
(where $z^A$ is the connecting vector of the deviating geodesics and $\frac{D}{D\lambda}$ is 
the absolute derivative).

Furthermore, Einstein introduced a semi-Riemannian metric $g_{AB}$ on spacetime, both to 
account for observers in spacetime having the ability to measure lengths and times if 
equipped with standard rods and clocks (paralleling the Minkowskian development of SR), and 
furthermore to geometrize the gravitational field.  For simplicity, he assumed a symmetric 
metric and that the aforementioned connection was the metric one \cite{EinGR}.  

\mbox{ }

This is not yet a gravitational theory: field equations remain to be found.  Einstein 
\cite{VEinstein} `derived' his field equations 
(EFE's)\fn{$\mbox{\tt\scriptsize T\normalsize\normalfont}_{AB}$ is the curved spacetime 
energy-momentum tensor of the matter content.  In this appendix, I keep $c$ and $G$ explicit.}
\be
\mbox{\tt G\normalfont}_{AB} \equiv \mbox{\tt{R}\normalfont}_{ AB } 
- \frac{1}{2} g_{AB}\mbox{\tt{R}\normalfont} = 
\left(
\frac{8\pi G}{c^4}
\right)
\mbox{\tt{T}\normalfont}_{AB}
\label{Vefes}
\ee
by demanding 

\mbox{ }

\noindent  {\bf GRP1} (the {\bf General Relativity Principle}): that all frames are equivalent 
embodied in spacetime general covariance [the field equations are to be a (spacetime) 
4-tensor equation].  

\mbox{ }

\noindent {\bf GR Newtonian Limit}: that the correct Newtonian limit be recovered in 
situations with low velocities $v \ll c$ and 
weak gravitational fields $\phi \ll c^2$.  Note that by (\ref{NewtR}) 
Poisson's equation of Newtonian gravity may now be written as 
\be
\mbox{\tt{R}\normalfont}_{00} = 4\pi G\rho \mbox{ } ,
\ee
which is suggestive that some curvature term should be equated to the energy-momentum causing 
the gravitation.

\mbox{ }

\noindent {\bf GR divergencelessness}: since $\mbox{\tt T\normalfont}_{AB}$ is conserved 
(divergenceless: $\nabla_A\mbox{\tt{T}\normalfont}^{AB} = 0$) and symmetric, this curvature 
term should also have these properties.  

\mbox{ }

\noindent Thus by the contracted Bianchi identity, the Einstein tensor 
$\mbox{\tt{G}\normalfont}_{AB}$ is a good choice of curvature term. 

The above considerations are all physical.  But in fact the following mathematical {\bf GR 
Cartan simplicities} \cite{VCartan} are also required to axiomatize GR: that 
$\mbox{\tt G\normalfont}_{AB}^{\mbox{\scriptsize trial\normalsize}}$ contains at most 
second-order derivatives and is linear in these.  The {\bf GR Lovelock simplicities} 
\cite{Lovelocktensor} eliminated the linearity assumption in dimension $n \leq 4$.    
One should note that throughout $\Lambda g_{AB}$ is an acceptable 
second term on the left hand side by all these considerations.  Such a $\Lambda$ is a 
\it cosmological constant \normalfont which is thus a theoretically-optional feature, the need 
for which is rather an issue of fitting cosmological observations.  

\mbox{ }

\noindent\Large{\bf Appendix B ADM Split of GR}\normalsize

\mbox{ }

\noindent The EFE's may also be obtained from variation of the 
Einstein--Hilbert action \cite{EH} 
\be
\mbox{\sffamily I\normalfont} = \int \textrm{d}^4x \sqrt{|g|}
(\mbox{\tt{R}\normalfont} + \mbox{\sffamily L\normalfont}_{\mbox{\scriptsize Matter\normalsize}}) 
\mbox{ } .
\label{VEinsteinHilbert}
\ee 
Arnowitt, Deser and Misner (ADM) \cite{ADM} presuppose the spacetime $g_{AB}$ and split it w.r.t a sequence of spatial slices, according to 
\be
g_{AB} = 
\left(
\begin{array}{ll} 
\beta_k\beta^k - \alpha^2   &    \beta_j    
\\
                                           \beta_i            &    h_{ij}
\end{array}
\right) 
\mbox{ } .
\label{VADMsplit}
\ee
Here, $h_{ij}$ is the metric induced on the spatial slice, 
the \it lapse \normalfont $\alpha$ is the change in proper time as one 
moves off the spatial surface and the \it shift \normalfont $\beta_i$ is is the displacement 
in identification of the spatial coordinates between two adjacent slices.

\begin{figure}[h] 
\centerline{\def\epsfsize#1#2{0.4#1}\epsffile{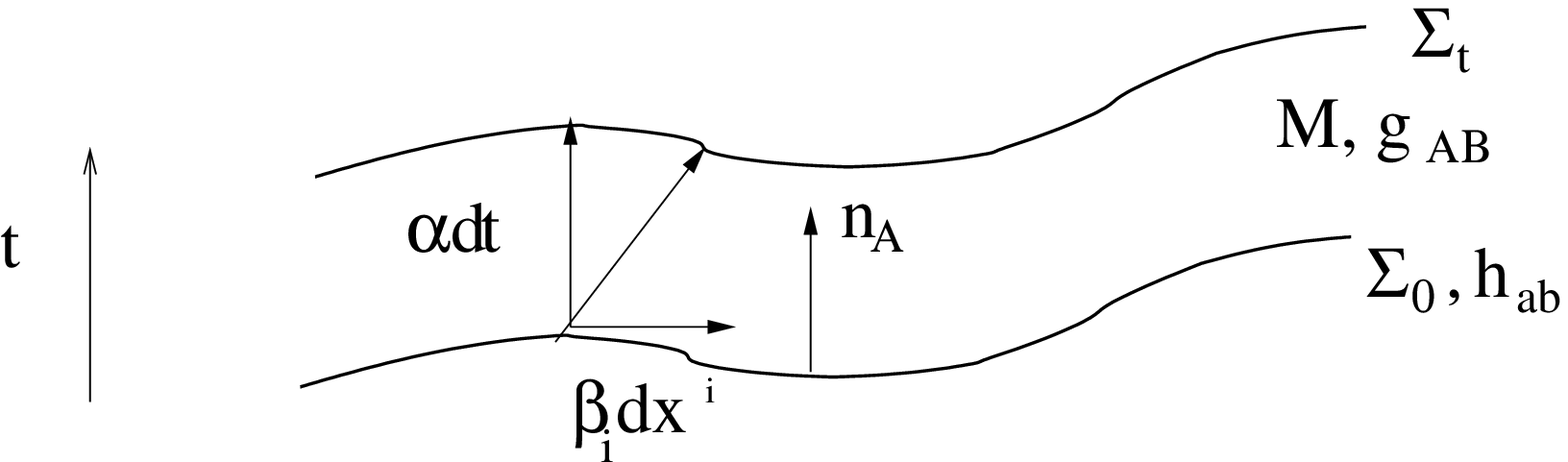}}
\caption[]{\label{TO3.ps}}
\end{figure}

For brevity I present just the vacuum case.  
Using Gauss' hypersurface geometry, (\ref{VEinsteinHilbert}) may be rewritten in split form as
$$
\mbox{\sffamily I\normalfont} = \int \textrm{d}t \int \textrm{d}^3x \alpha\sqrt{h}(  R + K\circ K - K^2)   
\mbox{ } 
$$
upon discarding a divergence, where 
$$
K_{ij} \equiv -\frac{1}{2\alpha}(\dot{h}_{ij} - \pounds_{\beta}h_{ij}) 
$$  
is the {\it extrinsic curvature}, and the dot is the 
derivative in the (time) direction perpendicular to the slice, $\frac{\pa}{\pa t}$.    
Using this last equation, one can arrive at the Lagrangian form of the split,
\be
\mbox{\sffamily I\normalfont} = 
\int \textrm{d}t \int \textrm{d}^3x\sqrt{h}\alpha
\left(
R + \frac{\mbox{\sffamily T\normalfont}^{\mbox{\scriptsize g\normalsize}}(\delta_{\check{\beta}} h_{ij})}{4\alpha^2}
\right) 
\mbox{ } , 
\mbox{\sffamily T\normalfont}^{\mbox{\scriptsize g\normalsize}} = (h^{ik}h^{jl} - h^{ij}h^{kl})\delta_{\check{\beta}} h_{ij}\delta_{\check{\beta}}h_{kl} 
\mbox{ } , 
\label{VBSWmethod}
\ee  
The BSW action of Sec 4 is the corresponding lapse-eliminated form.    

One may also pass to the Hamiltonian.  First, the conjugate momentum is
\be
p^{ab} \equiv \frac{  \pa\bar{\mbox{\sffamily L\normalfont}}  }{\pa\dot{h}_{ab}} = \frac{ \sqrt{h} }{ 2\alpha }
(h^{ac}h^{bd}  - h^{ab}h^{cd})\delta_{\check{\beta}}h_{bd} 
\mbox{ } .
\ee
Next, the Hamiltonian density is 
$$
\overline{    \mbox{\sffamily H\normalfont}    }(h_{ij}, p^{kl}; \alpha, p_{\alpha}; \beta_m, p^n_{\beta}) 
=  p\circ \dot{h} - 
\overline{    \mbox{\sffamily L\normalfont}    }(h_{ij}, \dot{h}_{kl}; \alpha, \dot{\alpha}; \beta_m, \dot{\beta}_n)   
= \int\textrm{d}^3x(\alpha{\cal H} + \beta^i{\cal H}_i) 
\mbox{ } ,  
$$
for
$$
\begin{array}{lr}
{\cal H} \equiv   \frac{1}{\sqrt{h}}
\left( 
p\circ p  - \frac{p^2}{2} 
\right) 
- \sqrt{h}R = 0 & \mbox{ } \mbox{ (Hamiltonian constraint)}  
\mbox{ } ,
\\
{\cal H}_i \equiv - 2 D_j{p^j}_i = 0 & \mbox{ } \mbox{ (momentum constraint)}.  
\end{array}
$$
Conventionally, $h_{ij}$, $\beta_i$ and $\alpha$ are regarded as canonical coordinates.  
There is no momentum associated with $\alpha$ nor $\beta_i$: these are Lagrange 
multiplier coordinates.  Thus the true gravitational d.o.f's in GR are contained in 
\be
\mbox{Riem} = \{\mbox{space of Riemannian 3-metrics $h_{ij}$}\} .
\ee
But the true d.o.f's are furthermore subjected to the Hamiltonian 
and momentum constraints.   

\mbox{ }

\noindent{\Large{\bf Appendix C: Interpretation of ADM formulation}}

\mbox{ }

\noindent Following on from App B, the Superspace interpretation part of Sec 4 is standard.  
I now consider Wheeler's interpretation of all this material.  He perceived 
\cite{geometrodynamica, WheelerGRT, Wheeler, MTW} that GR could be viewed as a theory of 
evolving 3-geometries: {\it geometrodynamics}.  
The central object of this scheme is the still-remaining vacuum Hamiltonian constraint ${\cal H}$.  
Whereas the momentum constraint is conceptually (if not technically) easy to deal with, ${\cal H}$ 
leads to the Problem of Time \cite{POT1, POT2} which plagues geometrodynamics.   
One question Wheeler asked \cite{Wheeler} is why ${\cal H}$ (or, strictly, the closely related 
Einstein--Hamilton--Jacobi equation) takes the form that it does, and whether arguments for this form could be made to 
rest on some `plausible first principles' without ever referring to the EFE's.   Contrary to the original attempted 
interpretation of vacuum geometrodynamics as a theory of everything \cite{geometrodynamica}, it is accepted that matter 
is to be `added on' rather than being an emergent property of gravitation alone.   Thus investigation of this question 
requires upgrading it to refer to the gravity-matter version of the equation as part of the required plausibility.  

\mbox{ }

The Hojman--Kucha\v{r}--Teitelboim (HKT) \cite{HKT} answer to this derives the form of 
${\cal H}$ from deformation algebra first principles.  The {\it deformation algebra} for a 
spacelike hypersurface is
\bea
\{              {\cal H}^{\mbox{\scriptsize d\normalsize}}(x),           {\cal H}^{\mbox{\scriptsize d\normalsize}}(y)          \} =
{\cal H}^{\mbox{\scriptsize d\normalsize}i}(x)\delta_{,i}(x,y) + {\cal H}^{\mbox{\scriptsize d\normalsize}i}(y)\delta_{,i}(x,y)\\
\label{DeformAlgebra1}
\{              {\cal H}_i^{\mbox{\scriptsize d\normalsize}}(x), {\cal H}^{\mbox{\scriptsize d\normalsize}}(y)                  \} = 
{\cal H}^{\mbox{\scriptsize d\normalsize}}(x)\delta_{,i}(x,y) \mbox{ } \mbox{ } \mbox{ }  \mbox{ } \mbox{ } \mbox{ } \mbox{ } \mbox{ } \mbox{ } \mbox{ } \mbox{ } \mbox{ } \\
\label{DeformAlgebra2}
\{              {\cal H}_i^{\mbox{\scriptsize d\normalsize}}(x), {\cal H}^{\mbox{\scriptsize d\normalsize}}_j(y)                \} = 
{\cal H}^{\mbox{\scriptsize d\normalsize}}_i(y)\delta_{,j}(x,y) + {\cal H}^{\mbox{\scriptsize d\normalsize}}_j(x)\delta_{,i}(x,y) 
\mbox{ } , 
\label{DeformAlgebra3}
\eea
It {\sl turns out} to be the Dirac Algebra of the constraints of GR.  

\begin{figure}[h]
\centerline{\def\epsfsize#1#2{0.4#1}\epsffile{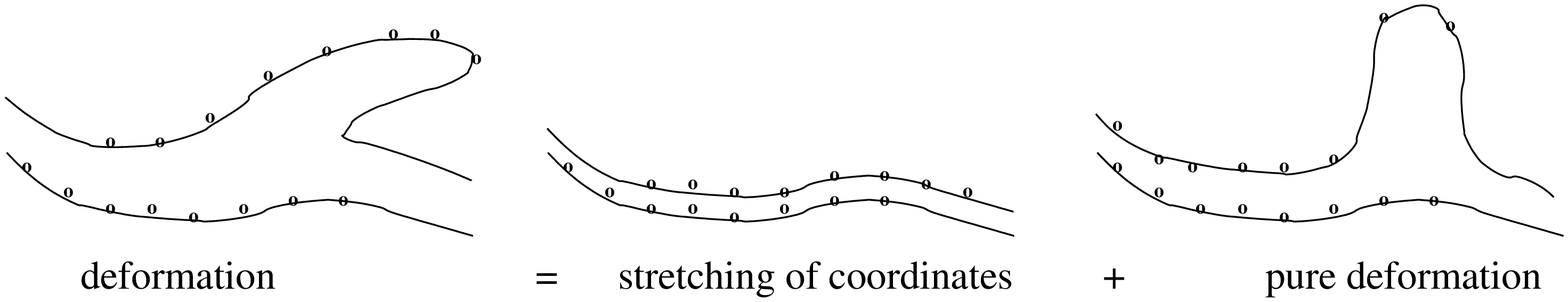}}
\caption[]{\label{TO7.ps}
\normalsize}
\end{figure}
${\cal H}^{\mbox{\scriptsize d\normalsize}}$ generates pure deformation (Fig 8), whereas 
${\cal H}^{\mbox{\scriptsize d\normalsize}}_i$ generates a stretching within the hypersurface 
itself.  In doing so, HKT are following Wheeler's advice of presupposing {\bf embeddability 
into spacetime} in order to answer his question, since HKT's first principles encapsulate 
embeddability.  HKT next demand the {\bf representation postulate}:  that the 
${\cal H}^{\mbox{\scriptsize trial\normalsize}}$,  
${\cal H}^{\mbox{\scriptsize trial\normalsize}}_i$ for a prospective gravitational theory 
close as the 
${\cal H}^{\mbox{\scriptsize d\normalsize}}$, ${\cal H}^{\mbox{\scriptsize d\normalsize}}_i$ 
do (i.e as the mathematical structure commonly known as the Dirac Algebra, but now regarded 
as {\sl emerging} as the deformation algebra).  See \cite{HKT} for further postulates and 
simplicities assumed.  As regards the plausibility criterion of robustness to the `adding on' 
of matter, Teitelboim \cite{Teitelboim} was able to include minimally-coupled scalars, 
electromagnetism and Yang--Mills theory into this approach.  

\mbox{ }

This chapter in large part addresses a distinct, 3-space approach principles answer to Wheeler's question.  

\mbox{ }

\noindent{\bf{\large Modelling assumptions}}

\mbox{ }

\noindent I need to explain and reasonably justify the modelling assumptions in this chapter.  
Throughout, compact without boundary (CWB) 3-spaces are used.  Einstein \cite{Ein34} argued 
for a closed universe; his arguments are based on simplicity and on attempting to realize 
Mach's principle (it is displeasing for `absolutist' boundary conditions at infinity to 
interfere with local physics).  Wheeler used the BSW formulation alongside CWB 3-spaces to 
conceptualize classical and quantum geometrodynamics \cite{W59, WheelerGRT, Wheeler, MTW} 
and to implement Mach's principle \cite{WheelerGRT}.  The conformal mathematics of the GR 
IVF was found to provide more adequate protection in the form of rigorous mathematical 
theorems than BSW's thin sandwich form.  This caused Wheeler later \cite{WIW} to shift the 
interpretation of Mach to be instead i.t.o constant mean curvature and conformal 3-geometry 
(see App D)  This led to the idea of `Wheeler--Einstein--Mach universes' \cite{Isenberg81}.  
A further simplicity argument for CWB geometries is then that the GR IVF is simplest for 
these \cite{CBY}.  

In addition to the above aesthetic arguments, it is of course crucial to know what observational 
restrictions can be placed on whether the universe is spatially CWB.  Whether the favoured cosmological model for the universe 
is open or closed has shifted around during the history of observational cosmology.  
One should note however that observationally-open universes may close on a larger-than-observed 
scale by gluing distinct patches together \cite{Lindquist} or by topological identification \cite{Levin}.  
Conversely, 3-spaces that look closed (by e.g multiple image or microwave background pattern 
criteria \cite{Levin}) could actually be open via hitherto unprobeably small holes leading to open 
regions.  So it appears that the aesthetic appeal of CWB spaces remains observationally unchecked.  

While spacetime is not always presupposed or emergent in this chapter, when it is, there is the additional consideration 
that the 3+1 split of it requires fixed spatial topology, time orientability and global hyperbolicity 
\cite{Wald}.  Time 
orientability is habitually supposed in the study of GR, i.e the existence of a well-behaved time function.  This includes 
disallowing regions with closed timelike curves.  Whereas many exact solutions in fact possess these, there is evidence of 
their classical instability in addition to conjectured quantum instability (Hawking's Chronology Protection Conjecture 
\cite{CPC}).  Global hyperbolicity, i.e the existence of a Cauchy surface $\Sigma$, is an assumption that includes the 
absense of naked singularities (the Penrose 1969 Cosmic Censorship Conjecture \cite{P6}), or indeed {\sl is} the more 
general Penrose 1979 Cosmic Censorship Conjecture \cite{P79}.  Whether GR is like this is both greatly important and, of 
course, unestablished.  The assumptions of this paragraph together imply spacetimes which are restricted to 
being of the form $\Sigma \times I$ for $I \subseteq \Re$ an interval.  

\mbox{ }

\noindent{\Large{\bf Appendix D York's Initial Value Formulation} }

\mbox{ }

\noindent Write ${\cal H}_i$ and ${\cal H}$ as
\be
D_ip^{ij\mbox{\scriptsize T\normalsize}} - D^j\frac{p}{3} = 0
\label{CODD}
\ee
\be
8\triangle\psi + \frac{1}{h}p \circ p\psi^{-7} - R\psi  + \frac{p^2}{h}\psi^5 = 0  
\mbox{ } , 
\label{LICHYORK}
\ee
where $h_{ij}$ and  
$p^{ij\mbox{\scriptsize T\normalsize}}  \equiv p^{ij} - \frac{p}{3}h^{ij}$ scale according to \cite{York}
\be
\widetilde{h}_{ij} = \psi^4h_{ij} 
\mbox{ } , \mbox{ } 
\widetilde{p}^{ij\mbox{\scriptsize T\normalsize}} = \psi^{-4}p^{ij\mbox{\scriptsize T\normalsize}}
\mbox{ } , \mbox{ } 
\ee
while
\be
\frac{p}{\sqrt{h}} = \mbox{spatial constant}
\ee
does not scale.  
This last condition involves making the choice to work on a constant mean curvature 
(CMC) slice.  The metric $h_{ij}$ used to write the equations down only corresponds to the 
physical metric up to the scale $\psi$. 
$p^{ij\mbox{\scriptsize T\normalsize}}$ is further decomposed according to 
$p^{ij\mbox{\scriptsize T\normalsize}} = p^{ij\mbox{\scriptsize TT\normalsize}} 
+ p^{ij\mbox{\scriptsize TL\normalsize}}$, 
$D_ip^{ij\mbox{\scriptsize TT\normalsize}} = 0$,  $p^{ij\mbox{\scriptsize TL\normalsize}} = 
2\left(D^{(i}W^{j)} - \frac{1}{3}h^{ij}D_kW^k\right)$
$p^{ij\mbox{\scriptsize TT\normalsize}}$ is also taken to be known. 

Solve (\ref{CODD}) for the potential $W^i$.  
This equation is conformally invariant and so decoupled from solving the well-behaved 
Lichnerowicz--York equation (\ref{LICHYORK}) [which is Lichnerowicz's original equation 
\cite{Lich} in the maximal ($p = 0$) subcase] for the physical scale factor $\psi$.  The 
generalization to include phenomenological and fundamental matter also works \cite{OY73, CBY}.  

Maintenance of CMC slicing in evolution requires the CMC LFE  
\be 
D^2N- N\left(R + \frac{p^2}{4h}\right) = \mbox{spatial constant}.
\label{CMCLFE}
\ee
to hold.  {\sl In GR, adopting this evolution is a gauge choice.}  Moreover, such a choice 
is not always possible.  

\mbox{ } 


\end{document}